\documentclass[twocolumn,twocolappendix]{aastex701}
\usepackage{natbib}
\usepackage{rotating}
\usepackage{multirow}
\usepackage{url}
\usepackage{amssymb}
\usepackage{comment}
\usepackage{adjustbox}
\usepackage[graphicx]{realboxes}
\usepackage{xspace}

\bibpunct{(}{)}{;}{a}{}{,} % to follow the A&A style

\usepackage[flushleft]{threeparttable}

\usepackage{caption}
\usepackage{graphicx}
\usepackage{subcaption}  % more modern than subfigure
\usepackage{hyperref}
\usepackage{booktabs}
\definecolor{green}{rgb}{0,0.5,0}

\newcommand{\HH}{\mbox{H$\rm _2$}}

\newcommand{\lya}{\mbox{${\rm Ly}\alpha$}}

\newcommand{\HI}{\ion{H}{1}}
\newcommand{\HII}{\ion{H}{2}}
\newcommand{\HeII}{\ion{He}{2}}

\newcommand{\CII}{\ion{C}{2}}
\newcommand{\CIV}{\ion{C}{4}}

\newcommand{\NI}{\ion{N}{1}}
\newcommand{\NII}{\ion{N}{2}}
\newcommand{\NV}{\ion{N}{5}}

\newcommand{\OI}{\ion{O}{1}}
\newcommand{\OII}{\ion{O}{2}}
\newcommand{\OIII}{\ion{O}{3}}

\newcommand{\SII}{\ion{S}{2}}
\newcommand{\SIII}{\ion{S}{3}}

\newcommand{\SiII}{\ion{Si}{2}}
\newcommand{\SiIV}{\ion{Si}{4}}

\newcommand{\MgII}{\ion{Mg}{2}}

\newcommand{\FeII}{\ion{Fe}{2}}
\newcommand{\FeIII}{\ion{Fe}{3}}

\newcommand{\ArIII}{\ion{Ar}{3}}

\newcommand{\MnII}{\ion{Mn}{2}}
\newcommand{\NiII}{\ion{Ni}{2}}
\newcommand{\PII}{\ion{P}{2}}

\begin{document}

\title{Unveiling Metal Mixing in a Grand-Design Spiral: \\ A UV-optical multiphase spatially resolved study of M83}
\shorttitle{Unveiling Metal Mixing in a Grand-Design Spiral}

\correspondingauthor{Adarsh Ranjan}

\author[0000-0001-9882-1576]{Adarsh Ranjan}
\affiliation{Space Telescope Science Institute,
3700 San Martin Drive, Baltimore, MD 21218, USA}
\email[show]{aranjan@stsci.edu}

\author[0000-0003-4372-2006]{Bethan L. James}
\affiliation{European Space Agency (ESA), ESA Office; Space Telescope Science Institute, 3700 San Martin Drive; Baltimore, MD 21218, USA}
\email{BETHAN.EMAIL@ADDRESS}

\author[0000-0003-4857-8699]{Svea Hernandez}
\affiliation{AURA for the European Space Agency, ESA Office, STScI,
3700 San Martin Drive, Baltimore, MD 21218, USA}
\email{SVEA.EMAIL@ADDRESS}

\author[0000-0001-9719-4080]{R. Rickards Vaught}
\affiliation{Space Telescope Science Institute,
3700 San Martin Drive, Baltimore, MD 21218, USA}
\email{RICKARDS-VAUGHT.EMAIL@ADDRESS}

\author[0000-0002-5320-2568]{Nimisha Kumari}
\affiliation{AURA for the European Space Agency, ESA Office, STScI,
3700 San Martin Drive, Baltimore, MD 21218, USA}
\email{NIMISHA.EMAIL@ADDRESS}

\author[0000-0002-9654-2108]{Alessandra Aloisi}
\affiliation{Space Telescope Science Institute,
3700 San Martin Drive, Baltimore, MD 21218, USA}
\affiliation{Astrophysics Division, Science Mission Directorate,
NASA Headquarters, 300 E Street SW, Washington, DC 20546, USA}
\email{ALESSANDRA.EMAIL@ADDRESS}

\author[0000-0002-6091-7924]{Peter Zeidler}
\affiliation{AURA for the European Space Agency, ESA Office, STScI,
3700 San Martin Drive, Baltimore, MD 21218, USA}
\email{PETER.EMAIL@ADDRESS}

%DATE (IMPORTANT)
\date{\today}

\begin{abstract}

We present a spatially resolved, multiphase study of chemical enrichment around young star clusters (YSCs) in the nearby grand-design spiral M83 by combining far-ultraviolet(UV) absorption-line spectroscopy from HST/COS with cospatial optical spectroscopy from VLT/MUSE and LBT/MODS. Our sample includes 18 YSCs spanning spectroscopic ages of $\sim$1–6 Myr and galactocentric radii out to R/R$_{25}$=0.56. Neutral (\HI) abundances were derived from UV absorption-line spectroscopy and compared with ionised (\HII) abundances from reddening-corrected optical emission lines. Because auroral lines are not detected in all regions, we develop and apply an empirical multi-zone electron-temperature(T$_e$) calibration based on strong-line diagnostics to estimate T$_e$ and derive reliable nebular abundances.

We measure oxygen(O), sulphur(S), nitrogen(N), and iron(Fe) abundance tracing enrichment from distinct nucleosynthetic channels. The $\alpha$-elements (O and S) exhibit similar behaviour, consistent with enrichment by core-collapse supernovae, whereas Fe shows weaker variations, reflecting its delayed production by Type Ia supernovae. Nitrogen displays the largest phase offset(ionized-neutral), with enhancements of up to $\Delta$N/H$\sim$1.5 dex and $\Delta$N/O$>$1.5 dex in the ionised gas relative to the neutral phase, indicating localised enrichment by massive stars and inefficient mixing between gas phases on Myr timescales. While the ionised gas exhibits signatures of feedback-regulated chemical enrichment and large-scale abundance gradients, corresponding trends are weak or absent in the neutral gas, consistent with metals remaining largely confined to the immediate star-forming environment during the earliest stages of cluster evolution in a massive grand-design spiral.

\end{abstract}

\keywords{\uat{Galaxy chemical evolution}{580} --- \uat{Galaxy abundances}{574} --- \uat{Galaxy evolution}{594} --- \uat{Interstellar medium}{847} --- \uat{Chemical enrichment}{225} --- \uat{Ultraviolet spectroscopy}{2284}}

\section{Introduction} 

%The need for studying gas within galaxies
The study of gas within galaxies is central to understanding how stars form, evolve, and influence their environments. Gas in its different phases - molecular (\HH), neutral (\HI), and ionised (\HII)\footnote{For this work, ionised gas will always refer to \HII\ regions surrounding the young star clusters (with ionisation energy, 13.6 - 40 eV), unless specified otherwise. Additionally, neutral gas will always mean the larger \HI\, gas envelope surrounding both the star clusters and the ionised region (with ionisation energy, $\lesssim$13.6 eV)} - not only provides the raw material for star formation but also acts as the reservoir through which metals and energy are cycled across galactic and intergalactic scales. Apart from providing the raw material for star formation, neutral gas also acts as the reservoir through which metals and energy are cycled across galactic and intergalactic scales. Stars are born in molecular clouds, where dust shields pre-stellar cores from radiation and enables cooling, fragmentation, and the eventual formation of stellar nurseries \citep[][]{Whitworth_Bate_2002, Hopkins_Lee_2016}. As they evolve, stars enrich their surroundings through feedback, winds, and supernovae, redistributing metals into the interstellar medium (ISM) and beyond, through processes such as galactic fountains that connect the ISM with the circumgalactic (CGM) and intergalactic medium \citep[IGM; ][]{Tumlinson2017, Krumholz2019}. Yet, the cycling of metals between different gas phases remains poorly constrained, with evidence that inhomogeneities in enrichment and mixing depend strongly on the interplay between star formation, feedback, and gas flows \citep[][]{James2013a, Ritter2015, Corlies2018}. Recent advances in spatially resolved spectroscopy, particularly with HST and JWST, now allow us to probe at unprecedented detail, processes such as the small-scale chemical inhomogeneities that shape galaxy evolution \citep[][]{Maragkoudakis2022, Hernandez2021, Hernandez2023}. Together, these studies highlight why examining gas across multiple phases is essential: it is the dynamic medium that regulates star formation, chemical enrichment, and the long-term evolution of galaxies. 

Building on the need to study gas across multiple phases, an equally critical question is how metals mix within and between these reservoirs. Metals preserve the imprint of stellar populations, enrichment timescales, and feedback processes \citep[see review by][]{Maiolino_Mannucci_2019}. Metals also regulate gas cooling and star formation by providing efficient radiative pathways and shaping the thermal balance of the ISM \citep[][]{Maiolino_Mannucci_2019, Kewley2019}. However, their spatial distribution is far from homogeneous, as demonstrated by both simulations and observations revealing small-scale mixing and chemical inhomogeneities across galactic environments \citep[][]{Ritter2015, Corlies2018, James2020}. Observations also show localised chemical inhomogeneities arising from events such as starburst-driven enrichment, inflows of metal-poor gas, and outflows that redistribute enriched material into the halo and beyond \citep[][]{James2009, James2013a, Kumari2017, Kumari2018, Bresolin2019}. These signatures encode the history of different stellar populations - massive stars enriching on $\sim$Myr timescales versus intermediate, low-mass stars enriching on $\sim$Gyr scales - and reveal the efficiency with which galaxies recycle metals back into new generations of stars. Neutral gas, which often contains the bulk of a galaxy's baryonic mass, is especially important since it can act as a long-term reservoir of metals, storing the record of previous star formation episodes \citep[][]{Lebouteiller2013, Sacchi2016, Zhang2018}. High-resolution simulations confirm that the rate of chemical mixing depends strongly on an element's nucleosynthetic origin, further emphasizing that disentangling the abundance patterns of different species is key to reconstructing feedback timescales of different stellar populations and the cycling of matter among stars, the ISM, and the CGM \citep[][]{Emerick2019, Emerick2020, Peroux2020}. Thus, tracing how individual metals are distributed across the molecular, neutral, and ionised phases surrounding a star cluster is essential for connecting stellar evolution with galaxy-scale processes of enrichment, outflows, and chemical evolution \citep[][]{James_2014, Hernandez2019, James2026}. 

%The need for spatially resolved observations to study the evolution of stars and gas within the galactic environment
High-resolution studies of individual star clusters and their environments show that chemical enrichment and gas dispersal are tied to stellar ages and cluster feedback, requiring parsec-scale views to connect stellar populations to their natal ISM \citep[][]{Larsen2011, Calzetti2015, Westmoquette2013a}. More recently, spatially resolved analyses of multiphase gas and stellar populations in a Blue Compact Dwarf (BCD) galaxy, NGC\,5253, have provided new evidence that local environments and enrichment pathways strongly shape the distribution of metals \citep[][]{Abril-Melgarejo2024}. In this context, focusing on the spatially resolved environment of young star clusters (hereafter, YSCs) and their immediate \HII\ and surrounding \HI\, regions in an actively star-forming, metal-rich, nearby grand-design spiral such as M83 is particularly powerful. This system offers a natural counterpart to the low-mass, low-metallicity BCD like NGC 5253, enabling us to explore \textit{how galaxy mass, metallicity, and stellar feedback strength regulate the efficiency of chemical enrichment and mixing}. Studying a high-metallicity, high-mass spiral complements the previous works in the literature related to chemically young systems, revealing how the same enrichment pathways can operate under vastly different physical conditions. Hence, to fully unravel the interplay between metals, stellar populations, and gas flows, there is a clear need for spatially resolved, multiphase observations across a broad range of environments that span both metallicity and stellar age. M83's nearly face-on geometry offers a unique `down-the-barrel' vantage point for tracing how freshly produced metals are dispersed within and above the galactic plane, providing a direct window into the mechanisms that regulate feedback-driven enrichment in massive spiral galaxies. Additionally, unlike NGC 5253, M83's evolved stellar populations and deeper gravitational potential provide a unique opportunity to examine whether enriched material can escape or instead remain trapped within different gas phases and/or the galactic disk.

\subsection{The M83 galaxy}
\begin{figure*}
\includegraphics[page=1,width=1.0\linewidth,trim={10 60 40 40},clip]{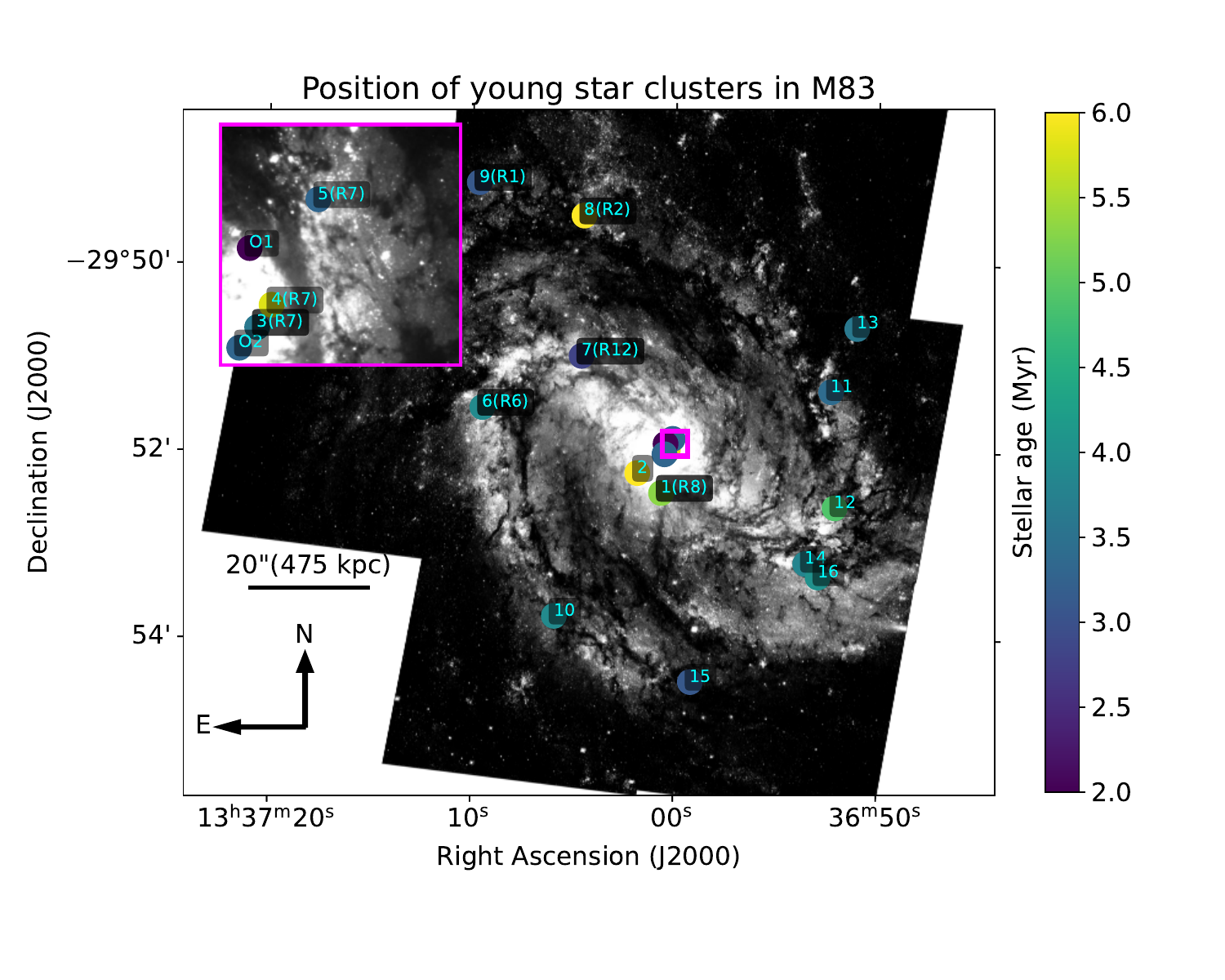}
\caption{HST (WFC3) image of M83 shown in greyscale. The position of the HST/COS observed young star clusters is overplotted as circles. The color fill of the circles represents the UV-spectra fitted average age (in Myr) of the younger stellar population. The coordinates (RA/Dec) are given in J2000 format. Note: The circle sizes do not represent the HST/COS aperture, which is too small for clear representation in this figure. All YSCs are labelled with numbers, `X', `OX', and `(RX)' representing YSCs with ID: `M83-X', `M83-POS-X', and `R(X)' respectively (See Table.~\ref{m83_basic_info}). The galactic center is highlighted with a magenta rectangle in the main figure, and a zoomed-in sub-figure in the top-left corner shows the labels for the 5 clusters within the galactic center. The sub-figure has a different flux normalization than the main image for clarity. The stellar ages shown here are values obtained from STARBURST99 models. For related stellar age uncertainty, please see Table~\ref{tab:3}. \label{fig:1}}
\end{figure*}

M83 (also known as NGC 5236, see Figure.~\ref{fig:1}) is a nearby, metal-rich, grand-design spiral galaxy (see Table.~\ref{tab:1} for details) that serves as a cornerstone for understanding the interplay between stars and gas in actively metal-rich star-forming environments. Optical spectroscopy has long established its high-metallicity \HII\ regions and young stellar populations \citep[][]{Bresolin2002, Bresolin2009, Bresolin2016}, while large IFU surveys have revealed the impact of stellar feedback on ionised gas pressures and outflows \citep[][]{Della_Bruna2022a, Della_Bruna2022b}. Far-ultraviolet spectroscopy of M83 and similar galaxies has demonstrated that a significant fraction of metals reside in the neutral ISM, highlighting the importance of probing multiple phases \citep[][]{Aloisi2003, Lebouteiller2009, James_2014}. Integrated-light studies of young massive clusters found supersolar metallicities and shallow metallicity gradients, consistent with sustained enrichment in the central disk \citep[][]{Hernandez2018}. Spatially resolved comparisons of oxygen abundance in the stellar, ionised, and neutral components show that it is well mixed on $\sim$100 pc scales, though gas and stellar gradients differ slightly, pointing to long mixing timescales \citep[][]{Hernandez2021}. More recent work has extended these analyses to multiphase diagnostics, underscoring the influence of local environments on abundance patterns. High-resolution ALMA surveys have mapped molecular structures across the disk \citep[][]{Koda2023, Leroy2021}, while JWST has now resolved warm molecular gas, ionised emission, and feedback around young clusters in the nucleus, directly connecting enrichment to star formation at parsec scales \citep[][]{Jones2025, Hernandez2025}. Together, these studies establish M83 as a benchmark system for tracing the small-scale physics of star formation and chemical evolution in a massive spiral galaxy.

\begin{table}[ht]
\centering
\caption{General Parameters for M83 \label{tab:1}}
\begin{threeparttable}
\begin{tabular}{ll}
\textbf{Parameter} & \textbf{Value} \\
\midrule
R.A. (J2000.0) & 204.253958$\rm ^{\circ}$ \\
Decl. (J2000.0) & -29.865417$\rm ^{\circ}$ \\
Distance\tnote{a} & 4.9 Mpc \\
Morphological type & SAB(s)c \\
$R_{25}$\tnote{b} & 6.44$^{\prime}$ (9.18 kpc) \\
Inclination\tnote{b} & 24$^{\circ}$ \\
Position angle\tnote{c} & 45$^{\circ}$ \\
Heliocentric radial velocity & 512.95 km s$^{-1}$ \\
\bottomrule
\end{tabular}
\begin{tablenotes}
\small
\item \textbf{Notes.} All parameters from the NASA Extragalactic Database (NED), except where noted.  
\item[a] \citep[][]{Jacobs2009} 
\item[b] \citep[][]{deVaucouleurs1991}
\item[c] \citep[][]{Comte1981}
\end{tablenotes}
\end{threeparttable}
\end{table}

We will follow up on these literature studies by conducting a detailed investigation of elemental abundances across the multiphase ISM (\HI, \HII\ regions) surrounding the YSCs in M83. Unlike previous single-phase or single-element analyses, we aim to look at the evolution of multiple elements that trace distinct nucleosynthetic pathways and release timescales, adding a fourth temporal dimension to the study of metal-mixing. Oxygen and sulphur are $\alpha$-elements mainly produced within core-collapse supernovae (CCNe) and released promptly on short ($\rm \sim$3-50 Myr) timescales, tracing rapid enrichment from massive stars\citep[][]{Zapartas2017, Kankare2021}. Iron, on the other hand, arises mainly from Type Ia supernovae on a much longer ($\rm \sim$100 Myr - 1 Gyr) timescale, providing a record for slower, cumulative enrichment. Since our star clusters are very young ($<$10 Myr), comparison with iron will provide a unique picture of the previous cycles of star formation. Additionally, iron is most depleted onto dust, providing a proxy for dust content within the medium \citep[see e.g.,][]{Roman-Duval2022a}. Nitrogen injection is more complicated. Nitrogen can be promptly injected via winds from massive stars\citep[see][]{Martins2024}, Wolf-Rayet (W-R) stars \citep[see][]{Schaerer1997, Westmoquette2013b} and Very Massive Stars \citep[VMS, see][]{Vink2023} on a very short ($\rm \sim$2-5 Myr) timescale. A secondary, more gradual channel in nitrogen enrichment is through the Asymptotic Giant Branch (AGB) stars after $\rm \sim$ 250 Myr \citep{Kobayashi2020}. By studying these elements together, we gain both spatial, temporal and morphological(stellar) perspectives on how metals are produced, transported and mixed within a metal-rich grand-design spiral galaxy. The literature is also rich with contrasting samples of YSCs from a BCD galaxy \citep[NGC 5253, see][]{Abril-Melgarejo2024} and a collection of low-metallicity star-forming galaxies from the CLASSY Survey \citep[][]{James2026}, which we will use in this work for a comparative analysis with a metal-rich grand-design spiral.

%Section description and cosmology
This work is further divided into five sections. Section~\ref{observations} describes the multi-wavelength observations used in this work along with the detailed process of data extraction. Section~\ref{data_analysis} describes the relevant analysis for different datasets performed to obtain the gas (chemical abundances and kinematics of \HI\ and \HII\ regions) and stellar properties within each YSC and their surrounding \HI, \HII\ regions. Section~\ref{sec:Results} describes the results of the analysis on the properties of gas and stars. The detailed discussion of these results along with physical conditions is written in Section~\ref{discussion}. Finally, we summarize our findings in Section~\ref{Conclusion}. For all cosmological calculations (e.g., luminosity distances), we follow the $\Lambda$CDM model \citep[][]{Planck2016} throughout the paper.

%-----------------------------------------------------------OBSERVATIONS / DATA REDUCTION ------------------------------------------------------------------------------------------------

\section{Observations and data extraction \label{observations}}

\subsection{FUV spectra from HST/COS}
\begin{figure*}[ht]
    \centering
    \begin{subfigure}{0.48\linewidth}
        \includegraphics[width=\linewidth,trim={10 0 10 20},clip]{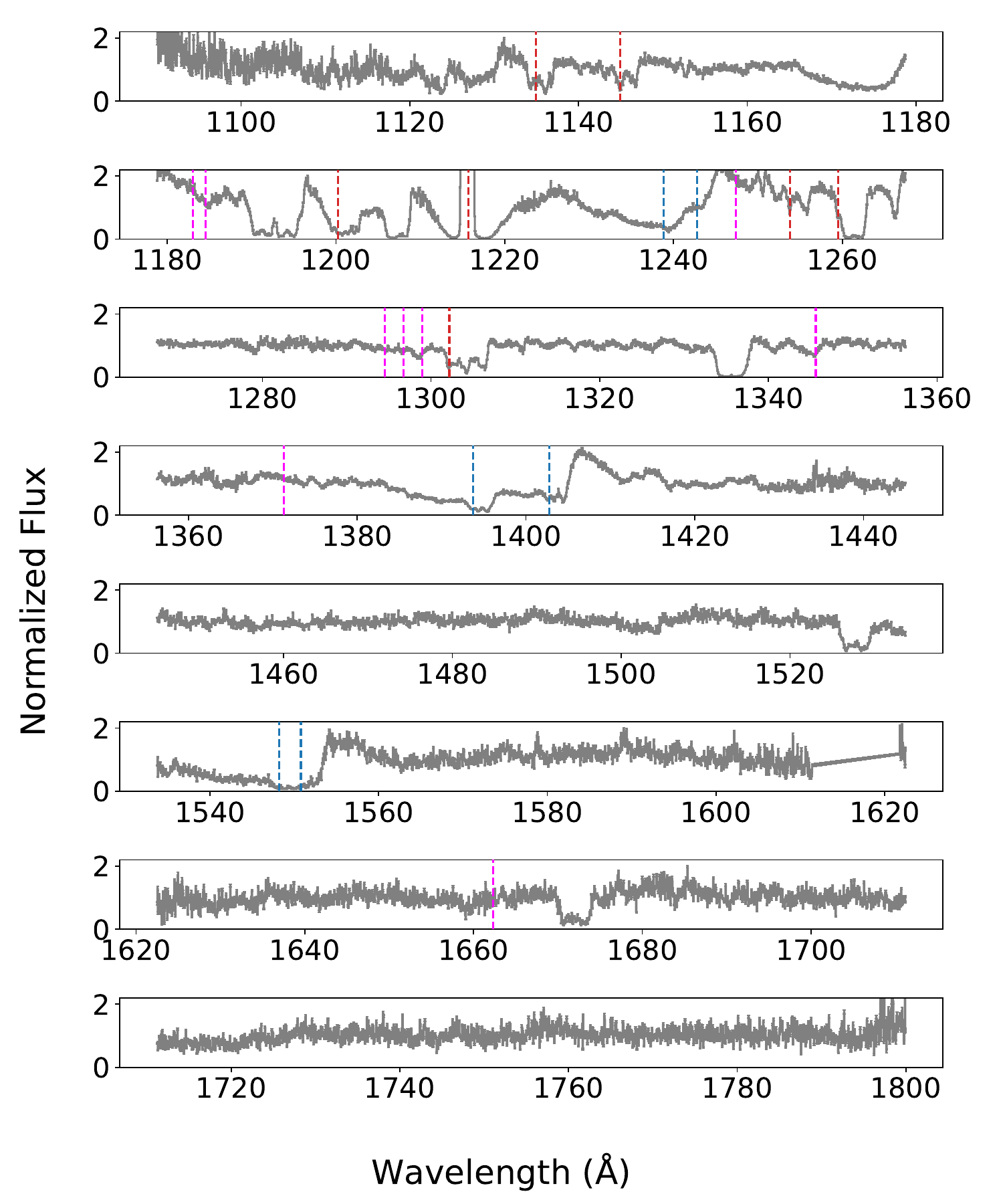}
        \caption{HST/COS Spectra}
        \label{fig:2_1}
    \end{subfigure}
    \vspace{0.1em} % space between rows
    \begin{subfigure}{0.48\linewidth}
        \includegraphics[width=\linewidth,trim={10 0 10 20},clip]{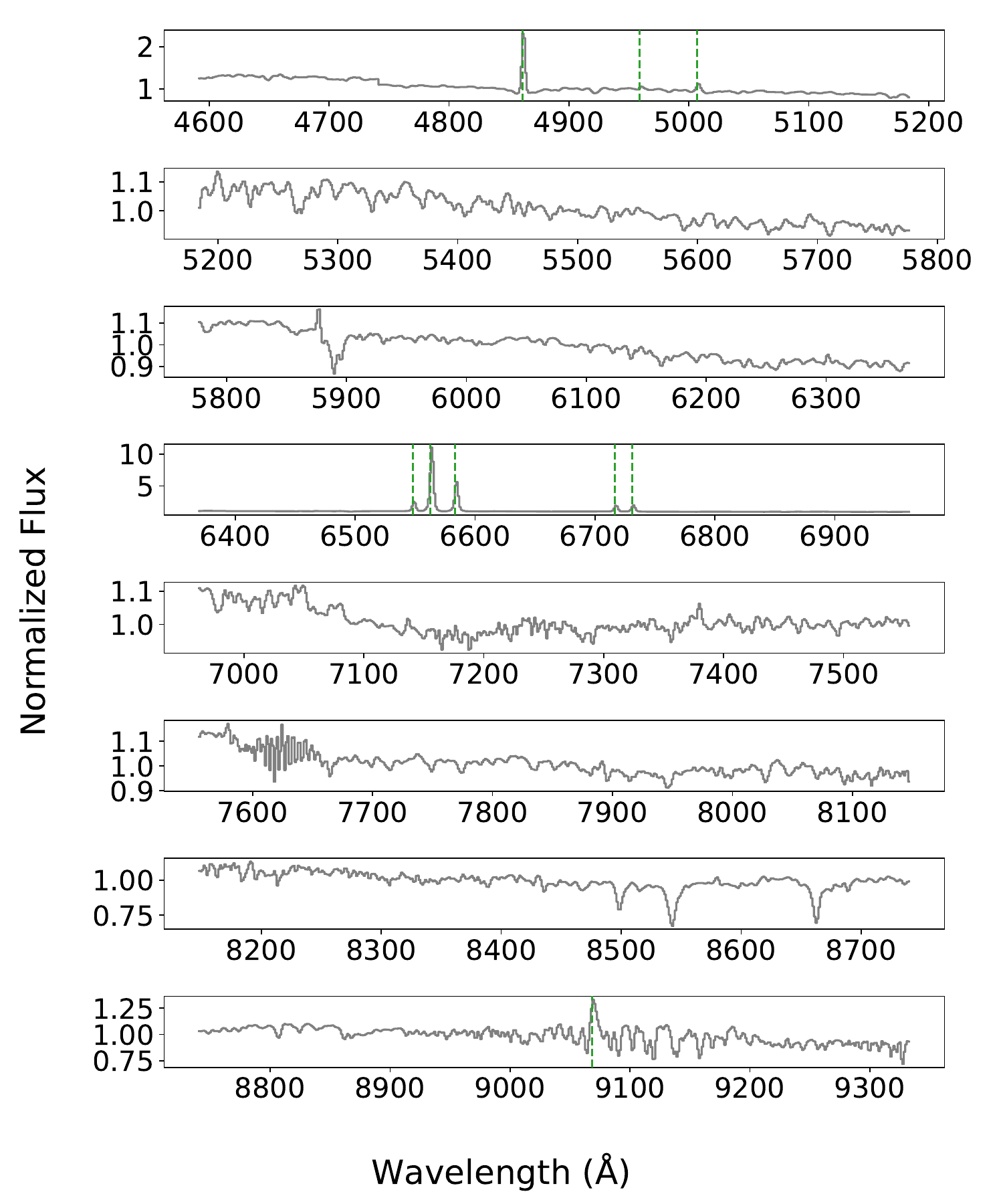}
        \caption{VLT/MUSE Spectra}
        \label{fig:2_3}
    \end{subfigure}
    \caption{HST/COS FUV spectra (left) and VLT/MUSE optical spectra (right) observed towards YSC M83-POS-1 in grey. The spaxels from VLT/MUSE IFU observations have been extracted within the COS aperture to make both spectra co-spatial. The blue, magenta, red and green dashed vertical lines show the location of the prominent lines from wind+ISM, stellar photospheric, ISM absorption, and nebular emission features, respectively. \label{fig:2}}
\end{figure*}

A set of 16 young star clusters was observed using HST/COS with the G130M/1291 and G160M/1623 settings as part of the main program for this study, PID: 14681 (PI: Aloisi). Two additional YSCs from the centre of M83 were observed with other programs (PIDs: 11579 and 15193, PI: Aloisi). We added these to the main program as complementary datasets. The observed data primarily cover the wavelength range of $\sim$1132-1800\AA\, except the two archival targets (called M83-POS1 and M83-POS2) with a slightly wider wavelength range of $\sim$1066-1800~\AA\ (see e.g. Fig.~\ref{fig:2_1}). With the availability of a background UV source (the bright star clusters), the observed spectra reveal absorption lines originating from neutral gas towards these YSCs. These lines give us access to different ions from within the neutral gas: \HI, \OI\, \NI, \FeII, \SII\, and \PII. In addition, the UV spectra give access to important stellar features that trace the physical properties of young stellar populations: e.g., strong stellar-wind lines such as \NV\,$\lambda$1240, \SiIV\,$\lambda$1400, and \CIV\,$\lambda$1550 exhibit broad, blue-shifted P-Cygni profiles from massive O and B stars, providing information about their stellar ages, metallicities and feedback strength. These and other similar features \citep[see table~1 of][for an exhaustive list]{Leitherer2011} can aid in the characterisation of the properties of the stellar population. For HST/COS observations, a source is considered extended if its full width at half maximum (FWHM) is larger than approximately 0.6" \citep[see][]{payne2026cosmic}. We note that most of our YSCs (except M83-7, 8, 9, 14, 15, and 16) are extended. Hence, the spectral FWHM (vFWHM) varies not only as a function of wavelength, but also as a function of the spatial extension of the source. For this, we calculated the spectral FWHM for each of the extended sources separately. The process of obtaining the spectral FWHM is described in detail in the Appendix: Section~\ref{varying_cos_res}. 

\subsection{Optical spectra from VLT/MUSE}
We use optical integral field unit (IFU) spectroscopic data from the Multi Unit Spectroscopic Explorer (MUSE) on the Very Large Telescope (VLT). The VLT/MUSE archival data cover 14 of the 18 YSCs in M83 observed with HST/COS (see, e.g., Fig.~\ref{fig:2}). The MUSE observations were taken from various ESO programs: 096.B-0057(A), 0101.B-0727(A) (PI Adamo), 097.B-0899(B) (PI Ibar), and 097.B-0640(A) (PI Gadotti), and combined to cover the spectral range: 4600-9380\AA\, with a spectral resolution of $\sim$1.2\AA ($\sim$40 km/s). Individual exposures cover a field of 1'$\times$1'; each spaxel is 0.2''$\times$0.2'' in extent \citep[see][for more information]{Long2022, Della_Bruna2022a, Della_Bruna2022b}. We locate the positions of the 14 YSCs from our COS sample within the MUSE data cubes and extract their optical spectra by summing the spectra within circular apertures of 2.5" diameter, centred on the HST/COS pointings. In M83, high metallicity leads to cooler gas, and hence, the auroral lines in optical spectra are weak. However, any detection of auroral lines is crucial to obtain the electron temperature within different phases of the \HII\ region via the `direct method'. As such, to prevent any erroneous removal of these weak features, we had to refrain from directly using the standard stellar fitting codes like \textbf{ppxf} \citep[see][]{Cappellari2023}: the uncertainty in the ppxf-fitted stellar continuum washes out the weak auroral lines. While ppxf was used as a reference for the stellar continuum, the final continuum fit near weak optical emission lines was fine-tuned by hand.

\subsection{Optical spectra from LBT/MODS}
We further complement the optical coverage of the YSCs using archival spectra from the Multi-Object Double Spectrograph on the Large Binocular Telescope \citep[LBT/MODS][]{Pogge2010}. The LBT/MODS data of the \HII\ regions in M83 were acquired on 21$^{st}$ of May, 2018. We use observations from 6 different pointings that align with our HST/COS targets. We primarily use the VLT/MUSE spectra for all cases where it is available, while taking complementary additional information from LBT/MODS spectra. Since LBT/MODS data cover a much wider spectral range ($\sim$3100-9900$\AA$) as compared with VLT/MUSE, we have access to more emission lines (e.g. [OII]$\lambda\lambda$3726,3728 and [SIII]$\lambda$9530).
%We primarily use VLT/MUSE pointings for all datasets where we detect auroral lines.
%and [\OII]$\lambda\lambda$7320,7321,7330,7331 quadruplet. 
%For all other cases, we use additional information from LBT spectra for the same target. 
Table.~\ref{m83_basic_info} shows the common pointings between LBT/MODS and VLT/MUSE spectra. More information about the observations and data reduction can be found in \citep[][]{Hernandez2021}. The wavelength solutions for LBT MODS spectra vary between the blue ($\lesssim$5500~\AA) and the red ($\gtrsim$5500~\AA) arms. While the multi-component fitting allows us to obtain a robust estimate of the total emission line flux, we refrain from using the kinematical information from LBT/MODS and always obtain the kinematical information of ionised gas from VLT/MUSE.
%Hence, they were fitted separately to avoid issues while tying the kinematic structure between different \HII\ region emission lines that fall within the red and the blue arms of the spectra. 
To ensure consistent flux calibration between VLT/MUSE and LBT/MODS, we rescaled the LBT/MODS spectra by matching their continuum level (using a multiplicative factor) to that of the corresponding VLT/MUSE spectra within the common spectral range. This correction accounts for small flux offsets arising from the different slit and IFU apertures and ensures that the line ratios are consistent between LBT and MUSE spectra. Similar to VLT, the stellar continuum levels were fine-tuned by hand for regions with weak emission lines after referring to an initial estimate of the stellar continuum from ppxf. We also ensured that possible Balmer absorption from H$\gamma$ was taken into account before fitting the [\OIII]$\lambda$4363 profile.  \\

\begin{table}[]
\centering
\setlength{\tabcolsep}{3pt}   % default is ~6pt
\begin{tabular}{lllll}
\hline
YSC        & RA          & Dec          & R/R$_{25}$ & E(B-V)$\rm _{MW}$ \\ 
\hline
M83-1(R8)  & 204.2527583 & -29.87391111 & 0.08   & 0.0674     \\
M83-2      & 204.2576792 & -29.87038333 & 0.06   & 0.0667     \\
M83-3(R7)  & 204.2517375 & -29.86662222 & 0.02   & 0.0667     \\
M83-4(R7)  & 204.2514333 & -29.86620556 & 0.02   & 0.0667     \\
M83-5(R7)  & 204.25045   & -29.86425    & 0.04   & 0.0666     \\
M83-6(R6)  & 204.2895625 & -29.85880833 & 0.34   & 0.0659     \\
M83-7(R12) & 204.2692667 & -29.84959167 & 0.21   & 0.0659     \\
M83-8(R2)  & 204.2688667 & -29.82460556 & 0.41   & 0.0654     \\
M83-9(R1)  & 204.2904083 & -29.81878333 & 0.56   & 0.0651     \\
M83-10     & 204.2746    & -29.89590556 & 0.34   & 0.0683     \\
M83-11     & 204.2179875 & -29.85573889 & 0.35   & 0.0674     \\
M83-12     & 204.2171125 & -29.87641111 & 0.36   & 0.0687     \\
M83-13     & 204.2127208 & -29.84446389 & 0.43   & 0.0671     \\
M83-14     & 204.2230583 & -29.886375   & 0.35   & 0.0695     \\
M83-15     & 204.2465708 & -29.90752222 & 0.4    & 0.0716     \\
M83-16     & 204.2204833 & -29.88874722 & 0.38   & 0.0697     \\
M83-POS-1  & 204.2519088 & -29.86516111 & 0.02   & 0.0666     \\
M83-POS-2  & 204.2521171 & -29.86700556 & 0.02   & 0.0667     \\
\hline
\end{tabular}
\caption{Information about YSCs in M83 studied in this work. The target IDs (with LBT/MODS target ID in brackets, if available) are given with their corresponding RA and Dec coordinates and the galactocentric radius of each YSCs studied in this work from M83 \citep[][]{Hernandez2021}. We also list here the galactic (MW) dust extinction in E(B-V) towards individual YSC sightlines calculated assuming the \citet{Gordon2023} Milky Way R(V) dependent dust model with R(V)=3.1.
\label{m83_basic_info} }
\end{table}

\section{Data analysis \label{data_analysis}}

This section describes the different analyses done on observations to obtain properties of young star clusters and the surrounding media (both ionised and neutral gas phases). For spectra obtained from all instruments: HST, VLT and LBT, we correct for foreground (Milky Way) dust extinction using the \citet{Gordon2023} Milky Way R(V) dependent model with (R(V)=3.1) and galactic E(B-V)$\rm _{MW}$ (values in Table.~\ref{m83_basic_info}) calculated uniquely for each YSC line of sight using the Schlegel, Finkbeiner \& Davis (1998) Galactic dust reddening maps \citep[see][]{Schlegel1998}.

\subsection{Obtaining stellar properties \label{stellar_fitting_main}}

Stellar age and metallicity for all YSCs in M83 studied here were initially reported in \citet{Hernandez2019}. They used the integrated-light method by \citet{Larsen2012} to obtain the stellar metallicities, while adopting ages inferred through photometric techniques. For our study, to simultaneously estimate stellar ages, metallicities, masses and reddening values, we fitted STARBURST99\citep[][]{Hawcroft2025} high-resolution stellar models to the HST/COS UV spectra using the stellar fitting code: SESAMME \citep[][]{Jones2023}. For this work, we adopted the inferred metallicities and photometric ages reported in \citet{Hernandez2019} as priors for our analysis.

A detailed description of literature studies about the stellar parameters, the method used in this study, as well as comparing the values obtained in this work with the values from literature, is given in Section.~\ref{stellar_fitting}. Figure~\ref{fig:4} shows the results of the single-population stellar fitting for the HST/COS spectra of YSC M83-POS-1. Table.~\ref{tab:3} lists the fitted stellar parameters for all YSCs. We infer a range of stellar properties: 1.6 Myr $<$ age $\lesssim$ 6 Myr, -0.27 $<$ [Z](=log10(Z/Z$\rm _{\odot}$)) $<$ 0.54, 0.01 $<$ dust (=E(B-V) in mag) $<$0.84 and $\rm 2.1\times\,10^{4}\,M_{\odot}\,<$ mass $\rm < 10\times\,10^{5}\,M_{\odot}$. We take Z$\rm _{\odot}$=0.0134 from \citet{Asplund+09}.

\begin{table*}[]
\centering
\begin{tabular}{lllllll}
\hline \\
YSC       & [Z]       & E(B-V)         & Age        & Mass        & L$_{Bol}$ & Wind mom  \\ 
\hline \\
\hline \\
M83-1             & 0.48$^{+0.06}_{-0.07  }$ & 0.35$^{+0.03}_{-0.00 }$ & 5.30$^{+9.70}_{-0.20 }$ & 5.70$^{+16.30}_{-0.40  }$ & 40.33$^{+0.26}_{-0.20 }$ & 29.09$^{+0.90}_{-0.62}$ \\
M83-2             & -0.15$^{+0.11}_{-0.10 }$ & 0.30$^{+0.01}_{-0.01 }$ & 6.00$^{+0.40}_{-0.40 }$ & 13.00$^{+1.00}_{-2.00  }$ & 40.93$^{+0.04}_{-0.02 }$ & 29.28$^{+0.07}_{-0.14}$ \\
M83-3             & 0.27$^{+0.06}_{-0.06  }$ & 0.20$^{+0.01}_{-0.01 }$ & 3.60$^{+0.50}_{-0.70 }$ & 63.00$^{+3.00}_{-28.00 }$ & 42.20$^{+0.13}_{-0.12 }$ & 31.65$^{+0.15}_{-0.24}$ \\
M83-4             & -0.04$^{+0.06}_{-0.07 }$ & 0.32$^{+0.01}_{-0.03 }$ & 5.80$^{+0.40}_{-0.60 }$ & 93.00$^{+7.00}_{-42.00 }$ & 41.65$^{+0.06}_{-0.05 }$ & 30.83$^{+0.22}_{-0.24}$ \\
M83-5             & -0.09$^{+0.46}_{-0.18 }$ & 0.40$^{+0.04}_{-0.03 }$ & 3.30$^{+0.80}_{-0.60 }$ & 8.50$^{+4.50}_{-2.50   }$ & 41.28$^{+0.11}_{-0.14 }$ & 30.72$^{+0.14}_{-0.26}$ \\
M83-6             & 0.08$^{+0.05}_{-0.05  }$ & 0.26$^{+0.00}_{-0.03 }$ & 3.90$^{+0.50}_{-0.50 }$ & 21.00$^{+1.00}_{-9.00  }$ & 41.68$^{+0.07}_{-0.07 }$ & 31.35$^{+0.34}_{-0.37}$ \\
M83-7             & 0.33$^{+0.04}_{-0.20  }$ & 0.33$^{+0.01}_{-0.02 }$ & 2.80$^{+1.10}_{-0.60 }$ & 28.00$^{+13.00}_{-6.00 }$ & 42.04$^{+0.07}_{-0.15 }$ & 31.82$^{+0.27}_{-0.39}$ \\
M83-8             & -0.07$^{+0.06}_{-0.07 }$ & 0.14$^{+0.03}_{-0.00 }$ & 6.00$^{+0.70}_{-0.30 }$ & 5.10$^{+3.10}_{-0.00   }$ & 40.63$^{+0.04}_{-0.04 }$ & 29.71$^{+0.11}_{-0.16}$ \\
M83-9             & -0.13$^{+0.14}_{-0.14 }$ & 0.10$^{+0.01}_{-0.01 }$ & 3.10$^{+0.70}_{-0.50 }$ & 2.30$^{+0.80}_{-0.20   }$ & 41.16$^{+0.07}_{-0.10 }$ & 30.59$^{+0.10}_{-0.17}$ \\
M83-10            & 0.21$^{+0.10}_{-0.11  }$ & 0.15$^{+0.00}_{-0.01 }$ & 3.90$^{+0.40}_{-0.70 }$ & 7.20$^{+0.10}_{-3.10   }$ & 41.13$^{+0.10}_{-0.08 }$ & 30.88$^{+0.27}_{-0.38}$ \\
M83-11            & 0.15$^{+0.15}_{-0.19  }$ & 0.11$^{+0.03}_{-0.00 }$ & 3.40$^{+1.30}_{-0.70 }$ & 3.70$^{+0.50}_{-1.50   }$ & 41.12$^{+0.17}_{-0.15 }$ & 30.53$^{+0.23}_{-0.37}$ \\
M83-12            & 0.13$^{+0.09}_{-0.09  }$ & 0.26$^{+0.00}_{-0.01 }$ & 4.90$^{+0.40}_{-0.70 }$ & 12.00$^{+0.00}_{-3.70  }$ & 41.05$^{+0.07}_{-0.08 }$ & 30.34$^{+0.13}_{-0.11}$ \\
M83-13            & 0.07$^{+0.08}_{-0.09  }$ & 0.19$^{+0.02}_{-0.02 }$ & 3.60$^{+0.60}_{-0.80 }$ & 5.30$^{+3.90}_{-0.70   }$ & 41.38$^{+0.13}_{-0.12 }$ & 31.26$^{+0.24}_{-0.28}$ \\
M83-14            & 0.15$^{+0.10}_{-0.09  }$ & 0.14$^{+0.01}_{-0.02 }$ & 3.80$^{+0.40}_{-0.70 }$ & 9.00$^{+0.80}_{-3.90   }$ & 41.42$^{+0.11}_{-0.08 }$ & 31.28$^{+0.18}_{-0.22}$ \\
M83-15            & 0.38$^{+0.13}_{-0.27  }$ & 0.24$^{+0.01}_{-0.03 }$ & 3.10$^{+0.70}_{-0.40 }$ & 6.80$^{+0.70}_{-2.00   }$ & 41.31$^{+0.08}_{-0.11 }$ & 31.18$^{+0.11}_{-0.14}$ \\
M83-16            & 0.17$^{+0.10}_{-0.08  }$ & 0.23$^{+0.02}_{-0.00 }$ & 4.00$^{+0.80}_{-0.40 }$ & 5.40$^{+2.40}_{-0.10   }$ & 41.27$^{+0.09}_{-0.08 }$ & 30.62$^{+0.19}_{-0.21}$ \\
M83-POS-1         & 0.46$^{+0.08}_{-0.07  }$ & 0.32$^{+0.00}_{-0.00 }$ & 2.00$^{+0.40}_{-0.40 }$ & 68.00$^{+1.00}_{-1.00  }$ & 42.48$^{+0.01}_{-0.02 }$ & 31.90$^{+0.03}_{-0.07}$ \\
M83-POS-2         & 0.10$^{+0.07}_{-0.06  }$ & 0.16$^{+0.01}_{-0.00 }$ & 3.30$^{+0.80}_{-0.60 }$ & 43.00$^{+30.00}_{-1.00 }$ & 42.34$^{+0.12}_{-0.12 }$ & 32.16$^{+0.20}_{-0.49}$ \\
\hline \\
\end{tabular}
\caption{Fitted Stellar Properties: Metallicity relative to solar ([Z]=log$_{10}$ (Z/Z$\rm _{\odot}$)), dust (E(B-V)), Age (Myr), Mass (10$\rm ^4$M$\rm _{\odot}$), Bolometric Luminosity, L$_{Bol}$, in logscale in [erg s$^{-1}$], and wind momentum rate in logscale in [g cm s$^{-2}$] for YSCs in M83. Details of the fitting can be found in Section.~\ref{stellar_fitting} \label{tab:3} }
\end{table*}

\subsection{Absorption-line fitting of neutral-gas species \label{hi_region_absorption_data_analysis_section}}
After normalising the stellar continuum using the fit obtained in Section.~\ref{stellar_fitting}, the HST/COS UV spectra are used to extract the column density of the ions present primarily within the neutral gas, i.e. \HI, \NI, \OI, \FeII\, and \SII. To this end, we fit multi-component, multi-ion Voigt profiles for the absorption lines using a custom-built Voigt profile fitting program, with tied kinematics (velocity, $v$, and dispersion, $\sigma$) and varying resolution (see Appendix, Section.~\ref{varying_cos_res}). Due to the complexity of the fit, we obtained the resolution and kinematics separately using strong and isolated \FeII\,($\lambda\lambda\lambda$1142, 1143, 1144) and \SII\,($\lambda\lambda$1250, 1253) lines first, before including other weaker, blended lines. We note that contamination from other ions (like \SiII\, and \MgII) was also taken into account to maximise the robustness of the fit obtained within complex absorption features. All star-cluster sightlines have contamination (neutral gas absorption features) from the Milky Way ISM within the spectra. Although due to proximity to us, these features will not vary between different YSC pointings within M83. We use two of the highest SNR spectra: towards M83-POS-1 and M83-POS-2 to obtain a robust multi-component Voigt profile model of the Milky Way ISM line contamination. We then include this contamination model while fitting the line profiles of different YSCs within M83. 

To account for uncertainties associated with continuum placement, we propagate the continuum scatter into the column density measurements using a simple equivalent-width–based approach. For each ion, we estimate the local continuum uncertainty using the standard deviation of the normalised flux in nearby line-free regions. This scatter is then converted into an uncertainty in the equivalent width, $\sigma(W_\lambda)$, by scaling with the pixel size and the number of pixels across the absorption feature. Assuming the linear (optically thin) regime, where $N \propto W_\lambda$, we translate this into an additional column density uncertainty, $\sigma_{N,\rm{std}} \approx N \times \sigma(W_\lambda)/W_\lambda$. This term is added in quadrature to the formal fitting uncertainty to obtain the total uncertainty in column density.

\begin{figure}
    \centering
    \includegraphics[width=1\linewidth,trim={0 0 0 0},clip]{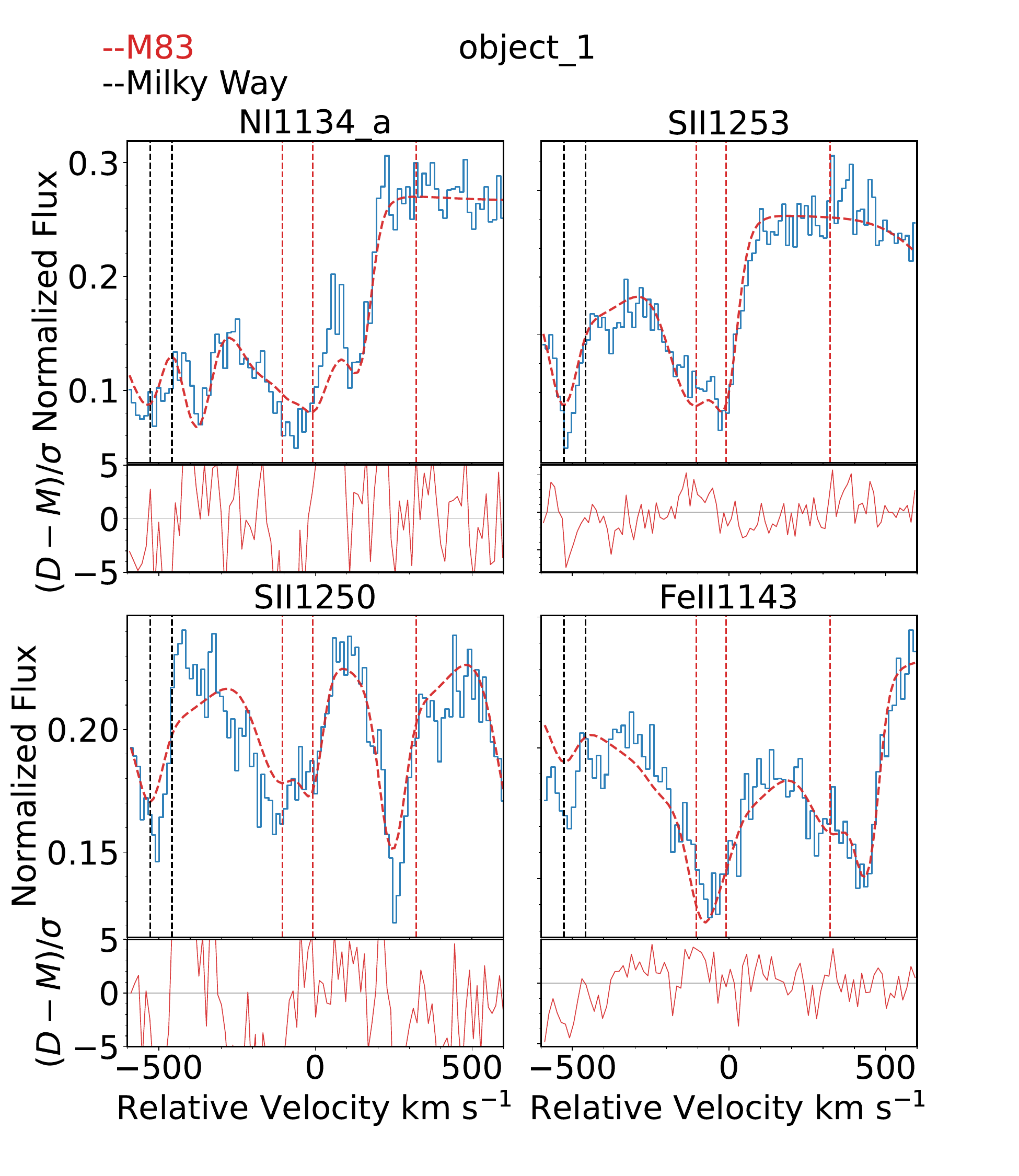}
    \caption{Low-ionisation metal lines associated with the neutral gas towards the young star cluster, M83-POS-1 of the M83 galaxy. The normalised HST/COS spectrum is shown in blue with the best-fit multi-component Voigt profile (M) overplotted in red. Red dashed vertical lines show the location of the different velocity components associated with the neutral gas in M83. Black vertical lines show the location of the different velocity components associated with the neutral gas in the Milky Way for the same species. The residuals ((D-M)/$\sigma$) are shown below each subplot in units of the standard deviation ($\rm \pm5\sigma$) obtained from the error spectrum. Each panel represents the total combined fit as a red dashed line, while the vertical lines show the specific positions of the subcomponents of the main ion within the subplot. For e.g., the NI1134\_a subplot represents the total fit for the region around \NI$\lambda$1134 with the MW contamination, \NI\ UV-absorption triplet, along with contribution from other lines like \FeII$\lambda$1133. Similarly, FeII1143 subplot shows a combined fit of \FeII$\lambda\lambda\lambda$1142,1143,1144. \label{fig:pos_1_abs}}
\end{figure}

Figure.~\ref{fig:pos_1_abs} shows the absorption-line fitting for the YSC M83-POS-1 with ISM line centres from the target shown in red vertical lines and those from the Milky Way shown in black vertical lines. We note that our column density estimates for different ions are consistent (within uncertainty) with \citet[][]{James_2014} for YSCs, M83-POS-1 and M83-POS-2. Our column density for \SII\, is also consistent with \citet{Hernandez2021} for all YSCs except M83-3 and M83-8. We note that these deviations can be accounted for by variation in the number of components used, continuum estimation, and blends with the \SiII\, transition with \SII.       

To ensure that the profiles are not saturated (intrinsically or otherwise), we estimate the line-center optical depth, $\tau$, of individual components within lines and only declare the column density as robust if $\tau\,\lesssim\,5$\footnote{For both thermally and pressure/collisionally broadened lines, the regime between $\tau\,\lesssim\,5$ is considered linear, i.e. equivalent width, W $\propto$ column density, N \citep[see][for detailed discussion]{Savage_Sembach_1991}}. We note that the width of absorption lines is degenerate between the turbulent velocity and the intrinsic temperature of the gas. Due to the moderate spectral resolution of our HST/COS spectra (R$\sim$20 km\,s$\rm ^{-1}$), we do not attempt to resolve that degeneracy and instead fit the total Doppler parameter ($b$-val) for each component as a single variable in the fit. The total fitted column density for different elements within neutral gas is given in Table.~\ref{tab1:absorption_col_den}. We further obtain the abundance, (X/H) of each element, X, as (X/H) = (log(N(X)) - log(N(H))) - (log(N(X)$_{\odot}$) - log(N(H)$_{\odot}$)). Here, N(X) and N(H) are the column densities of the element X and hydrogen, whereas N(X)$_{\odot}$ and N(H)$_{\odot}$ are the solar column densities of the same element X and hydrogen.

In the following subsections, we outline the fitting methodology applied to each ionic species used in this analysis, detailing the adopted line transitions, fitting constraints, and treatment of line saturation or blending effects specific to the individual transitions of those ionic species.

\subsubsection{Neutral (\HI) and molecular (\HH) hydrogen}
The neutral hydrogen column density is measured using the damped \lya\, profile. The neutral hydrogen column density for all YSCs except M83-5 is taken directly from \citet[][]{Hernandez2021}. For M83-5, due to the noisy spectra and the strong geocoronal emission, it is difficult to obtain a robust estimate of \HI\, column density, but we obtain a tentative limit on the \HI\, column density: $\sim$ log(N(\HI))$\sim$19.0, which is consistent with the noise in the spectra. 

It is possible to measure the diffuse \HH\, column density by fitting the rest-frame UV Lyman-Werner (LW) bands of molecular hydrogen within the covered spectral range. We note that this is only possible for YSCs: M83-POS1 and M83-POS2 due to wider spectral coverage (see section.~\ref{diffuse_h2_intro} for details). The total column density of hydrogen for these two YSCs, i.e. log[$N(H)$]=log(N(\HI)\, + 2*$N(\HH)$) is given in Table.~\ref{tab1:absorption_col_den}. Since the \HH\, column density contribution is small ($<$7\%) and we do not have \HH\, measurements for all clusters, we will not be including it for abundance calculations.

\subsubsection{Singly ionised iron, \FeII}
\FeII\,$\lambda$1133, 1142, 1143, 1144 are available for fitting within the HST/COS spectra and used for estimating the kinematics of neutral gas and the column density of \FeII. \FeII\,$\lambda$ 1133 is blended with \NI\,$\lambda$1133 and 1134, and as such, they were fitted simultaneously to obtain a robust column density estimate.

\subsubsection{Neutral nitrogen, \NI}
\NI\,$\lambda$1134 triplet and $\lambda$1200 are used to estimate the column density of neutral nitrogen. As previously mentioned, \NI\,$\lambda$1133 and 1134 profiles are blended with \FeII\,$\lambda$1133 and \NI\,$\lambda$1200 is blended with \MnII. We hold the \FeII\, fixed while simultaneously fitting the \MnII\, and \NI\, to resolve the influence of those blends. We primarily use the \NI\,$\lambda$1134 triplet lines for obtaining the \NI\, column density.

\subsubsection{Singly ionised sulphur, \SII}
\SII\,$\lambda$ 1250, 1253 and 1259 were used for estimating the kinematics of neutral gas and the column density of \SII. The profile for \SII\,$\lambda$1259 is blended with by \SiII\,$\lambda$1260. We had to include \SiII\, in our fits (with column density of \SiII\, informed with other isolated \SiII\, lines) to obtain a robust estimate of \SII\, column density. The transitions \SII\,$\lambda$1250 and 1253 are primarily used for obtaining kinematics and column density, while the \SII\,$\lambda$1259 is only used as an additional check for intrinsic saturation.

\subsubsection{Neutral oxygen, \OI}
\OI\,$\lambda$1302 is the most prominent oxygen line within the HST/COS spectra. Unfortunately, this line is either strongly saturated or contaminated by strong geocoronal emission. Hence, we could only use the weak, undetected \OI\,$\lambda$1355 line for obtaining an upper limit on neutral oxygen column density. Further, we use a technique discussed in \citet{James_aloisi_2018} to obtain the oxygen column density from \PII\, and \SII\, column densities. \PII\,$\lambda$1152 was used to obtain the \PII\, column density. We note that while \SII\ is a natural tracer for estimating \OI\ column density—given that oxygen, sulphur, and phosphorus all receive contributions from core-collapse supernovae and, to some extent, AGB stars—the interpretation is not entirely straightforward. Sulphur, although an $\alpha$-element, may also have an additional contribution from Type Ia supernovae, which can introduce subtle deviations from the nucleosynthetic behaviour expected for oxygen, and it can still be mildly affected by dust depletion. Phosphorus, on the other hand, is more susceptible to depletion onto dust grains. Moreover, the intrinsically weak \PII\ absorption feature: \PII\,$\lambda$1152 around a noisy part of the spectrum may lead to larger observational uncertainties, which can propagate into the inferred oxygen abundance. Taking these effects into account, we checked whether it would be ideal to use an average oxygen abundance derived from both \PII\ and \SII. However, we find that \PII-based oxygen abundances are unrealistically higher than \SII\ and hence, we refrain from using them in our analysis. Instead, similar to the approach by \citet{Hernandez2021}, we will use the sulphur-derived oxygen column density for this work. We note that for neutral gas around all YSCs in this study, the observed oxygen upper limits are consistent with the oxygen column density derived using the \citet{James_aloisi_2018} method.

\subsubsection{Ionisation Correction Factors for the Neutral Gas Column Densities}
\begin{figure}
\includegraphics[width=\linewidth]{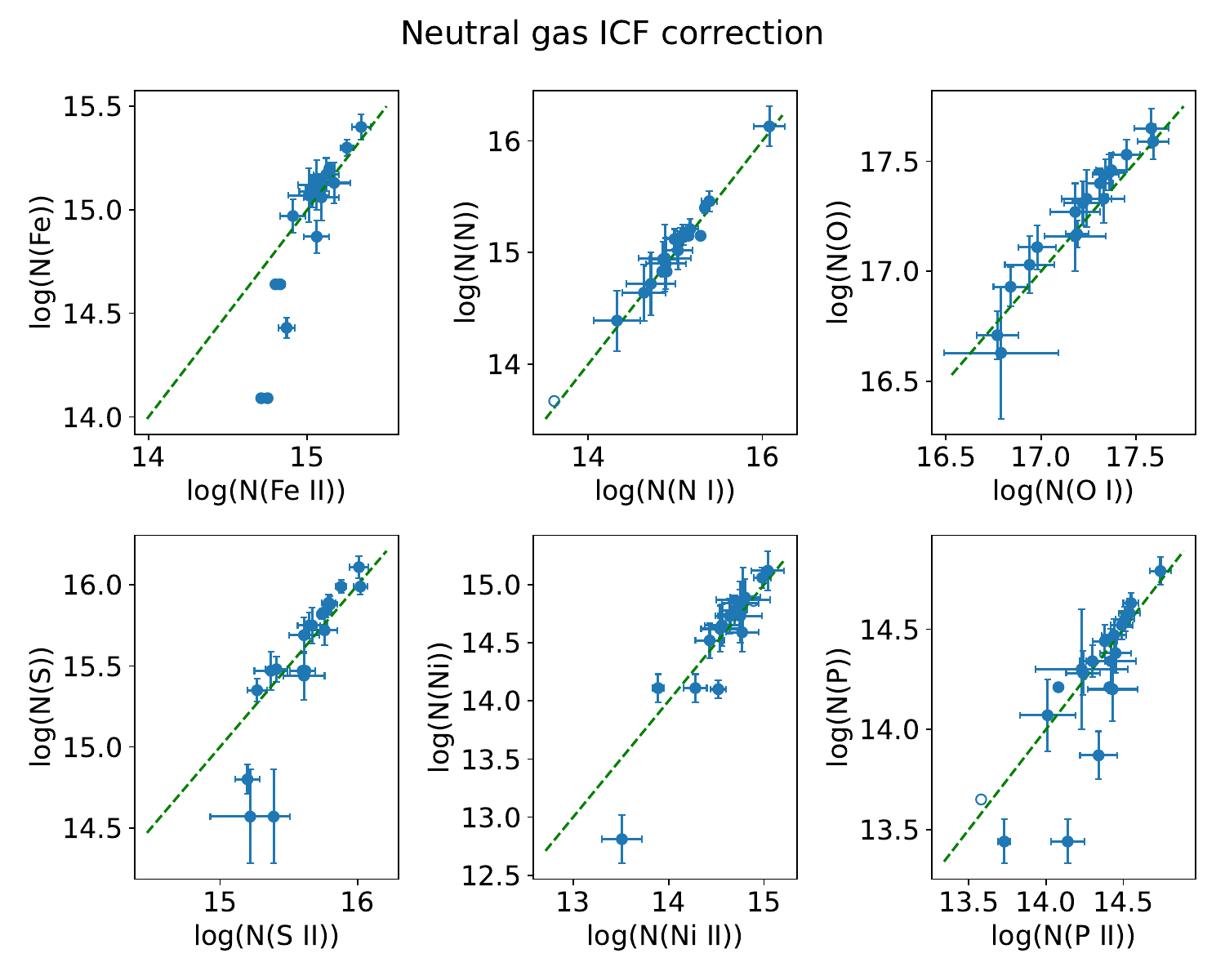}
\caption{Comparing the ICF-corrected elemental column densities (y-axis) to the observed ionic column densities (x-axis). Different subplots show different elements. The 1:1 line is shown in green.}
\label{fig:data4}
\end{figure}

Metals within neutral gas are primarily found in the ionisation state that closely follows the neutral gas conditions (with ionisation potentials $>$13.6 eV, e.g. oxygen in \OI, iron in \FeII). However, there can be some contribution within the neutral phase from other ionisation states of the elements that we do not observe, and also small amounts of that ion residing within the ionised gas. To account for the total column density of the elements originating from within neutral gas only, we need to add the contribution from other ionisation states and subtract the amount of the ion residing in the ionised gas. To account for this fraction, we use CLOUDY photoionisation modeling \citep[][]{Ferland1998} of the YSCs within M83. The detailed information about the specific models used in our work is given in \citet{James_2014, Hernandez2020, Hernandez2021}. We use these photoionisation models to obtain the ionisation correction factor (ICF, see Table.~\ref{tab1:absorption_icf_table}), and, in combination with the ionic column densities of the primary ion within neutral gas (e.g. log(N(\SII))), we obtain the elemental column density (e.g. log(N(S)) of all our fitted metals. We note that the median ICF correction for iron, nitrogen, oxygen, sulphur, nickel and phosphorus are 0.06, 0.06, 0.09, 0.08, 0.08 and 0.04 respectively. Figure.~\ref{fig:data4} shows a comparison between the observed ionic column density of the main ion within neutral gas and the ICF-corrected total column density of the element after applying the corrections. Table.~\ref{tab:data_icf} shows these ionisation-corrected elemental abundances of metals (i.e., oxygen, iron, sulphur and nitrogen) studied in this work. We note that the applied ionisation correction is not necessarily positive in direction. The total correction accounts for two competing effects: (i) contamination from ionised gas along the line of sight, which can contribute to the observed ionic column density and therefore leads to an overestimate of the neutral gas column density, and (ii) the presence of higher ionisation stages within the neutral gas (e.g. \FeIII, \NII) that are not directly observed, which leads to an underestimate of the total elemental column density. The former requires a downward correction, while the latter requires an upward correction. The net ICF therefore depends on the balance between these effects and can result in either an increase or a decrease in the inferred column density \citep[see also][]{James2026}. \\

\begin{table}[]
\centering
\setlength{\tabcolsep}{3pt} % apply only inside this group
\begin{tabular}{lllll}
\hline
YSC   & (N/H)   & (O/H)    & (S/H)   & (Fe/H)  \\
\hline
\hline
M83-1     & 6.04$\rm \pm$0.12 & 8.48$\rm \pm$0.12  & 6.94$\rm \pm$0.11 & 6.00$\rm \pm$0.11 \\
M83-2     & 6.13$\rm \pm$0.31 & 8.45$\rm \pm$0.15  & 6.87$\rm \pm$0.14 & 6.27$\rm \pm$0.11 \\
M83-3$^c$     & 8.08$\rm \pm$0.29 & 9.77$\rm \pm$0.25  & 7.86$\rm \pm$0.25 & 7.49$\rm \pm$0.24 \\
M83-4$^c$     & 7.09$\rm \pm$0.33 & 9.61$\rm \pm$0.26  & 7.89$\rm \pm$0.26 & 7.32$\rm \pm$0.22 \\
M83-5$^c$     & 6.67$\rm \pm$0.78 & 10.03$\rm \pm$0.33 & 8.47$\rm \pm$0.32 & 8.07$\rm \pm$0.33 \\
M83-6     & 6.44$\rm \pm$0.11 & 8.74$\rm \pm$0.10  & 7.16$\rm \pm$0.08 & 6.39$\rm \pm$0.08 \\
M83-7     & 6.07$\rm \pm$0.29 & 8.94$\rm \pm$0.12  & 7.34$\rm \pm$0.10 & 6.41$\rm \pm$0.14 \\
M83-8     & 6.49$\rm \pm$0.09 & 8.28$\rm \pm$0.09  & 6.70$\rm \pm$0.08 & 6.49$\rm \pm$0.04 \\
M83-9     & 6.38$\rm \pm$0.17 & 8.88$\rm \pm$0.10  & 7.28$\rm \pm$0.08 & 6.84$\rm \pm$0.09 \\
M83-10    & 6.38$\rm \pm$0.07 & 8.66$\rm \pm$0.07  & 7.08$\rm \pm$0.04 & 6.34$\rm \pm$0.04 \\
M83-11    & 6.92$\rm \pm$0.11 & 8.91$\rm \pm$0.12  & 7.28$\rm \pm$0.11 & 6.91$\rm \pm$0.08 \\
M83-12    & 6.42$\rm \pm$0.05 & 8.47$\rm \pm$0.08  & 6.91$\rm \pm$0.05 & 6.22$\rm \pm$0.04 \\
M83-13    & 6.71$\rm \pm$0.09 & 8.58$\rm \pm$0.13  & 7.00$\rm \pm$0.11 & 6.37$\rm \pm$0.12 \\
M83-14    & 6.47$\rm \pm$0.26 & 8.90$\rm \pm$0.17  & 7.29$\rm \pm$0.16 & 6.70$\rm \pm$0.16 \\
M83-15    & 5.35$\rm \pm$0.27 & 8.27$\rm \pm$0.11  & 6.71$\rm \pm$0.09 & 5.93$\rm \pm$0.09 \\
M83-16    & 6.95$\rm \pm$0.18 & 8.35$\rm \pm$0.07  & 6.81$\rm \pm$0.04 & 6.12$\rm \pm$0.04 \\
M83-POS-1$^c$ & 7.23$\rm \pm$0.04 & 9.25$\rm \pm$0.07  & 7.55$\rm \pm$0.04 & 6.72$\rm \pm$0.04 \\
M83-POS-2$^c$ & 7.92$\rm \pm$0.05 & 9.72$\rm \pm$0.30  & 7.66$\rm \pm$0.29 & 7.18$\rm \pm$0.03 \\
\hline
\end{tabular}
\caption{Ionisation-corrected elemental abundance, (X/H)$\rm _{HI}$ (= 12+log(X/H)) of the element X, in neutral gas for each YSC in M83 observed with HST/COS. $^c$: Clusters are within the galactic center, R/R$_{25}<$0.05 \label{tab:data_icf}}
\end{table}

\subsubsection{Depletion Correction for the Neutral Gas Column Densities}
In addition to applying ionisation corrections, we also explored dust depletion corrections for the neutral-gas elemental abundances. The methodology used to estimate depletion from the total observed column densities of neutral-phase ions (in our case \FeII, \SII, \OI, \NiII, and \MnII) is described in detail in Section~\ref{subsection_dust_neutral} \citep[see also][]{DeCia2016, DeCia2018, De_Cia_2024, Konstantopoulou2022, Konstantopoulou2024b}. In brief, this approach estimates the amount of depletion onto dust by examining the relative variations in the observed column densities of multiple neutral-gas ions. However, we find that the inferred depletion corrections exhibit substantial dispersion, reaching a median of $\sim$1 dex in log($N$) for iron, $\sim$0.39 dex in log($N$) for sulphur, and $\sim$0.12 dex in log($N$) for oxygen when observed \NiII\ column densities are included in the analysis. In addition, (i) depletion corrections cannot be derived for nitrogen within the adopted framework, and (ii) we do not apply depletion corrections to the ionised-gas abundances. While dust depletion is well characterised in the neutral ISM \citep[e.g.,][]{De_Cia_2024}, its behaviour in ionised gas remains significantly more uncertain due to the complex interplay of grain destruction, charging, and gas-phase recycling in high-radiation environments. As a result, it is not clear whether depletion prescriptions calibrated for neutral gas can be reliably extended to \HII\ regions. For these reasons, we do not use the depletion-corrected elemental abundances of neutral gas for analysis in this work. This choice also ensures a consistent, like-for-like comparison between the abundances derived for the neutral and ionised gas phases. Since we know that refractory elements (e.g., Fe) may still be affected by dust depletion and thus their gas-phase abundances could be systematically suppressed, we will discuss the effects of dust depletion on iron throughout the paper.

\subsection{Emission-line fitting of species from \HII\ regions \label{hii_region_emission_data_analysis_section}}

\begin{figure}
\includegraphics[width=\linewidth]{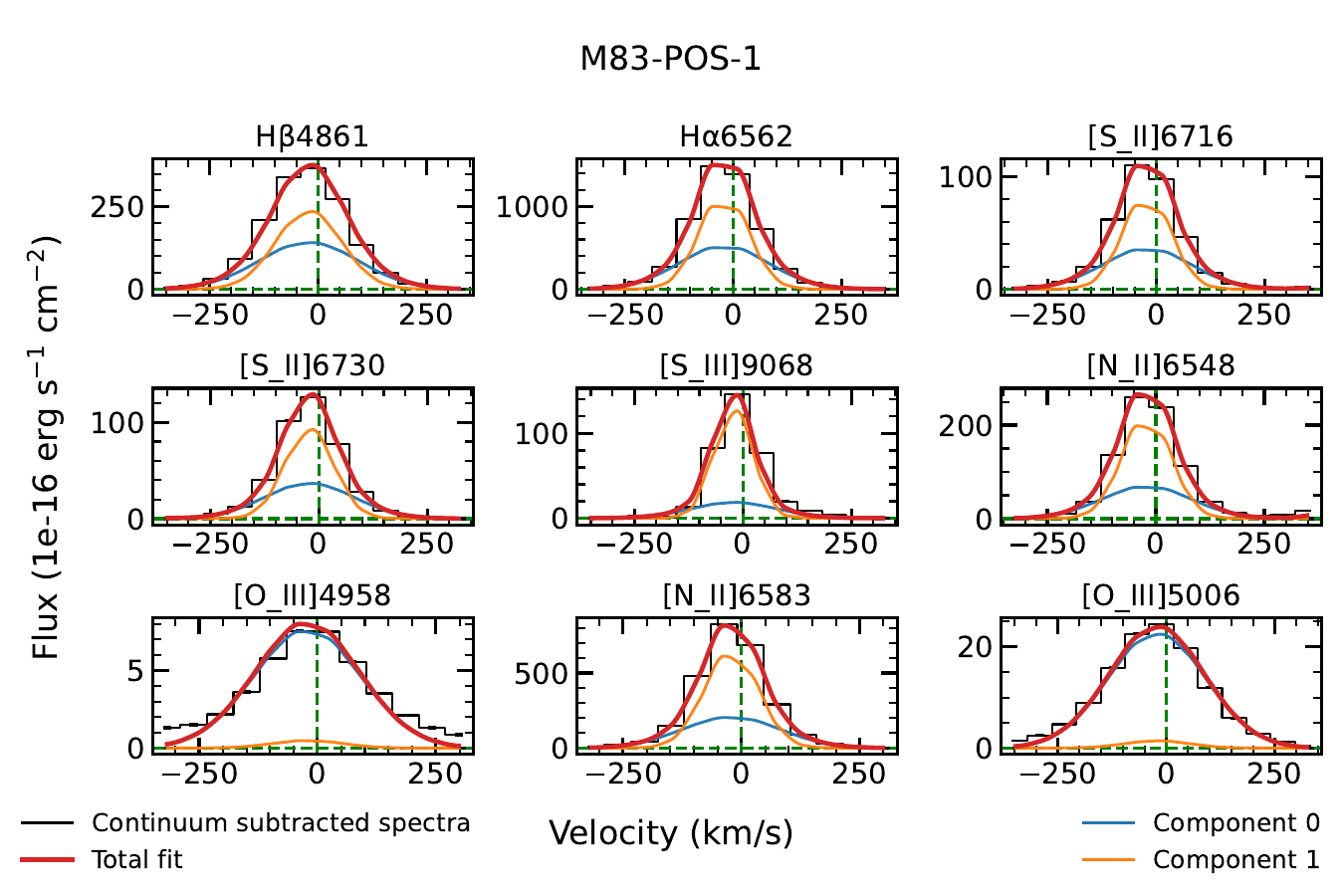}
\caption{Multi-component Gaussian fit (shown in red) to emission lines  (shown in black) observed in the spectra obtained from VLT/MUSE IFU spaxels of the regions surrounding the YSC M83-POS-1. The individual spaxels falling within the HST/COS aperture were summed to obtain the optical spectra seen here. Other colours show contributions from the individual components. \label{fig:emission_line_plot_obj_1}}
\end{figure}

To obtain the total observed flux for each species in the \HII\ region surrounding the YSCs in M83, we fit a multi-component Gaussian profile with tied relative velocity ($v$) and velocity dispersion ($\sigma$) across different ionised gas species (Balmer: H$\alpha$, H$\beta$, and other ions: O$^+$, O$^{++}$, S$^{+}$, S$^{++}$, N$^{+}$, Fe$^{++}$). To account for the effects of stellar Balmer absorption, we performed the spectral fitting using \href{https://pypi.org/project/ppxf/}{pPXF} \citep[][]{Cappellari2023} with MILES single stellar population (SSP) models assuming a Kroupa initial mass function (IMF)\citep[][]{Vazdekis2010}. We use this fitted model as a continuum for our emission line fitting, with localised continuum adjustments for fitting weak emission lines. We note that the multi-component fitting approach resolves distinct emission components within each spectral line, corresponding to separate gas clouds or regions along the line of sight. This method also incorporates minor wavelength adjustments when necessary to correct for calibration offsets and ensure that all components are accurately aligned in velocity space. We get an initial estimate of the flux within the main component of strong lines (like H$\alpha$) using results from \citet{Hernandez2021}. We use this estimate as a starting point for line fitting and add more spectral lines (auroral and others not available in the literature) as well as other additional components if required for polishing the fit. For VLT/MUSE, we note that a two-component fit seems sufficient, whereas we implement up to three components for fitting LBT/MODS spectra. The additional kinematic component required to fit the LBT/MODS spectra is most likely an artefact arising from the wavelength mismatch between the blue and red channels, which are separated by a dichroic beam splitter, rather than representing a physically distinct gas component. The dichroic crossover near 5650 \AA\, introduces a small reduction in throughput and can lead to residual calibration mismatches between the two channels. Consequently, while the LBT/MODS spectra provide robust integrated emission-line fluxes, all ionised-gas kinematic measurements presented in this work are derived from the VLT/MUSE observations, which offer a continuous spectral coverage without such channel-to-channel systematics. Figure~\ref{fig:emission_line_plot_obj_1} shows the emission-line fit results for the ionised gas surrounding the YSC, M83-POS-1. We then obtain the total flux (F$_\lambda$) of each ionised gas species observed by summing the flux from each component. Using the H$\alpha$/H$\beta$ Balmer line ratios, we obtain the amount of dust (E(B-V)) and correct for the dust extinction using the Case-B theoretical ratios to obtain the total intensity (I$_\lambda$) of each line. Tables~\ref{tab1:emission_line_flux}, \ref{tab2:emission_line_flux} show the reddening-corrected line intensities (I$_\lambda$) of the relevant lines that will further be used to calculate the abundance of nitrogen, oxygen, sulphur and iron in ionised (\HII) regions around the YSCs.  

\subsubsection{Obtaining the density and temperature of the \HII\ gas \label{subsection:denstemp}}

We derive ionic and elemental abundances (12+log(X/H)) in the \HII, regions using the nebular analysis code \href{https://pypi.org/project/pyneb/}{pyneb} \citep[see][]{Luridiana2013}. This requires measuring the intensity ratio of each diagnostic emission line to H$\beta$, together with an estimate of the local physical conditions of the gas, namely the electron density ($n_e$) and the electron temperature (T$_e$). These physical conditions are themselves determined from standard density- and temperature- sensitive line ratios, which are provided as inputs to \href{https://pypi.org/project/pyneb/}{pyneb}. In the following section, we describe the specific diagnostics used to infer $n_e$ and T$_e$ for each \HII\ region.

While different diagnostic emission line ratios can be used for calculating the electron density ($n_e$) \citep[see e.g.][for detailed discussion]{Mendez_Delgado2024}, we prominently use the highest S/N density-dependent line ratio available to us: [\SII]$\lambda$6731/$\lambda$6716, except for R(1) where use n$_e$ obtained from [\OII]$\lambda$3726/$\lambda$3729 ratio as the [\SII] ratio gives us n$_e\,<100\,cm^{-3}$. We also note that for most of our clusters, we derive densities that lie within the `allowed range' for the [\SII] and [\OII] ratio, $\rm 10^{2} < n_e < 10^{3.3}\, cm^{-3}$. For four other clusters, we find n$_e<100\,cm^{-3}$ with both ratios and hence assume, n$_e=100\,cm^{-3}$. We further note that we found negligible variations within abundance measurements with the mentioned change in density. For our YSCs, the measured densities range from 100 $\lesssim\,n_e\,\lesssim$ 550 cm$^{-3}$ (See Table.~\ref{tab:temperatures}).

The electron temperature of the ionised gas can, in principle, be measured from auroral-to-nebular line ratios such as [\NII]$\lambda$5755/$\lambda$6584, [\OIII]$\lambda$4363/$\lambda$5007, [\SIII]$\lambda$6312/$\lambda$9068, [\SII]$\lambda\lambda$4069,4076/$\lambda\lambda$6716,6730, and [\ArIII]$\lambda$5192/$\lambda$7135. However, because the M83 \HII\ regions studied here are metal-rich, efficient metal-line cooling lowers the electron temperature and makes these auroral lines intrinsically weak. We therefore detect only [\SIII]$\lambda$6312 in M83-7 and [\NII]$\lambda$5755 in five regions: M83-3, M83-R1, M83-R2, M83-R7, and M83-R12. The remaining auroral lines are not significantly detected.

This limited auroral-line coverage prevents direct temperature measurements ($T_e$) across all clusters and ionisation zones. Moreover, even when one temperature is measured, for example $T_e$([\NII]), converting it into $T_e$([\OII]), $T_e$([\OIII]), and $T_e$([\SIII]) requires empirical temperature relations. While temperature relations are widely used, different prescriptions in the literature can give different results and introduce additional systematic uncertainty \citep[see e.g.,][for detailed discussion]{Rickards_Vaught2024}. Since abundance measurements are extremely sensitive to temperature measurements, additional uncertainty always propagates significantly into elemental abundance estimates. Since our abundance analysis requires self-consistent temperatures for multiple ionisation zones, we avoid relying solely on these temperature-conversion relations.

Instead, we use the empirical temperature-calibration framework described in Appendix~\ref{temperature_fitting}. Briefly, we compile literature \HII\ regions in nearby star-forming galaxies \citep[from][]{Bresolin2002, Bresolin2005, Bresolin2009} with measured line intensities, electron temperatures, densities, and direct elemental abundances, focusing on the metal-rich regime relevant to M83. We then compare a suite of DESIRED strong-line metallicity calibrators \citep[see][]{Rosales-Ortega2026} against directly measured $T_e$([\OII]), $T_e$([\OIII]), and $T_e$([\SIII]) and identify the calibrators that best reproduce each ionisation-zone temperature. The resulting relations provide predicted temperatures for M83 regions without auroral-line detections, while remaining anchored to the direct-method temperature scale.

For \HII\ regions with significant auroral-line detections, we use the directly measured temperatures in the abundance calculations for that specific ionisation zone. For the remaining regions and zones, we adopt the predicted $T_e$([OII]), $T_e$([OIII]), and $T_e$([SIII]) values from the empirical relations derived in Appendix~\ref{temperature_fitting}. This approach allows us to obtain a consistent set of multiphase electron temperatures for all M83 \HII\ regions without applying uncertain single-zone temperature conversions. The final adopted temperatures and densities are listed in Table~\ref{tab:temperatures}.

\begin{table}[]
\raggedright
\setlength{\tabcolsep}{3pt} % apply only inside this group
\begin{tabular}{lllll}
\hline
YSC       & T$_e$(\NII)    & T$_e$(\SIII)   & T$_e$(\OIII)   & n$_e$(\SII)      \\
\hline
\hline
M83-1     & 7.8$\rm \pm$0.75 & 6.9$\rm \pm$0.66 & 5.9$\rm \pm$0.67 & 0.14$\rm \pm$0.10   \\
M83-2     & 7.2$\rm \pm$0.68 & 7.3$\rm \pm$0.7  & 6.5$\rm \pm$0.73 & 0.23$\rm \pm$0.09  \\
M83-3     & 6.8$\rm \pm$0.63 & 4.8$\rm \pm$0.55 & 5.3$\rm \pm$0.57 & 0.51$\rm \pm$0.04  \\
M83-4     & 7.0$\rm \pm$0.65 & 7.0$\rm \pm$0.66 & 5.3$\rm \pm$0.57 & 0.43$\rm \pm$0.02 \\
M83-5     & 6.9$\rm \pm$0.64 & 6.8$\rm \pm$0.64 & 5.1$\rm \pm$0.56 & 0.29$\rm \pm$0.02 \\
M83-6     & 8.2$\rm \pm$0.78 & 6.9$\rm \pm$0.66 & 6.1$\rm \pm$0.68 & 0.12$\rm \pm$0.09 \\
M83-7     & 6.8$\rm \pm$0.63 & 6.0$\rm \pm$0.26 & 4.6$\rm \pm$0.5  & 0.10$\rm \pm$0.01*   \\
M83-8     & --          & --          & --          & --          \\
M83-9     & --          & --          & --          & --          \\
M83-10    & 7.2$\rm \pm$0.7  & 6.6$\rm \pm$0.64 & 5.5$\rm \pm$0.64 & 0.11$\rm \pm$0.10 \\
M83-11    & 8.2$\rm \pm$0.82 & 7.2$\rm \pm$0.69 & 6.2$\rm \pm$0.68 & 0.17$\rm \pm$0.04 \\
M83-12    & 7.2$\rm \pm$0.71 & 6.6$\rm \pm$0.63 & 5.1$\rm \pm$0.61 & 0.10$\rm \pm$0.01*   \\
M83-13    & --         & --         & --         & --           \\
M83-14    & 7.2$\rm \pm$0.69 & 6.7$\rm \pm$0.64 & 5.1$\rm \pm$0.6  & 0.10$\rm \pm$0.01*   \\
M83-15    & --         & --         & --         & --           \\
M83-16    & 7.1$\rm \pm$0.7  & 6.9$\rm \pm$0.66 & 5.1$\rm \pm$0.61 & 0.10$\rm \pm$0.09  \\
M83-POS-1 & 5.8$\rm \pm$0.54 & 6.4$\rm \pm$0.61 & 4.2$\rm \pm$0.46 & 0.5$\rm \pm$0.03  \\
M83-POS-2 & 6.9$\rm \pm$0.64 & 6.9$\rm \pm$0.65 & 5.2$\rm \pm$0.56 & 0.43$\rm \pm$0.04 \\
R(1)      & 7.1$\rm \pm$0.67 & 8.3$\rm \pm$1.1  & 6.9$\rm \pm$0.75 & 0.43$\rm \pm$0.23$^1$  \\
R(2)      & 6.3$\rm \pm$0.59 & 6.9$\rm \pm$1.0  & 5.2$\rm \pm$0.57 & 0.10$\rm \pm$0.01*   \\
R(6)      & 7.2$\rm \pm$0.78 & 7.1$\rm \pm$0.72 & 5.7$\rm \pm$0.87 & 0.11$\rm \pm$0.1   \\
R(7)      & 7.0$\rm \pm$0.65 & 6.9$\rm \pm$0.65 & 6.4$\rm \pm$1.1  & 0.36$\rm \pm$0.04 \\
R(8)      & 6.2$\rm \pm$0.84 & 6.5$\rm \pm$0.69 & 4.5$\rm \pm$0.86 & 0.22$\rm \pm$0.21  \\
R(12)     & 7.0$\rm \pm$1.2  & 6.1$\rm \pm$0.71 & 4.9$\rm \pm$1.1  & 0.25$\rm \pm$0.24   \\
\hline
\end{tabular}
\caption{Electron temperatures (T$\rm _e$[\SIII], [\OIII], [\NII] in 10$\rm ^3$K) and density (n$\rm _e$[\SII] in 10$\rm ^3$cm$\rm ^{-3}$) of \HII\ regions around YSCs in M83. Temperatures are derived from the strong line calibration relation (see Section.~\ref{temperature_fitting} for details). Density is always derived from the [\SII]$\lambda\lambda$6731/6716 doublet ratio except for R1. $^1$: Density was obtained using [\OII] ratio instead of [\SII]. *: Density obtained from both [\SII] and [\OII] ratios are $<$100 cm$^{-3}$, so we assume n$_e$=100$\rm \pm$10 cm$^{-3}$. \label{tab:temperatures}}
\end{table}

\subsubsection{Oxygen Abundance in \HII\ regions}

The density and electron temperatures obtained in the previous section were used to calculate ionic and elemental abundances for all \HII\ regions surrounding the YSCs in M83. In \HII\ regions, the oxygen abundance is primarily obtained using the sum of its first (O$^+$) and second (O$^{++}$) ionisation states. 

In the rest-frame optical, O$^+$ is traced by the [\OII] $\lambda\lambda$7320, 7321, 7330, 7331 red quadruplet, and, when the blue part of the spectrum is available (as in the LBT/MODS data), also by the stronger [\OII]$\lambda\lambda$3726, 3728 blue doublet. The red doublet was detected at $\rm >3\sigma$ significance in only three \HII\ regions (M83-7, M83-10, and M83-14), whereas the blue doublet is robustly detected in all LBT/MODS spectra. We therefore derived O$^+$ abundances using the [\OII]$\lambda\lambda$3726, 3728 lines from LBT/MODS. 

In rest-frame optical, O$^{++}$ is observed using the [\OIII]$\lambda\lambda$4958, 5007 doublet. The doublet is robustly detected ($>3\sigma$) within all \HII\ regions in VLT/MUSE spectra. To ensure consistency between instruments and to correct for aperture effects, we rescaled the LBT/MODS spectra by matching their continuum level to that of the corresponding VLT/MUSE spectra within the common spectral range free of nebular emission ($\sim$4740-4840\AA) with a multiplicative factor. This correction accounts for small flux offsets arising from the different slit and IFU apertures but does not alter the line ratios within each instrument. Each ionic abundance was derived using its own instrument's H$\beta$ measurement-i.e., [\OII]/H$\beta$ from LBT/MODS and [\OIII]/H$\beta$ from VLT/MUSE- further minimising aperture-driven biases. After applying the continuum rescaling, we were able to obtain reliable O$^+$ abundances for seven additional \HII\ regions across M83. In our sample, we obtain a robust oxygen abundance (O$^+$ + O$^{++}$) within 10 \HII\ regions around YSCs in M83 (see Table.~\ref{tab:emission_abundance}). 

The abundance of O$^+$ and O$^{++}$ are also crucial in determining the ionisation correction factor (ICF) to include missing contribution from higher ionisation states \citep[see e.g.][]{Stasinska1990, Thuan1995, Rodriguez2002, Rodriguez_Rubin_2005, Izotov2006b}. Since the correction factor changes not only from element to element, but also as a function of total gas-phase metallicity, the impact of using upper limits on O$^+$ abundance also affects the ionisation correction of different elements uniquely. We discuss the impact of ionic oxygen abundance and oxygen ratio in different ionisation correction factors in the literature in detail in Appendix~\ref {hii_icf_estimation}. For simplicity and uniformity, we chose to use the ICF correction by \citet{Izotov2006b} for final abundances of iron, sulphur and nitrogen listed in Table.~\ref{tab:emission_abundance}.

\subsubsection{Iron in \HII\ regions}
Iron abundance is calculated using the Fe$^{++}$ line ([\FeIII]$\lambda$4658) with additional contribution from other ions (like Fe$^{+}$, Fe$^{3+}$ and Fe$^{4+}$) being accounted for through ICFs. Different ICFs are used in the literature to obtain the correct iron abundance \citep[see e.g.][for detailed discussion]{Mendez_Delgado2024}. We list the different ICF corrections for iron in detail in the Table.~\ref{tab:emission_abundance_icf} within Appendix~\ref {hii_icf_estimation}. Figure.~\ref{DESIRED_icf_correction_m83} also shows the impact of different ICF corrections on iron abundance.[\FeIII]$\lambda$4658 is robustly detected within 13 \HII\ regions. We further note that an upper limit on O$^{+}$ abundance will yield a lower limit on iron abundance after applying ionisation correction. In the case of M83-11, both Fe$^{++}$ and O$^{+}$ abundances are limits, so we declare the calculated abundance as tentative($\sim$). The average ionisation correction for iron abundance for all YSCs studied in this work is $\sim$0.3 dex. We note that we have robust iron abundance calculated for 8 \HII\ regions around YSCs as shown in Table.~\ref{tab:emission_abundance}.

\subsubsection{Sulphur in \HII\ regions}
Sulphur abundance is primarily obtained within \HII\ regions using the summed abundance of its first, S$^{+}$ and second, S$^{++}$ ionisation states. S$^{+}$ is robustly detected in all \HII\ regions using the strong [\SII]$\lambda\lambda$6716, 6730 doublet. The S$^{++}$ transition, [\SIII]$\lambda$9068, is relatively weak and also located within the noisy part of the optical spectrum in both VLT/MUSE and LBT/MODS spectra. However, we robustly obtain the [\SIII]$\lambda$9068 line intensity for all except three \HII\ regions (M83-1 and M83-2 in VLT/MUSE and M83-R6 in LBT/MODS). We note that the ICF correction that yields total sulphur abundance from S$^{+}$ + S$^{++}$ varies as a function not only of the O$^{+}$ abundance but also of the O$^{+}$/(O$^{+}$+O$^{++}$) ratio. Hence, we ensure that the limits and uncertainties within the O$^{+}$ abundance are propagated carefully onto the total sulphur abundance. Limits from the O$^{+}$ and S$^{++}$ measurements in each case are carefully propagated into the sulphur abundance. The average ionisation correction for sulphur abundance for all YSCs studied in this work is negligible: $\sim$0.01 dex. We note that we have robust sulphur abundance calculated for 10 \HII\ regions around YSCs as shown in Table.~\ref{tab:emission_abundance}.

\subsubsection{Nitrogen in \HII\ regions}
Nitrogen abundance is calculated using N$^{+}$ line ([\NII]$\lambda$6583\footnote{We fit the [\NII]$\lambda\lambda$6548,6583 doublet, but use the stronger line, i.e. [\NII]$\lambda$6583 for abundance calculations}) with additional contribution from other nitrogen ions being included using an ionisation correction factor from \citep[][]{Izotov2006b} (see their eq.~18). We note that their ICF prescriptions are consistent with the models from \citet{Stasinska1990}. [\NII]$\lambda$6583 is a strong line that is robustly detected within all \HII\ regions. However, upper limits from the O$^{+}$ abundance measurements lead to a lower limit on nitrogen abundance. The average ionisation correction for nitrogen abundance for all YSCs studied in this work is $\sim$0.2 dex. Using the detections of O$^{+}$ abundance, we obtain a robust nitrogen abundance for 10 \HII\ regions around YSCs as shown in Table~\ref{tab:emission_abundance}.

\begin{table}[]
\centering
{\setlength{\tabcolsep}{3pt} % apply only inside this group
\begin{tabular}{lllll}
\hline
YSC       & (N/H)        & (O/H)           & (S/H)           & (Fe/H)          \\
\hline
\hline
M83-1     & 7.70$\rm \pm$0.23  & 8.70$\rm \pm$0.63*    & 6.94$\rm \pm$0.22*    & 7.52$\rm \pm$0.30     \\
M83-2     & 7.76$\rm \pm$0.18  & $\rm <$9.30 & $\rm <$6.94 & 7.72$\rm \pm$0.29     \\
M83-3     & 8.20$\rm \pm$0.24  & 8.35$\rm \pm$0.26*    & 7.17$\rm \pm$0.15     & 7.27$\rm \pm$0.28     \\
M83-4     & 8.18$\rm \pm$0.27  & 8.35$\rm \pm$0.26*    & 6.85$\rm \pm$0.10     & 7.36$\rm \pm$0.31     \\
M83-5     & 8.25$\rm \pm$0.30  & 8.35$\rm \pm$0.26*    & 6.81$\rm \pm$0.12     & 7.67$\rm \pm$0.35     \\
M83-6     & 8.00$\rm \pm$0.35  & 8.61$\rm \pm$0.31*    & 6.55$\rm \pm$0.14     & $\rm <$8.53 \\
M83-7     & 8.07$\rm \pm$0.28  & 8.44$\rm \pm$0.33     & 6.90$\rm \pm$0.07     & 6.97$\rm \pm$0.32     \\
M83-8*    & 8.13$\rm \pm$0.21* & 8.79$\rm \pm$0.25*    & 7.04$\rm \pm$0.13*    & 6.98$\rm \pm$0.38*    \\
M83-9*    & 7.94$\rm \pm$0.18* & 8.66$\rm \pm$0.22*    & 6.95$\rm \pm$0.11*    & 6.57$\rm \pm$0.44*    \\
M83-10    & 7.78$\rm \pm$0.22  & 8.51$\rm \pm$0.36     & 6.74$\rm \pm$0.10     & $\rm <$7.27 \\
M83-11    & 7.75$\rm \pm$0.20  & $\rm <$8.79 & 6.57$\rm \pm$0.10     & $\rm <$7.88 \\
M83-12    & 7.87$\rm \pm$0.29  & $\rm <$8.84 & 6.69$\rm \pm$0.12     & 7.29$\rm \pm$0.40     \\
M83-13    & --           & --              & --              & --              \\
M83-14    & 7.95$\rm \pm$0.29  & 8.43$\rm \pm$0.37     & 6.74$\rm \pm$0.11     & 7.41$\rm \pm$0.34     \\
M83-15    & --           & --              & --              & --              \\
M83-16    & 8.09$\rm \pm$0.34  & $\rm <$8.71 & 6.84$\rm \pm$0.13     & 7.58$\rm \pm$0.41     \\
M83-POS-1 & 8.43$\rm \pm$0.31  & $\rm <$9.40 & 7.14$\rm \pm$0.11     & 7.08$\rm \pm$0.35     \\
M83-POS-2 & 8.27$\rm \pm$0.27  & $\rm <$8.64 & 6.84$\rm \pm$0.13     & 7.69$\rm \pm$0.32     \\    
\hline
\end{tabular}
\caption{Elemental abundance, (X/H)$\rm _{HII}$ = 12+log(X/H) of the element, X, in \HII\ regions (ionised gas) around the same YSCs in M83. Most measurements are using spectra from VLT/MUSE. We use VLT/MUSE spectra throughout, other than those indicated by an asterisk(*), for which we use LBT/MODS measurements. \label{tab:emission_abundance}}}
\end{table}

%=================================== ABSORPTION ANALYSIS ===================================

\section{Results \label{sec:Results}}

In this section, we will describe the main results obtained by comparing properties of the YSCs within M83, the surrounding gas ionised by the radiation from the stars (\HII\ regions), and the larger enveloping neutral gas (\HI\, regions). To aid the eye, in all future plots in this paper, we mark the stellar properties as diamonds, neutral gas abundances as squares, and ionised gas abundances as circles. The plots showing differences in abundance between ionised and neutral phases ($\rm \Delta$=(X/H)$\rm _{HII}$ - (X/H)$\rm _{HI}$) are plotted as stars. Unless specified otherwise, the trends are always fitted as a straight line. The statistics are always calculated using the detections unless specified otherwise. The fitted trends shown throughout this work were obtained by fitting a first-order polynomial of the form $y = mx + b$ using an iterative weighted least-squares procedure. Measurement uncertainties in both variables were incorporated by minimising the residuals with an effective uncertainty given by
\begin{equation}
\sigma_{\rm eff} = \sqrt{\sigma_y^2 + m^2\sigma_x^2},
\end{equation}
where $\sigma_x$ and $\sigma_y$ are the uncertainties in the independent and dependent variables, respectively. The fitting procedure was iterated until the slope converged. The reported Pearson correlation coefficients ($r$) were computed using inverse-variance weighting, while the statistical significance of the correlation ($p$-value) was estimated using a two-sided permutation test with 10,000 random realisations. The uncertainties on the fitted slope ($m$) and intercept ($b$) were derived from the covariance matrix of the weighted fit. In addition, an optional Monte Carlo bootstrap analysis was performed to verify the robustness of the fitted trends against measurement uncertainties. The details about the trends (fitted straight line slope, $m$, intercept, $b$, and the Pearson correlation coefficient, $r$, and significance of correlation, $p$-value) are showcased as legends in all relevant figures. For any case where the Pearson correlation is statistically significant ($p$-value$<$0.05), the legends are marked as bold.

\subsection{The galactic center \label{galaxy_center_results}}

Before moving on to discuss elemental abundances, it is important to assess the distribution of hydrogen within the galactic centre. This specific discussion is inspired by M83 literature. While studying the neutral gas around YSCs within the galactic centre, \citet{Hernandez2021} found that there is a significant absence of neutral hydrogen (\HI) within the ISM in the centre of M83. Furthermore, \citet{Hernandez2023} looked at the distribution of warm H$\rm _2$ gas using the infrared emission lines observed with JWST and found that 75\% of the total molecular gas mass is contained in the warm H$\rm _2$ component, with this warmer H$\rm _2$ emission concentrated near the optical nucleus. From the YSCs studied in this work, 5 of them fall within the galactic centre of M83(R/R$\rm _{25}$<0.05), including M83-POS1 and M83-POS2, for whom we have the cold molecular (diffuse H$\rm _2$) gas column density.

\begin{figure}
    \centering
    \includegraphics[width=0.5\textwidth]{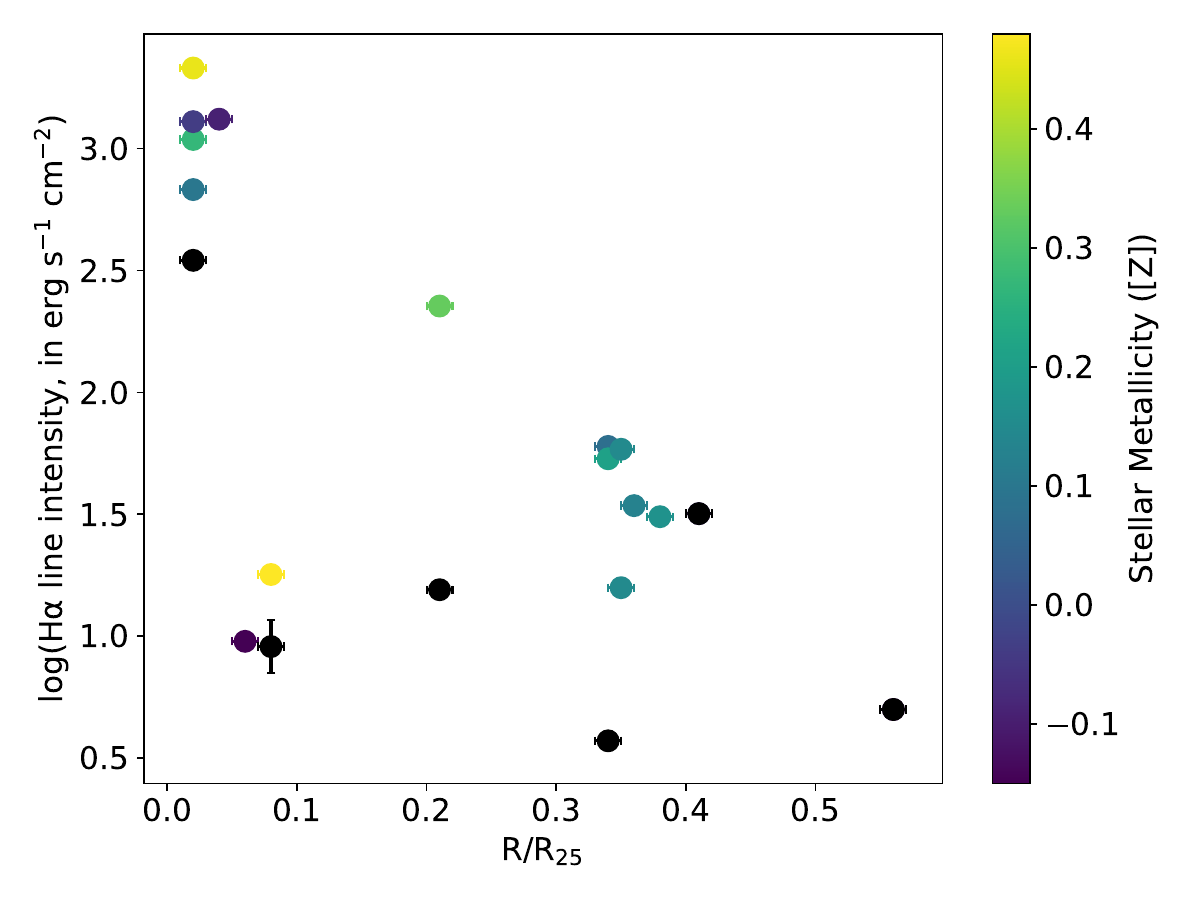}
    \includegraphics[width=0.5\textwidth]{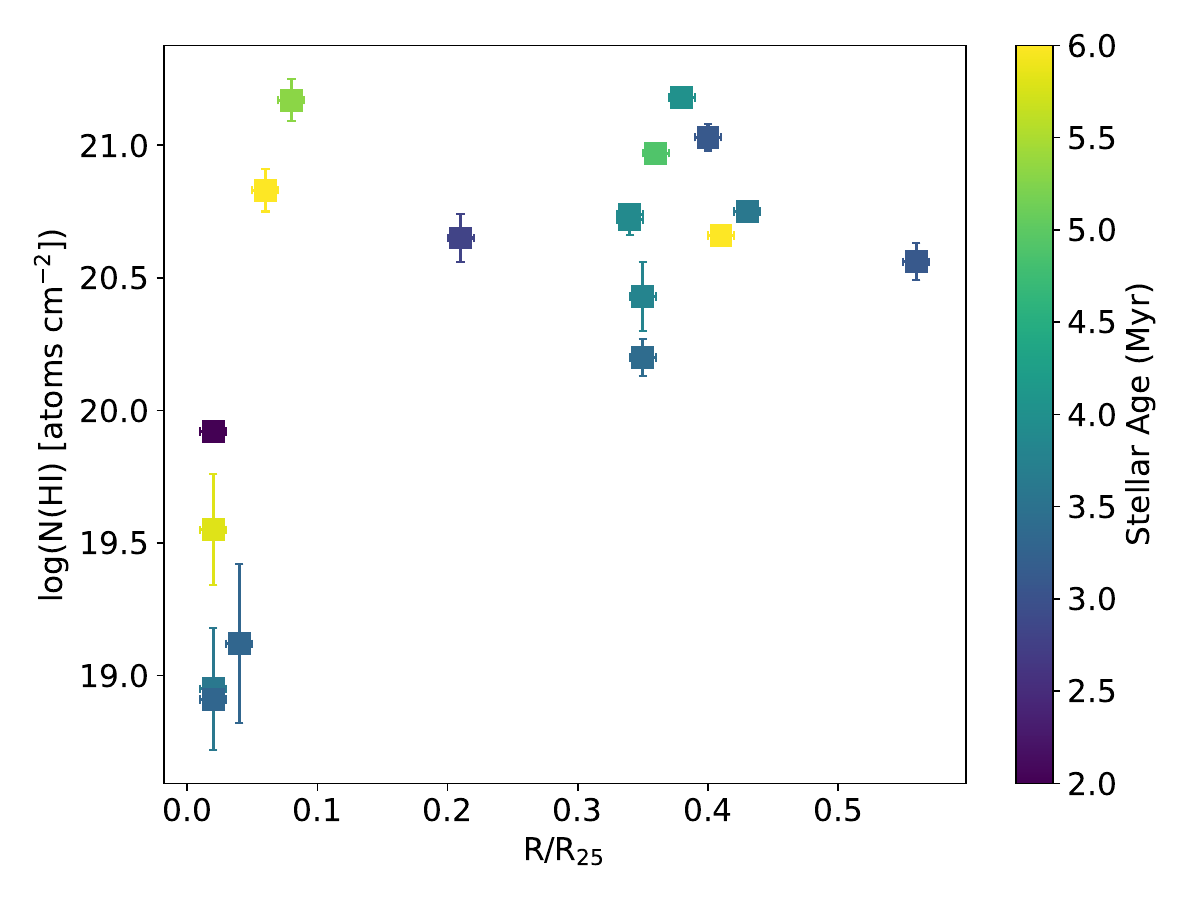}
    \caption{Distribution of hydrogen as a function of galactocentric distance (R/R$\rm _{25}$). Top subplot: For ionised \HII\ region, the H$\rm \alpha$ intensity is shown (in logscale in erg s$\rm ^{-1}$ cm$\rm ^{-2}$). Bottom subplot: For neutral gas, the \HI\, column density is shown (in logscale in atoms cm$\rm ^{-2}$). The top and bottom subplots are colored with stellar metallicity ([Z]) and stellar ages (in Myr) respectively. \label{hydrogen_distance_abs_em}}
\end{figure}

We report that diffuse cold molecular gas constitutes $\lesssim$7\% of the total hydrogen content along the line of sight of these YSCs located within the galactic centre. This further strengthens the idea that most of the H$\rm _2$ is within the warm gas. This is consistent with the findings of \citet{Sextl2025}, who show that the central dust cavity of M83 is nearly devoid of both atomic (\HI) and molecular gas, indicating a general depletion of ISM material across all phases in this region. Figure.~\ref{hydrogen_distance_abs_em} also shows a comparison of H$\rm \alpha$ intensity within the ionised region alongside the \HI\, column density within the neutral gas. We can clearly see a low column density of \HI\, in neutral gas around YSCs within the galactic centre and a significantly enhanced H$\rm \alpha$ intensity in \HII\ regions around those same YSCs compared to the YSCs in the disk. We do not see any specific patterns that stand out for the galactic centre in both stellar metallicity and stellar ages as a function of galactocentric distance, while we do note a spike in the \HI\, ionising flux within the galactic centre as calculated from SB99 models. To further understand the physical conditions responsible for this deviation, we will need to look at mm-based observations of CO that trace the dense molecular gas. We refrain from discussing this further as it deviates from the primary objective of this work. However, due to the extreme behaviour in N(\HI) we see, we will exclude the YSCs in the galactic centre from neutral gas and multi-phase discussion. On the other hand, while solely analysing the ionised phase, elemental abundances can be used. Hence, in all subsequent ionised-gas plots, data points corresponding to the galactic nucleus are included. This choice follows \citet{Hernandez2021}, who also excluded YSCs from M83's galactic nucleus when studying multi-phase trends. They noted that this decision was due to the complex physical conditions \citep[see also][]{Sextl2025} in the nucleus, where part of the observed absorption from their metallicity tracers (e.g., \SII) likely traces CO-dark molecular gas (i.e., warm \HH) rather than the classical neutral ISM.

\subsection{The \HII\ region \label{emission_results}}

The enrichment of the ISM is highly dependent on both the evolutionary stage of the stellar population and the timescales for chemical mixing between phases (ionised, \HII\ and neutral, \HI). Metals released by massive stars during starburst episodes are first injected into the hot ionised gas on short $\sim$Myr timescales \citep[][]{Kunth_Sargent_1986, Westmoquette2013b}, while their mixing into the neutral phase can take up to a billion (10$^{9}$) years \citep[][]{Lebouteiller2013}. This leads to observable spatial and temporal variations in elemental abundances-particularly for nitrogen, which can be released on very short timescales through the powerful winds of massive (O, B) stars\citep[see][]{Martins2024}, very massive stars \citep[VMSs, 100-1000 M$_{\odot}$, $\lesssim$2 Myr, see][]{Vink2023}, and Wolf-Rayet \citep[W-R stars, $\sim$2-5 Myr, see][]{Schaerer1997, Westmoquette2013b}, and later enriched over $\gtrsim$100 Myr by asymptotic giant branch (AGB) stars \citep{Kobayashi2020}. Spatially mapped ionised regions also allow us to directly capture localised, short-lived chemical signatures-such as anomalously high N/O ratios coinciding with clusters containing W-R stars \citep[][]{James2009, Kumari2018, Westmoquette2013b} - before they are diluted or homogenised across larger ISM reservoirs. $\alpha$-elements (S and O) are produced primarily by CCSNe and mix within $\sim$50 Myr. Iron is predominantly supplied by Type Ia SNe over gigayear (10$^9$ yr) timescales \citep[][]{Matteucci2012, Kobayashi2020}, far longer than the ages of the YSCs probed here. As such, different elemental abundances serve as clocks of galactic chemical evolution. However, understanding how different galactic environments affect the chemical mixing of these elements is still an open question.

Probing elemental abundances in \HII\ regions around YSCs in M83 is essential for capturing the time-dependent and multiphase nature of chemical enrichment in star-forming galaxies. Spatially resolving systems on a cluster-by-cluster basis has already demonstrated the importance of this approach. \citet{Hernandez2021} compared stellar, neutral, and ionised gas metallicities on $\sim$100 pc scales, finding consistent values outside the nucleus but also showing that metals mix on timescales longer than the lifetimes of massive stars. \citet{Abril-Melgarejo2024} studied star-forming clusters within the BCD, NGC 5253, which is from the same galaxy group as M83, but significantly lower in metallicity and far smaller in mass. They demonstrated that metal enrichment is strongly age-dependent: N/O and (N/H) increase with age in the neutral phase but decrease in the ionised gas, with offsets between the phases shrinking on $\sim$ 10-15 Myr timescales as WN-type Wolf-Rayet stars inject nitrogen first into the ionised phase ($\sim$2-5 Myr) before it cools and mixes into the neutral ISM. However, the timescales reported by \citet{Abril-Melgarejo2024} stem from their adoption of photometric cluster ages from \citet{Calzetti2015}, rather than spectroscopically constrained ones. Adopting spectroscopic ages derived by \citet{Hernandez2026} for the same COS sample as that by \citet{Abril-Melgarejo2024}, the enrichment timescale in NGC 5253 is likely much shorter ($\sim8$ Myr) than previously inferred photometric cluster ages, which are more similar to YSCs in M83 studied in this work.

\subsubsection{(O/H) vs (X/H) in ionised gas \label{emission_results_1}}
We begin by comparing our oxygen abundances with the abundance of other elements within ionised gas. The correlations (or lack thereof) between (O/H) and other (X/H) ratios in the ionised gas can reveal not only whether elements share common production channels and enrichment timescales, but also whether metallicity itself influences the efficiency of metal mixing within the ionised gas \citep[see e.g.][]{Abril-Melgarejo2024, James2026}. 

\begin{figure}
    \centering
    \includegraphics[width=0.5\textwidth]{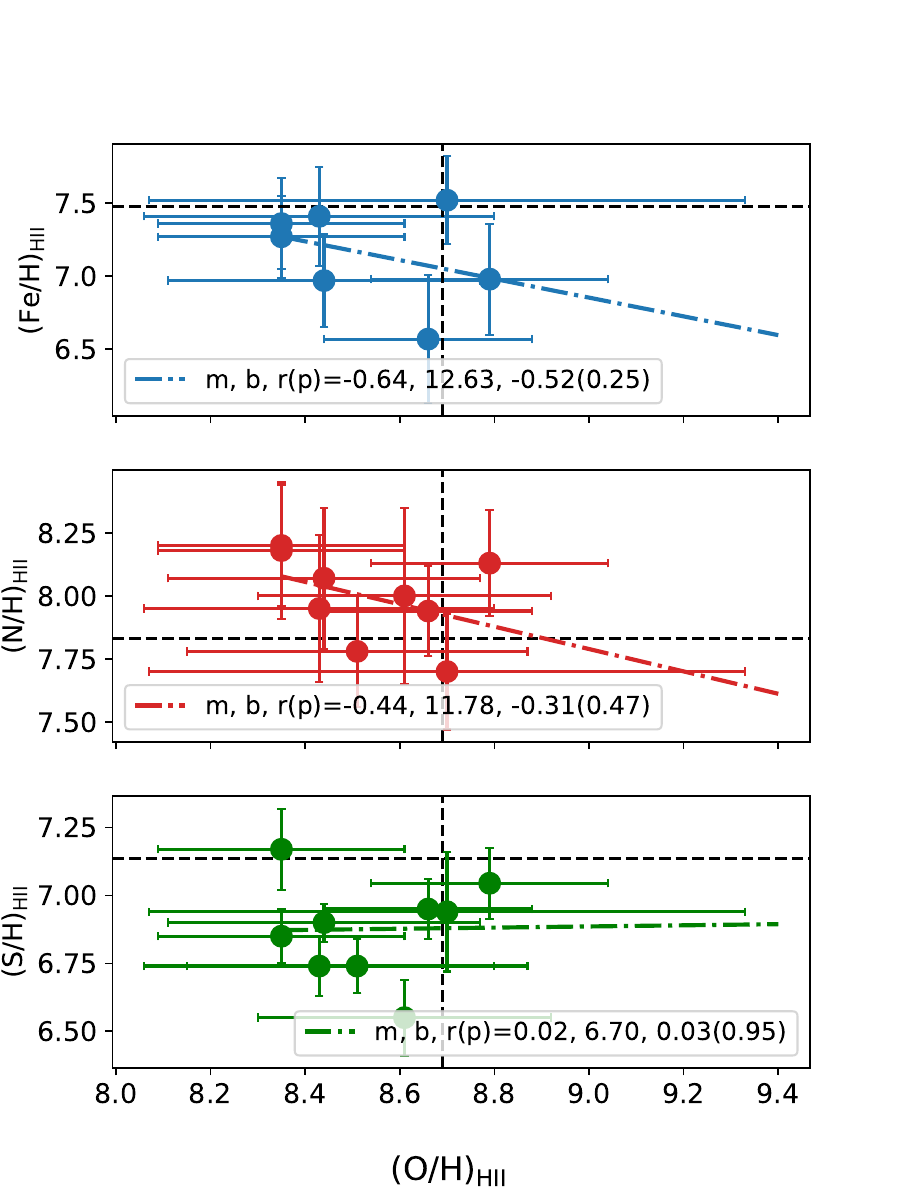}
    \caption{Elemental abundances (12+log(X/H)) of X=iron, nitrogen,  and sulphur plotted as a function of oxygen in ionised gas (\HII\ region) around YSCs in M83. Only detections are shown here for clarity. The legend for each subplot shows the statistics: fitted line slope, $m$, intercept, $b$, Pearson correlation coefficient, $r$ and significance, $p$-value in brackets. The colored dashed line represents the best line fit, and the horizontal/vertical black dashed line represent solar abundance for each element. The legends appear in boldface, and the best-fit line is thicker if the Pearson correlation is significant, i.e. $p$-value$\lesssim$0.05. \label{hii_abundance_oxygen_vs_others}}
\end{figure}

Figure.~\ref{hii_abundance_oxygen_vs_others} shows the elemental abundances, (X/H)$\rm _{HII}$ (where X = Fe, N and S) for \HII\ regions as a function of their oxygen abundance, (O/H)$\rm _{HII}$. (O/H) shows mild anti-correlation with (Fe/H). IFS studies of dwarf galaxies show that spatial variations in Fe can be linked to shocks, inflows, and local starburst episodes \citep[][]{James2010, James2013a, Kumari2017} - which could explain the very large scatter that we see within Fe abundances in the ionised gas. We see no correlation with (N/H). However, we note that nitrogen and oxygen are consistently close to solar, with one YSC having super-solar oxygen and nitrogen abundance. This coincides with the well-established increase of N/O with (O/H) at high metallicity, reflecting secondary nitrogen production and coupled chemical enrichment in ionised gas \citep[e.g.,][]{Bresolin2005, Bresolin2009, Bresolin2012}. This behaviour suggests either linked enrichment pathways or comparable mixing timescales for elements produced through different nucleosynthetic channels. 
%There is also an influence of oxygen on ICF correction of (N/H); however, this is small: $\sim$ 0.2 and unlikely to be the primary reason for the correlation.

(O/H) shows no correlation with (S/H). Since both O and S are $\rm \alpha$-process products of massive star core-collapse SNe (CCSNe) and should broadly track each other in the ionised phase. Hence, we note that our analysis of the (O/H)-(S/H) correlation warrants more data points. 

We further compare the relative elemental abundance patterns measured around the YSCs with the solar abundance pattern. Since the median oxygen abundance of our sample is close to solar, differences in O-X primarily reflect changes in the abundance of element X relative to oxygen. Specifically, ${\rm O-X}>({\rm O-X}){\odot}$ corresponds to an X/O ratio below the solar value, whereas ${\rm O-X}<({\rm O-X}){\odot}$ corresponds to an X/O ratio above solar. Because oxygen is produced almost exclusively by core-collapse supernovae and is only weakly depleted onto dust, it provides a robust reference element for comparing the enrichment histories of other species. 

For sulphur, we measure ${\rm O-S}=1.70\pm0.13$, compared to the solar value of 1.56. Since both oxygen and sulphur are $\alpha$-elements synthesised predominantly in massive stars and released by core-collapse supernovae, their abundance ratio is expected to remain nearly constant over the short ($<10$ Myr) lifetimes of our YSCs. The small positive offset of $0.14\pm0.13$ dex is therefore consistent with only a modest reduction in the gas-phase S/O ratio, likely reflecting a combination of mild sulphur depletion and measurement uncertainties, rather than a distinct nucleosynthetic signature.

For iron, we find ${\rm O-Fe}=1.47\pm0.16$, compared to the solar value of 1.22, indicating a gas-phase Fe/O ratio approximately 0.25 dex below solar. Because iron is highly refractory, this offset is most naturally explained by depletion of Fe onto dust grains, since young stars are yet to produce any iron.

In contrast, nitrogen exhibits a significantly different behaviour. We measure ${\rm O-N}=0.58\pm0.13$, compared to the solar value of 0.86, corresponding to an N/O ratio approximately 0.28 dex above solar. Since both oxygen and nitrogen are only weakly affected by dust depletion, this offset is unlikely to arise from differential depletion and instead reflects genuine nitrogen enrichment. This interpretation is consistent with the localised chemical enrichment expected around young massive clusters, where nitrogen can be rapidly injected into the surrounding ISM through the winds of massive stars on timescales of $\lesssim5$ Myr, well before the delayed contribution from AGB stars. The elevated N/O ratios therefore support the picture that our spatially resolved observations are capturing localised, short-lived enrichment signatures around young stellar populations before they become diluted into the larger ISM, consistent with previous observations of nitrogen-enhanced \HII\, regions associated with massive stars \citep[see e.g.,][]{James2009, Kumari2018, Westmoquette2013b, Abril-Melgarejo2024}.

Taken together, these results highlight the need to probe more data points to confirm sulphur and oxygen coupling as is expected from their nucleosynthetic origins; nitrogen abundance seems to anti-correlate with oxygen in ionised gas surrounding YSCs. Iron, on the other hand, is decoupled from oxygen. Similar decoupling of Fe from $\alpha$-elements has been reported in both local star-forming galaxies \citep[][]{Abril-Melgarejo2024, James2026} and in theoretical enrichment models \citep[see][]{Webster2015, Emerick2019}. To further probe enrichment timescales by young stars, we need to study the relation of elemental abundances as a function of the average stellar age of the YSCs within M83.

\subsubsection{ionised (X/H)$\rm _{HII}$ as a function of stellar age \label{emission_results_2}}
\begin{figure}[ht]
    \centering
    \begin{subfigure}{0.48\textwidth}
        \centering
        \includegraphics[width=\textwidth,trim={0 0 0 0},clip]{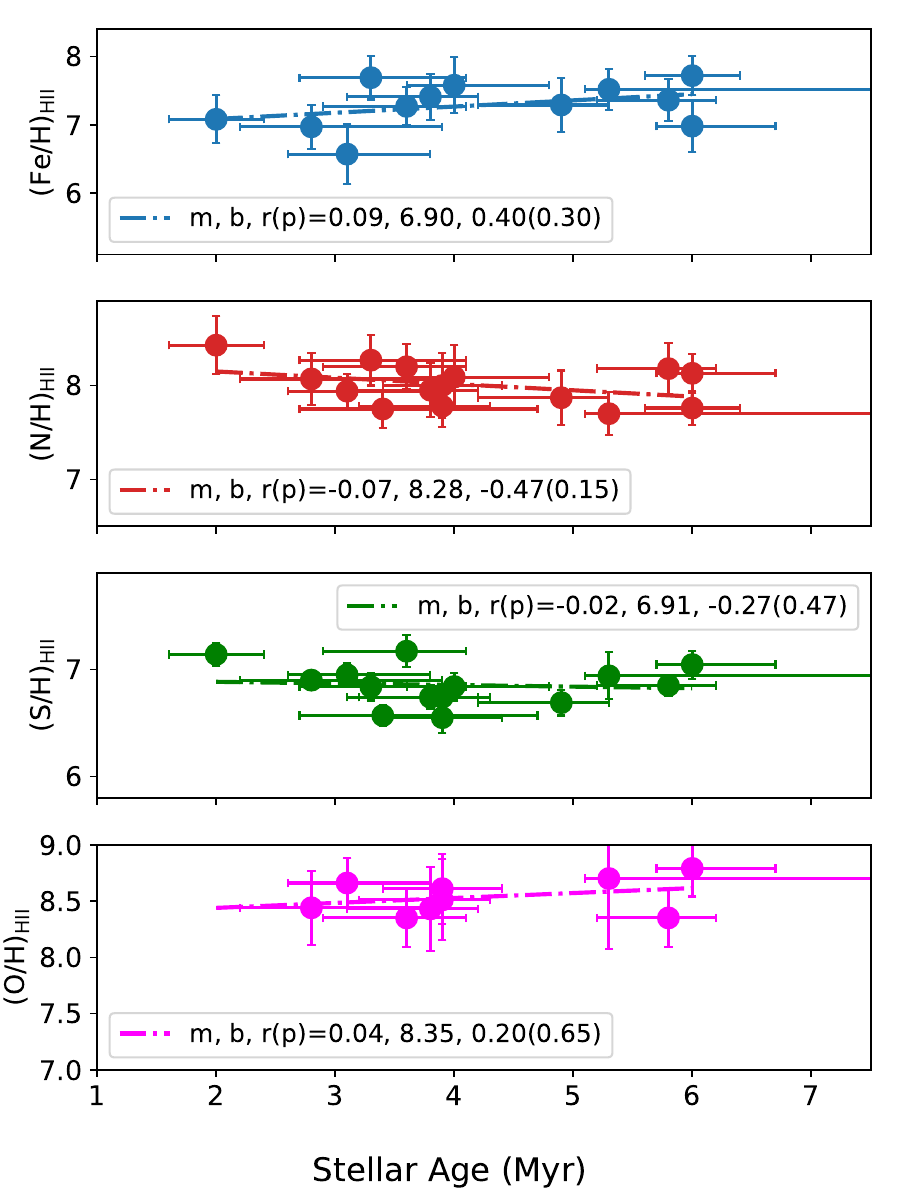}
    \end{subfigure}
    \caption{Iron (Fe/H), nitrogen (N/H), sulphur (S/H) and oxygen(O/H) abundance in \HII\ regions around YSCs shown as a function of the stellar age (Myr) of the YSCs. The colored dashed lines represent the weighted best-fit for each subplot. Legends are the same as Figure.~\ref{hii_abundance_oxygen_vs_others}.   \label{hii_abundance_st_age_fig}.}
\end{figure}

In this section, we examine how the abundances of different elements vary with the best-fitting stellar age of the natal YSCs. Figure~\ref{hii_abundance_st_age_fig} presents the correlations of (Fe/H), (S/H), (O/H), and (N/H) in the ionised gas with stellar age. We find that (Fe/H) slightly increases with cluster age, but the trend is only mildly significant (Pearson $r$ = 0.4, $p$-value = 0.3). Fe is primarily produced through Type Ia SNe (mixing within $\sim$1 Gyr) and hence it is unlikely to have any trend with stellar age within these young stellar lifetimes. $\alpha$-elements (oxygen, sulphur), in contrast, show no significant correlation with stellar age. Nitrogen displays moderate anti-correlation with stellar age (Pearson, $r$ = -0.47, $p$-value = 0.15). The negative correlation might hint at fresh injection of nitrogen by massive stars within the youngest clusters. 

\citet{Abril-Melgarejo2024} find strongly falling (N/H), (O/H) in \HII\ regions in NGC5253. $\alpha$-enrichment is tied mainly to CCSNe, which begin to contribute significantly only after $\sim$3-5 Myr and continue over tens of Myr. Hence, the statistics seen here might need revision after additional observations of YSCs with stellar age $\gtrsim$10 Myr.

\subsubsection{ionised gas kinematics \label{emission_results_3}}

\begin{figure}
    \centering
    \includegraphics[width=0.5\textwidth]{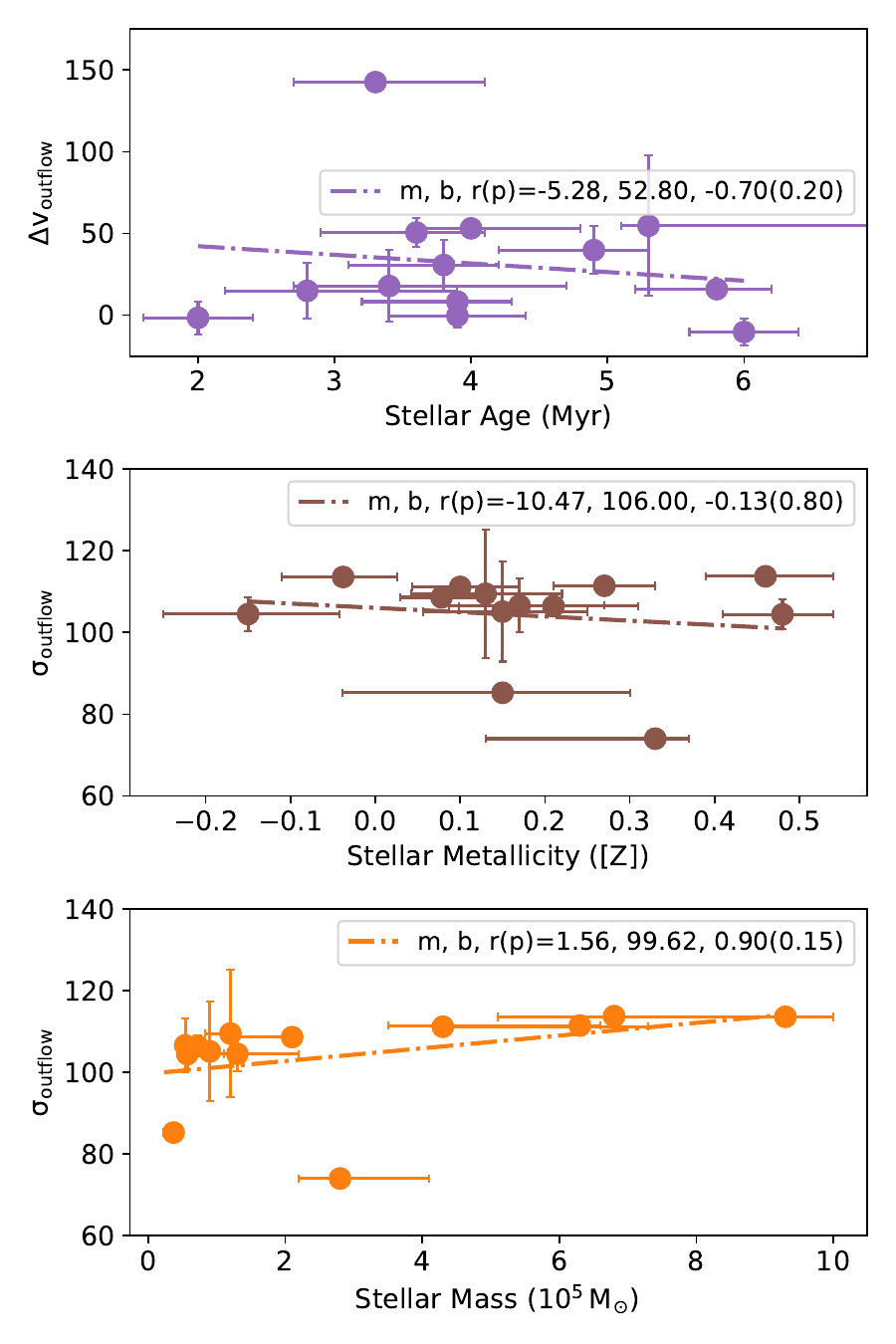}
    \caption{Kinematic properties of \HII\ region gas ($\sigma_{\rm outflow}$ and $\Delta v_{\rm outflow}$, see exact definition in sub-section-\ref{emission_results_3}) plotted against different stellar properties. The colored dashed lines represent the best line fit for each subplot. Legends are the same as Figure.~\ref{hii_abundance_oxygen_vs_others}.   \label{hii_region_kinamatics_1}}
\end{figure}

For a massive spiral, the gravitational potential might be strong enough to bind the gas within the \HII\ region as compared with a blue compact dwarf, where the dispersion is quicker for similar feedback. Alternatively, metal-rich environments also have much stronger feedback, injecting a lot more metals into the surrounding ionised regions within a smaller timescale \citep[see e.g.][and references therein]{Mokiem2007}. Hence, to investigate the role of feedback and outflows in shaping the ionised ISM, we examined the kinematic properties of \HII\ regions relative to their natal stellar populations (see Figure~\ref{hii_region_kinamatics_1}). Specifically, we measured the velocity offsets of the main (strongest flux) and outflowing (widest profile) ionised \HII\ region components from strong and prominent optical emission lines (e.g. H$\alpha$, H$\beta$, [\OIII], [\NII], and [\SII], see Figure.~\ref{fig:emission_line_plot_obj_1}) with respect to the stellar velocity ($\Delta v_{\rm main}\,=v_{\rm main}-v_{\rm stellar}$, $\Delta v_{\rm outflow}\,=v_{\rm outflow}-v_{\rm stellar}$), as well as their respective velocity dispersions ($\sigma_{\rm main}$, $\sigma_{\rm outflow}$).

We do not find any significant correlation between kinematics: ($\Delta v_{main}$, $\sigma_{main}$, $\Delta v_{outflow}$ and $\sigma_{outflow}$) and stellar mass or metallicity. We see weak anti-correlations between stellar age and kinematics (both v and $\sigma$) and a weak correlation of outflow with stellar mass. This hints at turbulent environments being related to younger, more massive star clusters. However, we note that drawing firm conclusions about ionised gas kinematics will require more observed data points of ionised gas surrounding YSCs.

\subsection{The \HI\, regions \label{absorption_results}}

Neutral (\HI) gas phase holds critical information about how freshly synthesised metals mix and cycle through the interstellar medium (ISM) on cluster-to-galaxy scales. To this end, it is important to study the evolution of neutral gas around each YSC. \citet{James_2014} performed a spatially resolved study of neutral gas around YSCs within star-forming galaxies using HST/COS spectroscopy. They show that a substantial fraction of metals (including N, O, S, and Fe) reside in the neutral phase and are often overlooked when focusing solely on the \HII\ regions. Similarly, \citet{Hernandez2021} directly compared stellar, ionised, and neutral metallicities in M83 and found that while metals appear well mixed on $\sim$100 pc scales outside the nucleus, caution is needed in molecular-gas-dominated regions since typical neutral-gas metallicity tracers (e.g., \SII) may provide inaccurate abundances, as these may trace multiple components of the ISM, neutral and molecular gas phases. Most recently, \citet{Abril-Melgarejo2024} revealed that, in BCDs, the enrichment proceeds differentially, i.e. N/O and (N/H) increase with age in the neutral gas but decrease in the ionised phase, reflecting a two-step process in which massive stars enrich the ionised medium on $\sim$2-5 Myr timescales, followed by delayed mixing into the cold neutral phase over $\sim$10-15 Myr. Their results highlight that neutral gas abundances show strong age correlations (e.g. (N/H), N/O) and distinct radial behaviours (e.g. (Fe/H) anticorrelated with radius), while (O/H) remains relatively flat. This is evidence that different elements trace different nucleosynthetic channels and mixing efficiencies. This broader picture, echoed in CLASSY results \citep[][]{James2026} where offsets between ionised and neutral abundances vanish only after $\sim$20-30 Myr, further underscores why probing the neutral gas in M83 is essential, i.e. it captures the delayed imprint of stellar yields, and thereby the true timescales of metal cycling. Additionally, studying neutral gas also gives important information on the previous generations of stars and their yields. In this section, we report the results related to the abundances of elements within neutral gas around YSCs in M83.

\subsubsection{(O/H) vs (X/H) in neutral gas \label{absorption_results_2}}

\begin{figure}
    \centering
    \includegraphics[width=0.5\textwidth]{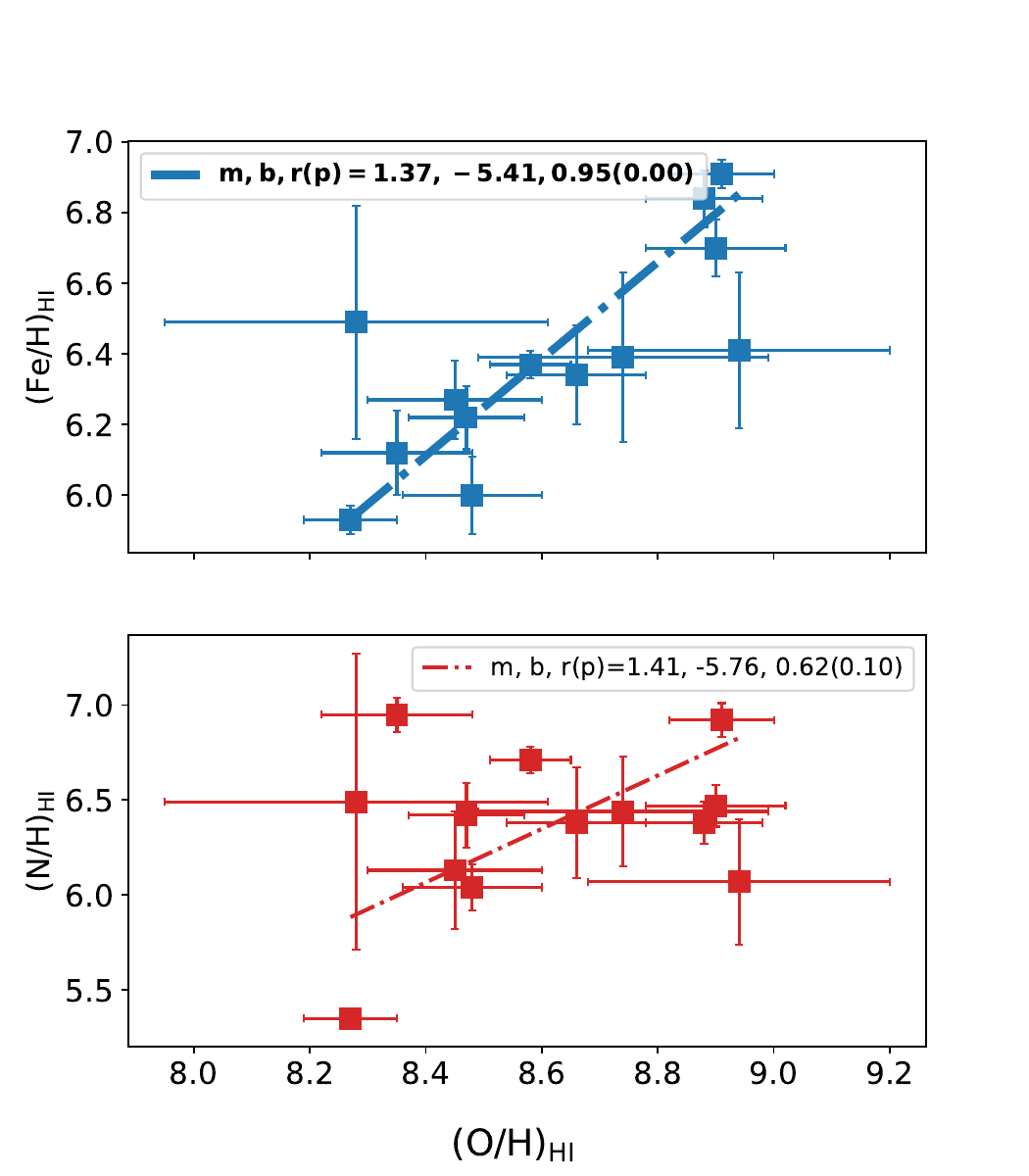}
    \caption{Comparing elemental abundances of iron ((Fe/H), blue) and nitrogen ((N/H), red) to that of oxygen (O/H) in neutral gas (\HI\, region) around YSCs in M83. The legends are the same as Figure.~\ref{hii_abundance_oxygen_vs_others}. The points from the galactic centre (R/R$_{25}<$0.05) have been removed for clarity. \label{elemntal_abun_vs_oxygen_neutral}}
\end{figure}

The correlations between elemental abundances and other parameters (e.g. stellar age and dust) in the neutral gas of M83 provide critical insight into the cooling and mixing timescales of metals, and the influence of dust depletion. By examining oxygen abundance against iron and nitrogen, we can distinguish between the effects of mixing after $\rm \alpha$-element pathways, delayed Type Ia and AGB contributions, and element-specific depletion effects. Figure.~\ref{elemntal_abun_vs_oxygen_neutral} shows the comparison between oxygen abundance (O/H) and other elemental abundances, (X/H)$\rm _{HI}$ in the neutral gas (where X=Fe and N).

It is well known that $\rm \alpha$-element ratios remain relatively constant in local star-forming systems, from metal-rich spirals like M83 to low-metallicity dwarfs and CLASSY galaxies \citep[see e.g.][]{Izotov2006b, Croxall2016, Berg2019, Arellano-Cordova2024}. This is also consistent with chemical evolution models in which O and S co-evolve through massive star yields, and with IFU studies showing limited small-scale inhomogeneity in $\rm \alpha$-element distributions of the ionised gas \citep[][]{Ritter2015, Corlies2018}. Yet, we cannot address this connection in our neutral gas sample as the oxygen column density is derived directly from sulphur. 

We note that the O-Fe relation is positive and significant (Pearson $r$ = 0.95, $p$-value $<$ 0.01). For a nearly solar oxygen abundance but an iron abundance lower by $\sim$1 dex, the average abundance ratio in the neutral ISM of M83 is O$-$Fe $=2.25\pm0.33$ dex, substantially higher than the solar value of 1.22 dex. This corresponds to a gas-phase Fe/O ratio that is approximately one dex below the solar abundance pattern, indicating a significant deficiency of gas-phase iron relative to oxygen. Since oxygen is only weakly depleted onto dust whereas iron is highly refractory, the elevated O$-$Fe ratio is most naturally explained by the strong depletion of iron onto dust grains in the neutral ISM. A secondary contribution from the delayed nucleosynthetic production of iron, primarily by Type Ia supernovae, cannot be excluded; however, the magnitude of the observed offset is consistent with dust depletion being the dominant effect (see Figure.~\ref{fig:data5}). Unlike sulphur, iron is not an $\alpha$-element and therefore traces a fundamentally different chemical enrichment history. The observed correlation between Fe/H and O/H suggests that the iron contributed by previous generations of stars has become well mixed throughout the neutral ISM. Although Fe/H scales with O/H, the systematically elevated O$-$Fe ratio relative to the solar abundance pattern indicates that the gas-phase iron abundance remains significantly suppressed. Consequently, iron behaves as a ``slow-clock'' element whose gas-phase abundance records the cumulative effects of long-term chemical enrichment, dust depletion, and multiphase mixing within the neutral ISM.  

Nitrogen also exhibits a positive correlation with oxygen in the neutral phase (Pearson, $r = 0.62$, $p = 0.10$). While the oxygen abundance is nearly solar, the nitrogen abundance is lower by approximately 1.2 dex, resulting in a mean abundance offset of O$-$N $= 2.28 \pm 0.34$ dex. This is substantially larger than the solar value of 0.86 dex and indicates a pronounced nitrogen deficiency relative to oxygen in the neutral ISM of M83. Unlike O and S, which are prompt CCSNe products, N has dual origins: primary N is produced in massive stars on $\sim$2-5 Myr timescales \citep[][]{Martins2024}, while secondary N arises from AGB stars on delayed ($\sim$250 Myr) timescales and with metallicity dependence \citep[see][]{Johnson2023}. \citet{Johnson2023} further showed that AGB yields are essential to reproduce the observed N/O-(O/H) relations in galaxies and that secondary production dominates the rise of N/O once metallicity builds up. There is a significantly large O-N offset in M83's neutral gas. \citet{Abril-Melgarejo2024} reports such an offset in NGC5253 YSCs and state that the offset diminishes within $\sim$8Myr. They note that the winds of W-R stars mix the nitrogen into the neutral gas within such a time frame. Hence, for further analysis, we need to look at the nitrogen offset as a function of stellar age.

Together, these correlations highlight not only the layered enrichment history of M83's neutral gas, but also the strong chemical homogeneity across elements within the neutral ISM despite their distinct nucleosynthetic origins. Iron also tracks oxygen, but only in neutral gas, reflecting that the SNe injection has probably mixed well with the larger neutral ISM but is susceptible to dust depletion within both phases \citep[][]{De_Cia_2024}. Nitrogen correlates strongly with oxygen but remains significantly under-abundant in the neutral phase relative to the solar abundance as well as the ionised phase.

\subsubsection{Neutral (X/H)$\rm _{HI}$ as a function of stellar age \label{absorption_results_3}}

\begin{figure}[ht]
    \centering
    \begin{subfigure}{0.48\textwidth}
        \centering
        \includegraphics[width=\textwidth]{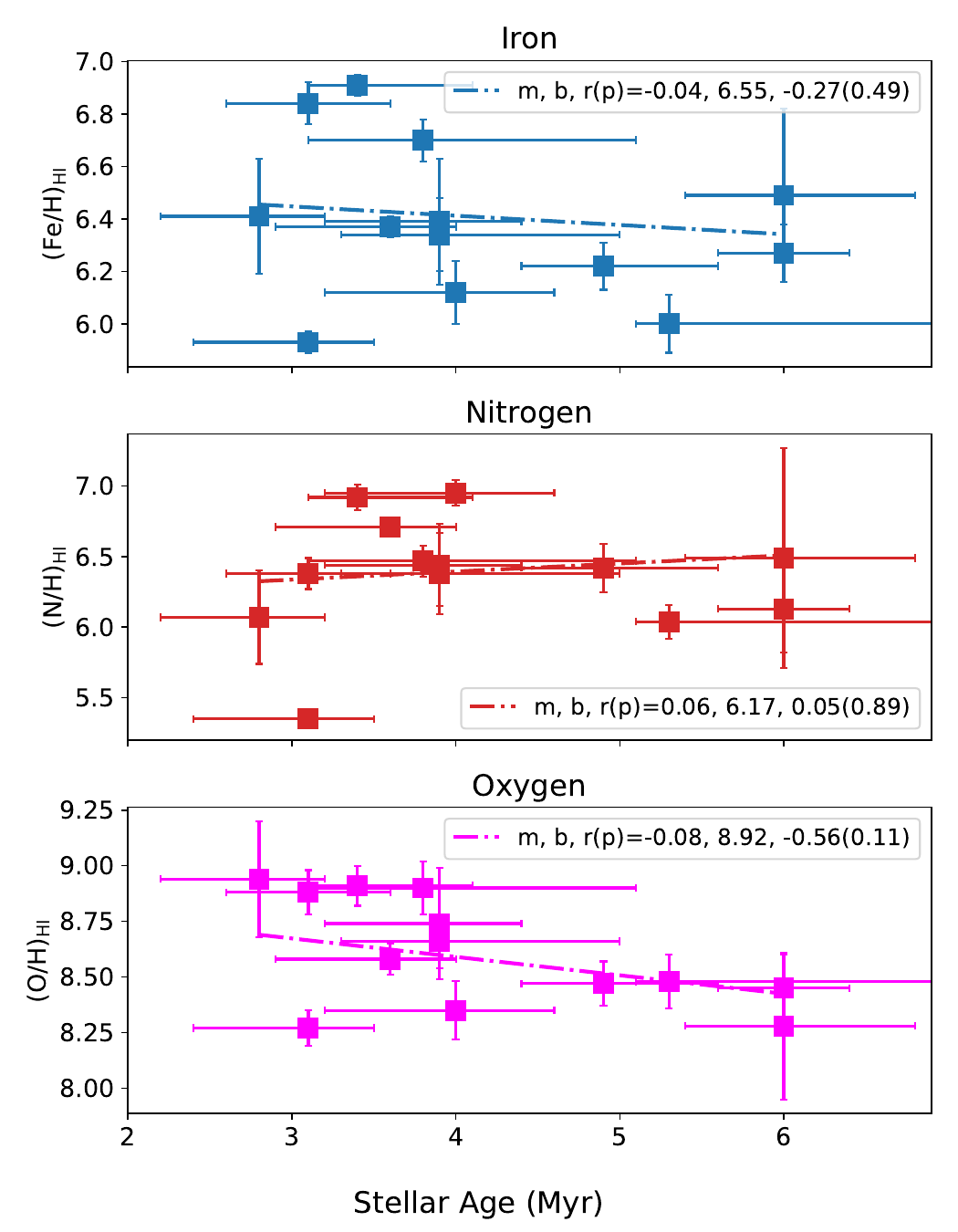}
    \end{subfigure}
    \caption{Neutral gas elemental abundances of iron ((Fe/H), blue), nitrogen ((N/H), red), sulphur ((S/H), green), and oxygen ((O/H), pink) as a function of the stellar age of their natal-star clusters in M83. The legends are the same as Figure.~\ref{hii_abundance_oxygen_vs_others}.   \label{elemntal_abun_vs_age_neutral}.}
\end{figure}

Building on the elemental correlations discussed above, we now consider how neutral-phase abundances of different elements vary with the ages of their natal young star clusters ($\sim$1-6 Myr) in M83. Figure.~\ref{elemntal_abun_vs_age_neutral} shows the neutral gas elemental abundances ((X/H)$\rm _{HI}$) plotted as a function of the mean stellar age of the YSCs that they surround. Weak to moderate anti-correlations are observed for (O/H), (S/H), and (Fe/H) with stellar age, while (N/H) shows no significant dependence. This is expected as the timescale is too short for elements from the current stellar population to have mixed well within neutral gas. \citet{Abril-Melgarejo2024} note that the elements from the YSCs in NGC5253 start mixing with the neutral gas within $\sim$8Myr. However, YSCs in M83 are younger ($\sim$6Myr) with a different environment. Hence, we investigate the mixing within M83's neutral gas. 

Oxygen shows anti-correlation with age (Pearson $r$ = -0.56, $p$-value = 0.11). This is in contrast with what was found in NGC5253 by \citet{Abril-Melgarejo2024}. The decline with age seen here is insignificant and hence cannot be interpreted in the context of young stellar enrichment. The neutral gas traces a long-lived reservoir that retains the imprint of previous generations of star formation, whereas the stellar ages we measure correspond to the currently ionising clusters. Thus, even if the youngest clusters are actively enriching their surrounding ionised gas, the neutral phase will primarily reflect older enrichment cycles, leading to little or no dependence on the present-day cluster ages. Iron exhibits a weak anti-correlation trend (Pearson $r$ = -0.27, $p$-value = 0.49), with younger clusters showing higher neutral (Fe/H). Although given the short timescale, this cannot be attributed to any known iron enrichment mechanism, especially when $alpha$-elements do not follow the same trend. Nitrogen in neutral gas shows no correlation (Pearson $r$ = 0.05, $p$-value = 0.89) with stellar age. Since the trends are only moderately significant, we refrain from drawing conclusions from it. We further note that a larger sample of YSCs with higher stellar age ($\gtrsim$10Myr) would help provide a more robust statistical sample to study the neutral gas abundance trend with stellar age.

The large O-N offset in M83's neutral gas cannot be produced by the current YSCs alone: nitrogen from massive stars is released on $\sim$5 Myr timescales, while significant oxygen enrichment from CCSNe builds up over much longer ($\gtrsim$50 Myr) timescales, and our clusters are only $\lesssim$6 Myr old. If the neutral gas reflects enrichment from earlier generations with enough time to mix ($\gtrsim$100 Myr), the nitrogen deficit could indicate either intrinsically weak nitrogen-producing channels (e.g., few O-B/VMS/W-R/AGB stars) or it is possible that CCSNe-driven feedback expelled the N-enriched material out of the neutral gas or heated much of it, leaving an oxygen-dominated signature in the cold neutral reservoir. Distinguishing between these possibilities requires knowledge of the prior star-formation history and multi-phase outflow constraints, which is beyond the scope of this work.

We further note that nitrogen is more enriched relative to $\alpha$-elements within ionised gas. This indicates that the nitrogen enrichment from current young stars has been mixed with ionised gas, but not the neutral gas. One possible contributing factor is the deeper gravitational potential wells of massive spirals, which confine feedback and create more compact, tightly bound ionised regions. This could make it harder for nitrogen-rich material injected by current-generation massive stellar winds to disperse widely into the neutral ISM.

\subsubsection{The impact of dust depletion \label{absorption_results_1}}

Dust plays a central role in regulating the physical and chemical state of the ISM. It enables \HH\, formation, shields dense gas from ionising radiation, and provides cooling pathways that promote star formation \citep[][]{Vidali2013, Whitworth_Bate_2002, Hopkins_Lee_2016}. By absorbing UV/optical light and re-emitting in the infrared, dust also shapes galaxy SEDs \citep[][]{Gordon1997, Dole2006}, while PAHs contribute to the thermal balance of the gas \citep[][]{Wolfire1995, Tielens2008}. In metal-rich systems like M83, dust content broadly traces metallicity, yet the detailed coupling between dust and individual elemental abundances remains uncertain. Measuring neutral-phase abundances as a function of dust provides a direct test of how depletion, stellar yields, and mixing timescales shape the observed chemical composition \citep[][]{DeCia2018, De_Cia_2024}. Thus, the impact of dust depletion of elemental columns and the trends of (X/H) and X/O ratios with dust extinction E(B-V) reveal which elements are most susceptible to depletion, which are minimally affected, and which trace longer-term nucleosynthetic enrichment.

We estimate the dust depletion strength in the neutral ISM of M83 using the [Fe/Zn]\footnote{We do not directly measure Zn abundances in this work. The notation [Fe/Zn] is retained following the historical convention established by \citet{Ledoux2002}; here, the depletion strength is inferred from a combination of elements with different refractory properties.} prescription of \citet{De_Cia_2024} and \citet{Konstantopoulou2022}. Applying the full depletion prescription increases the median Fe-, S-, and O-abundances by approximately 1.1, 0.44, and 0.13 dex, respectively, demonstrating that refractory elements are substantially more affected by dust than oxygen. However, the depletion corrections depend sensitively on the measured column densities of the refractory species, particularly Fe, Ni, and Mn. In our data, the Mn absorption lines fall within a relatively noisy spectral region, introducing additional uncertainty into the inferred depletion strength. Furthermore, the empirical coefficients adopted from \citet{Konstantopoulou2022} were calibrated using the Milky Way ISM and damped Lyman-$\alpha$ absorbers, whose physical conditions and metallicities may not fully represent those of M83. Furthermore, nitrogen is excluded from this depletion analysis because no corresponding prescription is currently available \citep{De_Cia_2024, Konstantopoulou2024a}. Consequently, we do not use the depletion-corrected abundances in the subsequent analysis and instead adopt only the ionisation-corrected abundances. 

While carbon could give us more information about depletion onto graphite and PAH grains (actively being studied using JWST; see e.g. \citet[][]{Hernandez2023, Jones2025}, we could not constrain the total column density of carbon due to the prominent \CII\, lines being saturated. This has also been a common limitation in previous neutral gas UV-absorption studies \citep[see e.g.][]{James_2014, Ranjan2022}. Rather than relying on the full depletion corrections, we instead use the inferred dust reddening, $E(B-V)$, derived from the depletion formalism of \citet{DeCia2018}, as an empirical tracer of the dust column. We then examine how the ionisation-corrected abundances and abundance ratios vary with $E(B-V)$, allowing us to assess the influence of dust on individual elements while avoiding direct reliance on the more uncertain depletion-corrected abundances.

\begin{figure}
    \centering
    \includegraphics[width=0.5\textwidth]{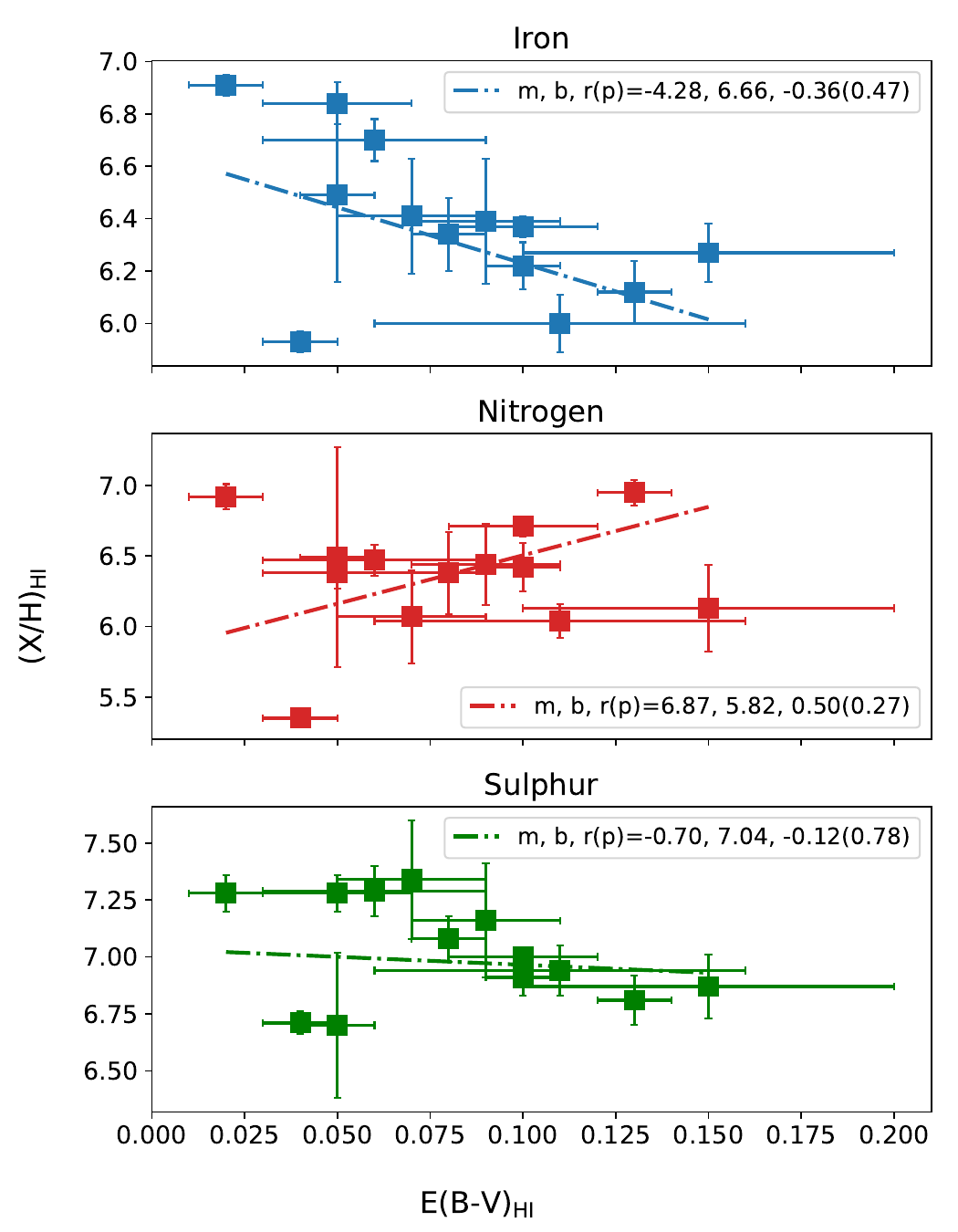}
    \caption{Abundance of different elements: iron ((Fe/H), blue), nitrogen ((N/H), red), and sulphur ((S/H), green) as a function of the amount of dust (E(B-V)) in the neutral gas (\HI\, region) around YSCs in M83. The legends are the same as Figure.~\ref{hii_abundance_oxygen_vs_others}.  \label{elemental_abundance_vs_dust_neutral}}
\end{figure}

Figure~\ref{elemental_abundance_vs_dust_neutral} shows how the ionisation-corrected neutral-phase abundances of Fe, S, O, and N vary with E(B-V). We find anti-correlations for (Fe/H) (Pearson $r$ = -0.36, $p$-value = 0.47) and (S/H) (Pearson $r$ = -0.12, $p$-value = 0.78). The anti-correlations confirm that higher dust columns are associated with higher depletion. By contrast, nitrogen shows correlation with E(B-V) (Pearson $r$ = 0.5, $p$-value = 0.27). While the trends for Fe and S align with the expectation that dust-rich sightlines will deplete, the lack of correlation for (N/H) might indicate nitrogen's insensitivity to dust.

\begin{figure}
    \centering
    \includegraphics[width=0.5\textwidth]{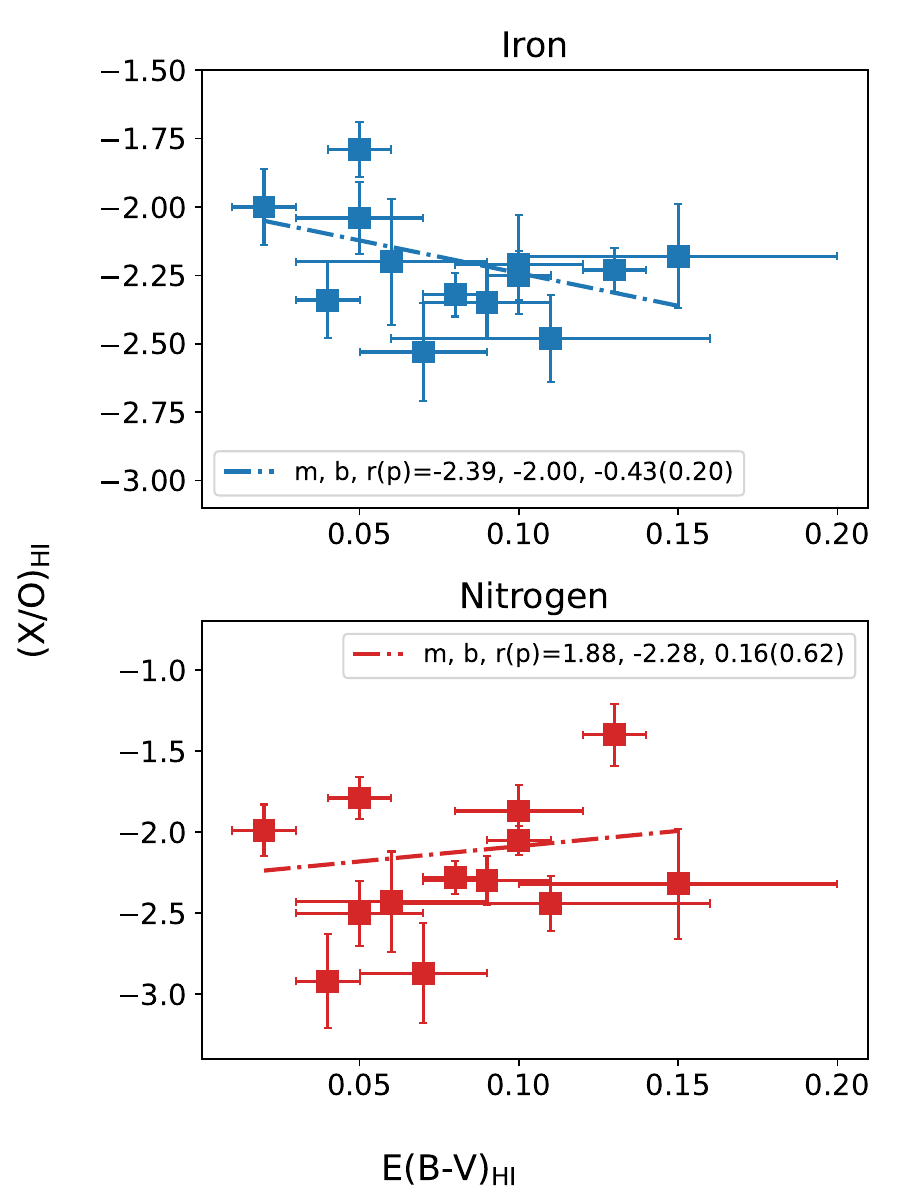}
    \caption{Depletion of different elements relative to oxygen: iron (Fe/O, blue) and nitrogen (N/O, red) as a function of the amount of dust (E(B-V)) in the neutral gas (\HI\, region) around YSCs in M83. The legends are the same as Figure.~\ref{hii_abundance_oxygen_vs_others}.   \label{elemental_depletion_vs_dust_neutral}}
\end{figure}

Since our system is unique with much higher metallicity than average galaxies, we perform an alternate test for dust depletion within our YSCs. For this test, we compare the ratio of elemental abundances of elements: Fe and N relative to O, which is the least depleted onto dust. Figure~\ref{elemental_depletion_vs_dust_neutral} presents the ratios Fe/O and N/O as functions of E(B-V). We observe anti-correlation for Fe/O with dust (Pearson $r$ = -0.43, $p$-value = 0.2), while correlation with N/O with dust (Pearson $r$ = 0.16, $p$-value = 0.62), indicating that while iron depletes more strongly than oxygen, nitrogen does not show a strong dust depletion signature. Although we note that there is a significant scatter within the plots, making the trends weak. Furthermore, nitrogen is not well studied in the literature \citep[see][for more details]{De_Cia_2024}. Since the correlations for dust with nitrogen are insignificant, we refrain from drawing any firm conclusions. In summary, neutral-phase abundances in M83 demonstrate that Fe strongly track dust content, consistent with their susceptibility to depletion (see Figure~\ref{fig:data5}) and their co-evolution with metallicity, whereas the trend with nitrogen remain ambiguous.

%======================================= DISCUSSIONS ============================================================

\section{Discussion \label{discussion}} 

\subsection{Comparing neutral and ionised gas \label{hi_hii_results}}

Direct, spatially resolved comparisons of neutral- and ionised- gas abundances offer unique insights that cannot be obtained from global galaxy-wide metallicity measurements. Most studies of chemical enrichment focus on bright \HII\, regions accessible via optical emission-line spectroscopy \citep[e.g.,][]{Croxall2016, Berg2019, Arellano-Cordova2024}, yet metals reside across multiple phases, and the neutral ISM can contain a large fraction of a galaxy's baryonic mass \citep[][]{Lebouteiller2013, James_2014}. Such multiphase measurements require combining far-UV absorption-line and optical emission-line spectroscopy at matched spatial scales and therefore remain rare. Recent work shows that abundance offsets between neutral and ionised gas can evolve with stellar age \citep[][]{Abril-Melgarejo2024}, and that metals may appear well mixed on $\sim$100 pc scales while still exhibiting phase-dependent gradients \citep[][]{Hernandez2021}. In M83 specifically, optical and UV spectroscopy reveal high metallicity, stellar-driven feedback from massive stars, and flattened gradients in the ionised phase \citep[][]{Bresolin2002, Bresolin2009, Bresolin2016}, while IFU studies highlight the role of stellar winds and gas kinematics in shaping the ionised medium \citep[][]{Della_Bruna2022a, Della_Bruna2022b}. However, despite extensive characterisation of the ionised gas around young clusters, the larger, chemically significant neutral reservoir surrounding these regions has only been sparsely probed \citep[see][]{James_2014, Hernandez2021, James2026}.

Studying neutral and ionised gas together at the level of individual star clusters is therefore unique: it directly links stellar populations to their surrounding multiphase ISM, avoids the dilution effects inherent in integrated galaxy spectra \citep[see e.g.,][]{Arellano-Cordova2024}, and, when separated element wise, provides crucial constraints on enrichment pathways, dust depletion, and metal mixing timescales \citep[][]{James_2014, Hernandez2021, Abril-Melgarejo2024}. Literature shows that while neutral gas is generally metal-poor compared to ionised gas, the offsets depend on environment, gas flows, and feedback processes \citep[][]{Lebouteiller2013, Emerick2019, Arabsalmani2023}. Moreover, chemical enrichment is strongly element- and phase- dependent. Hence, we present here results of a systematic, spatially resolved comparison of elements: Fe, N, S, and O in both \HII\ and \HI\, phases around YSCs in M83.

\subsubsection{Ionised-neutral elemental abundance comparison around YSCs in M83  \label{ab_em_disc_1_0}}

\begin{figure}
    \centering
    \includegraphics[width=0.5\textwidth]{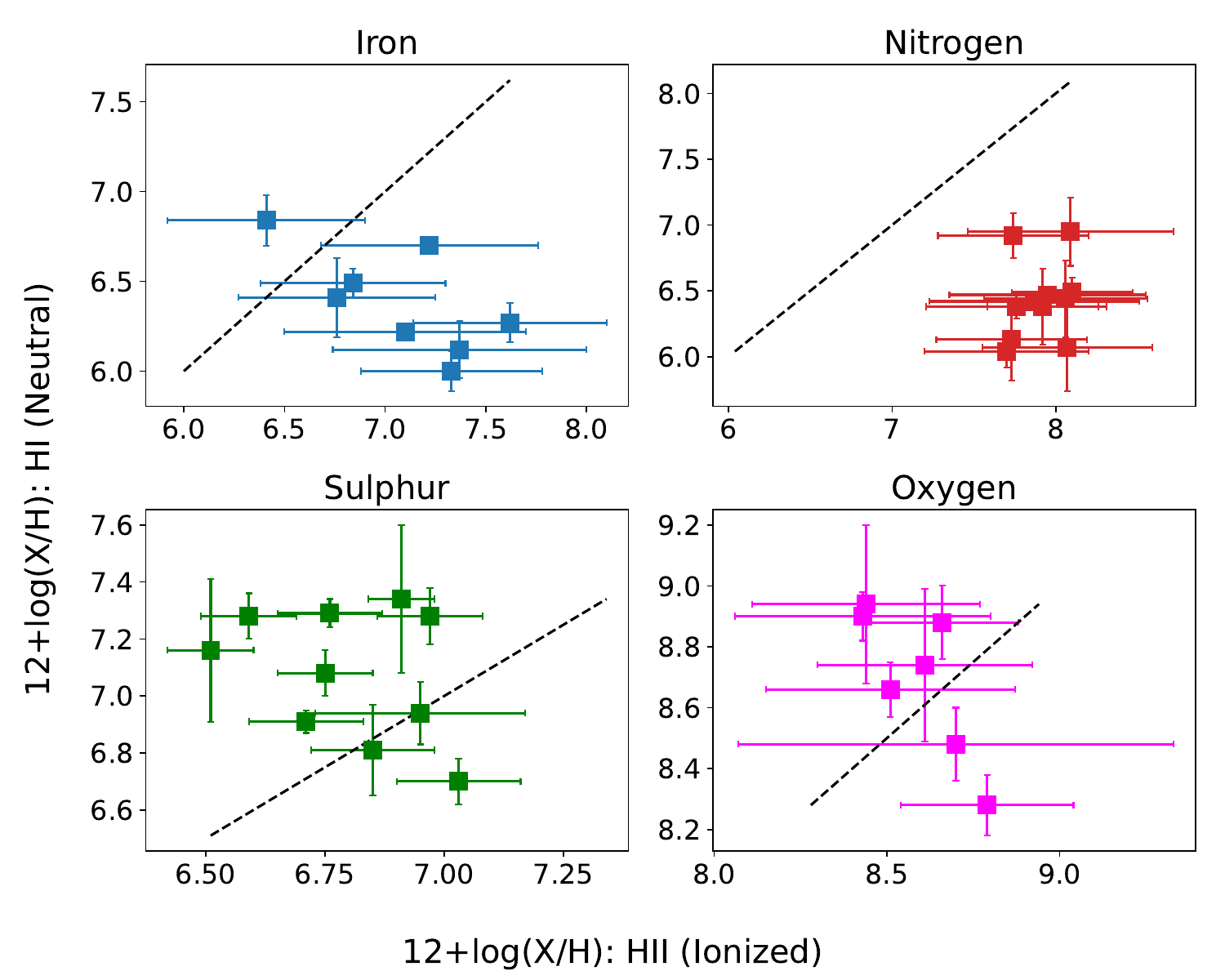}
    \caption{Elemental abundance in ionised gas relative to neutral within YSCs in M83. The black dashed line shows the 1:1 relation. We have removed all points from the galactic centre. Other legends are the same as Figure.~\ref{hii_abundance_oxygen_vs_others}.   \label{x_h_ionised_vs_neutral}}
\end{figure}

We compared Fe, N, S, and O abundances in the neutral (\HI) and ionised (\HII) gas surrounding the YSCs in M83 (see Figure.~\ref{x_h_ionised_vs_neutral}). Because the clusters are $\lesssim$10 Myr old, the two phases trace different chemical timescales: the ionised gas reflects recent enrichment from young stars and their winds, and the earliest CCSNe, whereas the neutral gas retains the imprint of older stellar populations. Nitrogen shows a clear and systematic enhancement in the ionised phase relative to the neutral phase. This is expected: nitrogen is injected on very short timescales ($\lesssim$5 Myr) by massive stellar winds, but its incorporation into the surrounding neutral reservoir is slower. The offset, therefore, reflects incomplete cross-phase mixing in these young environments. 

$\rm \alpha$-elements show no definite enhancement in either phase, but significant variation within individual YSCs. This can arise naturally from how the two measurements sample different gas phases and timescales. 
%The ionised-gas abundances are derived from temperature calibrations that reproduce literature direct-method abundances, suggesting that the observed offsets are unlikely to arise from systematic errors in the abundance determination itself.} In addition, the COS neutral-gas measurement is effectively a pencil-beam column through the disk (often averaging multiple structures along the line of sight), while the optical \HII\ abundances are emission-weighted toward the brightest ionised components, which may be locally diluted or geometrically biased.
Iron shows significant enhancement ($\sim$1 dex) in the ionised phase relative to the neutral phase. Iron enhancements in the ionised ISM, however, are unlikely to be caused by Type Ia SNe. Thus, this excess in the ionised phase possibly represents the stronger depletion of iron onto dust within the neutral ISM. Indeed, our analysis of dust depletion (see Fig.~\ref{fig:data5}, Section~\ref{subsection_dust_neutral}) also shows similar depletion of iron.

\subsubsection{Phase-dependent enrichment trends as a function of ionised gas metallicity  \label{ab_em_disc_1}}

\begin{figure}
    \centering
    \includegraphics[width=0.5\textwidth]{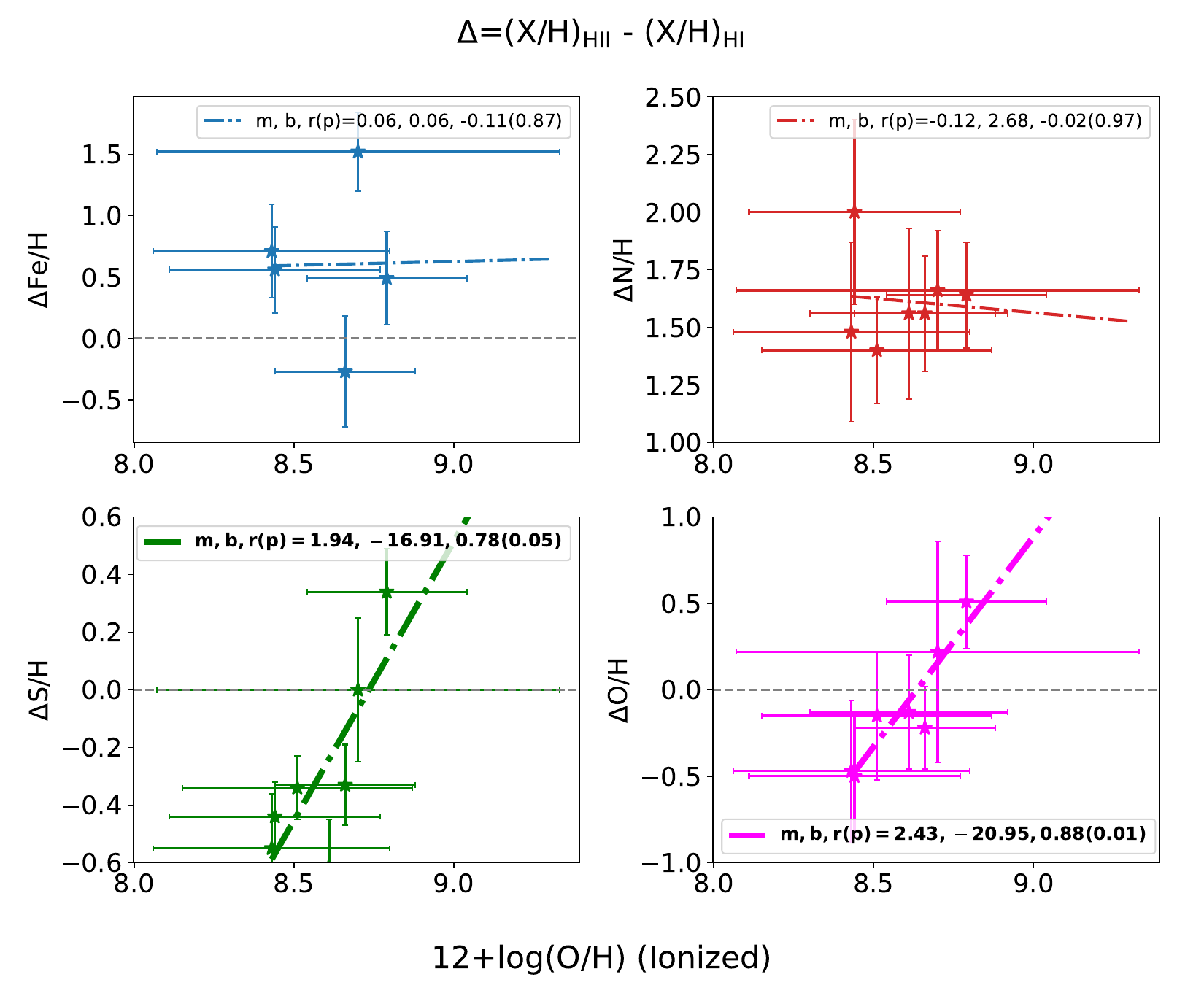}
    \caption{Elemental excesses in ionised gas relative to neutral gas, $\rm \Delta$X/H plotted as a function of ionised gas oxygen abundance, 12+log(O/H). We have removed all the upper limits, 12+log(O/H)$>$9, for clarity. The legends are the same as Figure.~\ref{hii_abundance_oxygen_vs_others}.   \label{delta_x_h_vs_ionised_o_h}}
\end{figure}

To examine how enrichment differs between the neutral and ionised gas surrounding the YSCs in M83, we compare the abundance offsets $\Delta$X/H = (X/H)$\rm_{ionised}$ - (X/H)$\rm_{neutral}$ for Fe, N, S, and O as a function of the ionised gas-phase oxygen abundance, 12+log(O/H)$\rm_{ionised}$ in Figure.~\ref{delta_x_h_vs_ionised_o_h}. Oxygen abundance offset ($\Delta$O/H) in Figure.~\ref{delta_x_h_vs_ionised_o_h} shows a strong, significant positive correlation with ionised oxygen abundance (Pearson $r$ = 0.88, $p$-value = 0.01). This indicates that oxygen enrichment is increasingly `locked' into the ionised phase at high metallicity. Sulphur also shows a strong correlation(Pearson $r$ = 0.78, $p$-value = 0.05), suggesting that both $\rm \alpha$-elements are produced predominantly through a similar nucleosynthetic process. The variance in alpha-element abundances between the neutral and ionised phases is stochastic, ranging up to $\sim$0.5 dex excess in neutral gas at lower ionised oxygen abundance to $\sim$0.5 dex excess in the ionised gas at higher ionised oxygen abundance. The slight excess towards neutral gas could likely be caused by enrichment from the CGM as discussed in section~\ref{ab_em_disc_1_0}. Furthermore, the excess within ionised gas could be fresh enrichment from young stars. However, we refrain from discussing it further due to the lack of a larger sample.       

Nitrogen abundance offset ($\Delta$N/H) shows no significant correlation with ionised oxygen abundance, indicating that nitrogen mixing in the ionised region is dependent on factors other than ionised gas metallicity. We also check the abundance offset correlation with stellar metallicity and, while it is weak, the correlation is relatively strongest with nitrogen. In addition, the absolute values confirm that N is systematically and uniquely enhanced ($\gtrsim$1 dex) in the ionised gas relative to the neutral gas \citep[see also][]{Abril-Melgarejo2024, James2026}. There are several ways that N can be enriched quickly ($\sim$10Myr). We checked for various signatures for this: massive stars \citep[O,B][primary signature, O stars: p-cygni profiles of \NV\,$\lambda\lambda$1238,1242, \SiIV\,$\lambda\lambda$1393,1402, \CIV\,$\lambda\lambda$1548,1550; B stars: \OI$\lambda$7774 triplet]{Smith2014, Leitherer2020}, Wolf-Rayet stars \citep[W-R][primary signatures: broad emission of \CIV\,$\lambda$5808 and \HeII\,$\lambda$1640 line]{Lopez-Sanchez2007, Crowther2007review, Berg2024}, very massive stars \citep[VMS][primary signature: broad emission of \HeII\,$\lambda$1640 line]{Martins2022}. We only see strong p-cygni profiles of \NV\,$\lambda\lambda$1238,1242, \SiIV\,$\lambda\lambda$1393,1402, \CIV\,$\lambda\lambda$1548,1550, indicating that, if this enrichment is from the current generation of stars, the nitrogen excess with the YSCs in our work is probably caused by O, B-type stars.

Iron abundance offset ($\Delta$Fe/H) shows no correlation with oxygen (Pearson $r$ = -0.11, $p$-value = 0.87). We also note that iron is significantly enhanced within the ionised phase as compared to the neutral. Although this excess should not be confused with the production or mixing mechanism. Instead, this excess might be hinting towards the fact that iron is significantly depleted onto dust within the neutral gas.

\subsubsection{Multi-phase abundance offsets as a function of stellar age  \label{ab_em_disc_2}}

Having established that phase-dependent offsets encode element-specific enrichment, the natural next step is to examine how these excesses evolve with stellar age, which sets the clocks for distinct channels. By comparing $\rm \Delta$X/H with stellar age, we can assess how quickly metals from the current stellar life-cycle are being released and enriching the ionised gas. Any change in $\rm \Delta$X/H with stellar age reflects the timescale over which mixing, cooling, and phase exchange redistribute these elements across the multiphase ISM. In a system like M83-where clustered star formation, feedback-driven kinematics, and dust-rich environments shape the ISM structure \citep[][]{Della_Bruna2022a, Della_Bruna2022b, Hernandez2023}, the time dependence of $\rm \Delta$X/H provides a clean, chronological probe of how enrichment proceeds and how efficiently metals transfer between phases.

\begin{figure}
    \centering
    \includegraphics[width=0.5\textwidth]{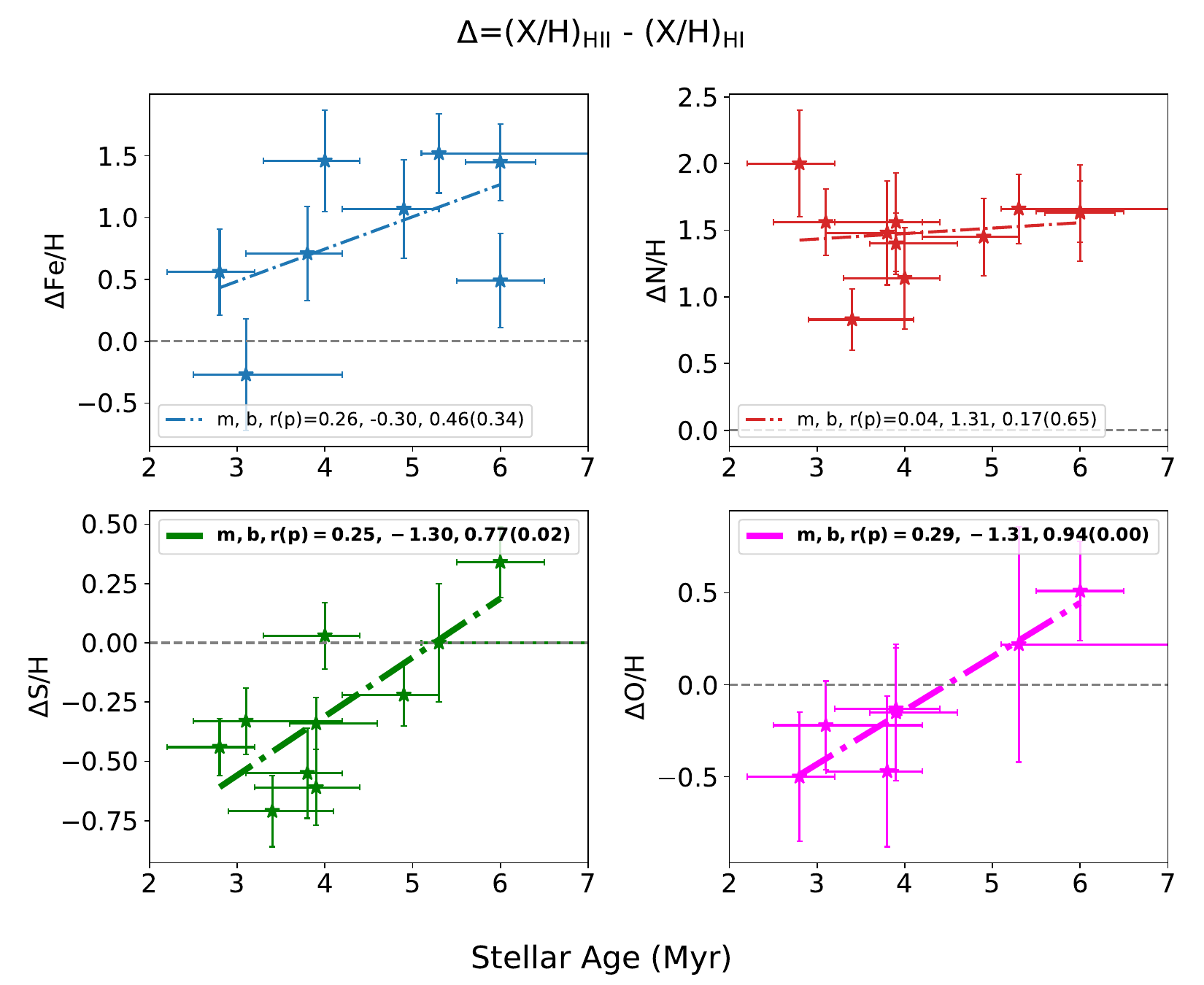}
    \caption{Elemental excesses in ionised gas relative to neutral gas, $\rm \Delta$X/H plotted as a function of average stellar age of the natal star-cluster. The legends are the same as Figure.~\ref{hii_abundance_oxygen_vs_others}.   \label{delta_x_h_vs_st_age}}
\end{figure}

In Figure.~\ref{delta_x_h_vs_st_age}, we examine how the ionised-neutral abundance offsets (excess) vary with stellar age. For $\alpha$-elements: sulphur and oxygen, we find strong positive correlations with stellar age (Pearson $r=0.77,0.94$, $p$-values $=0.02,<0.01$), which is exactly consistent with the onset of CCSNe enrichment at $\sim$3-5 Myr for typical metallicities \citep[][]{Leitherer2014}. Iron excess shows an insignificant correlation with age (Pearson $r=0.46$, $p-value=0.34$) while maintaining an average offset of $\sim$1 dex. This excess is most likely a consequence of depletion effects.  

Nitrogen shows a weak correlation and excess in ionised gas within all star clusters. The youngest cluster at $\sim$3Myr has the highest ionised to neutral offset. The average offset between ionised and neutral phases in M83 is large ($\Delta$N/H $\sim$ 1.5 dex), indicating that the ionised gas remains N-rich in the entire age range seen here (1-6 Myr). Nitrogen shows a strong and early phase separation driven by fast enrichment from massive O, B stars (see discussion in the previous section), but only weak age evolution within the limited age range sampled here.

\subsubsection{Elemental abundances within phases as a function of galactocentric radius \label{ab_em_disc_3}}

\begin{figure}[ht]
    \centering
    \begin{subfigure}{0.48\textwidth}
        \centering
        \includegraphics[width=\linewidth]{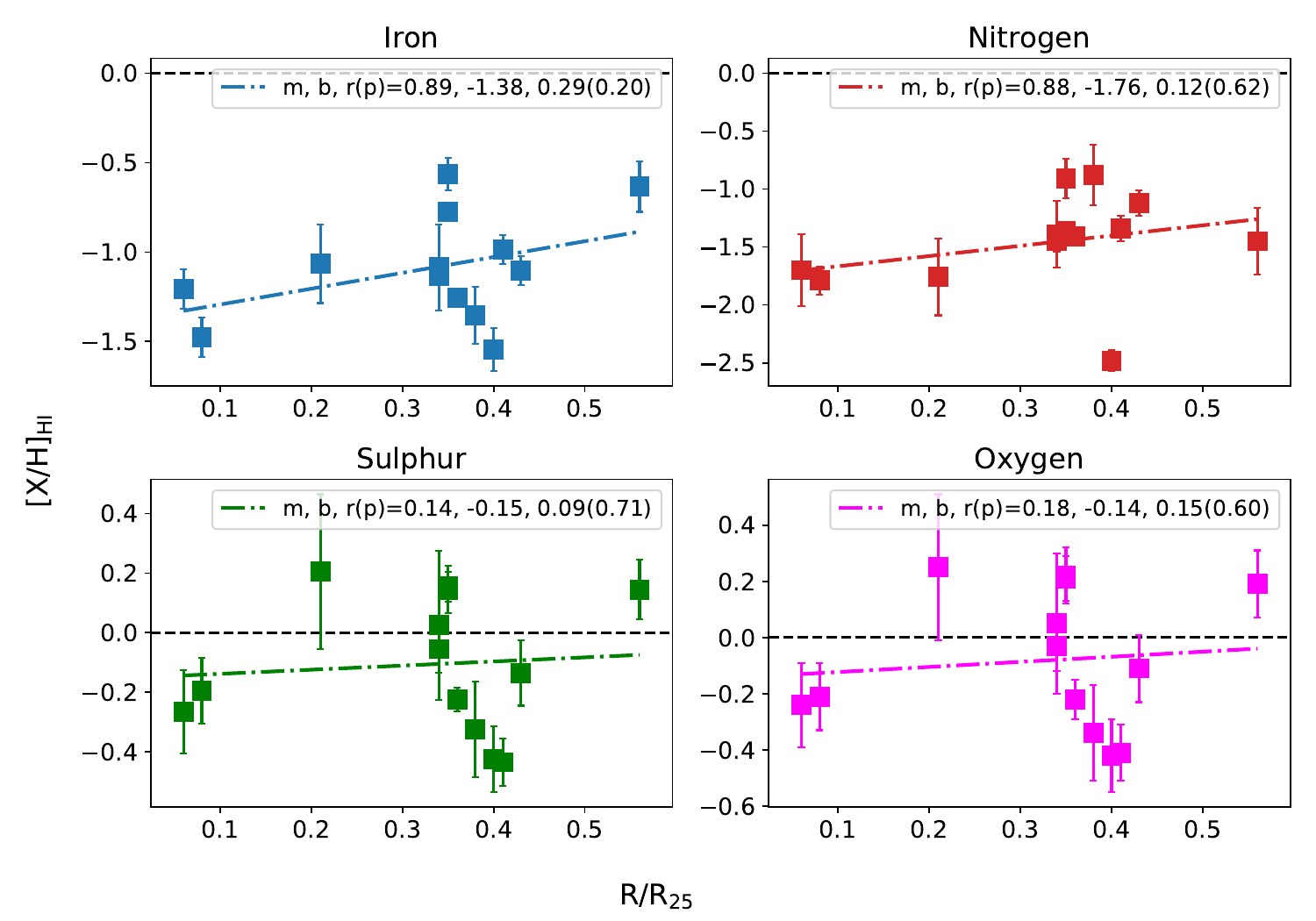}
    \end{subfigure}
    \begin{subfigure}{0.48\textwidth}
        \centering
        \includegraphics[width=\linewidth]{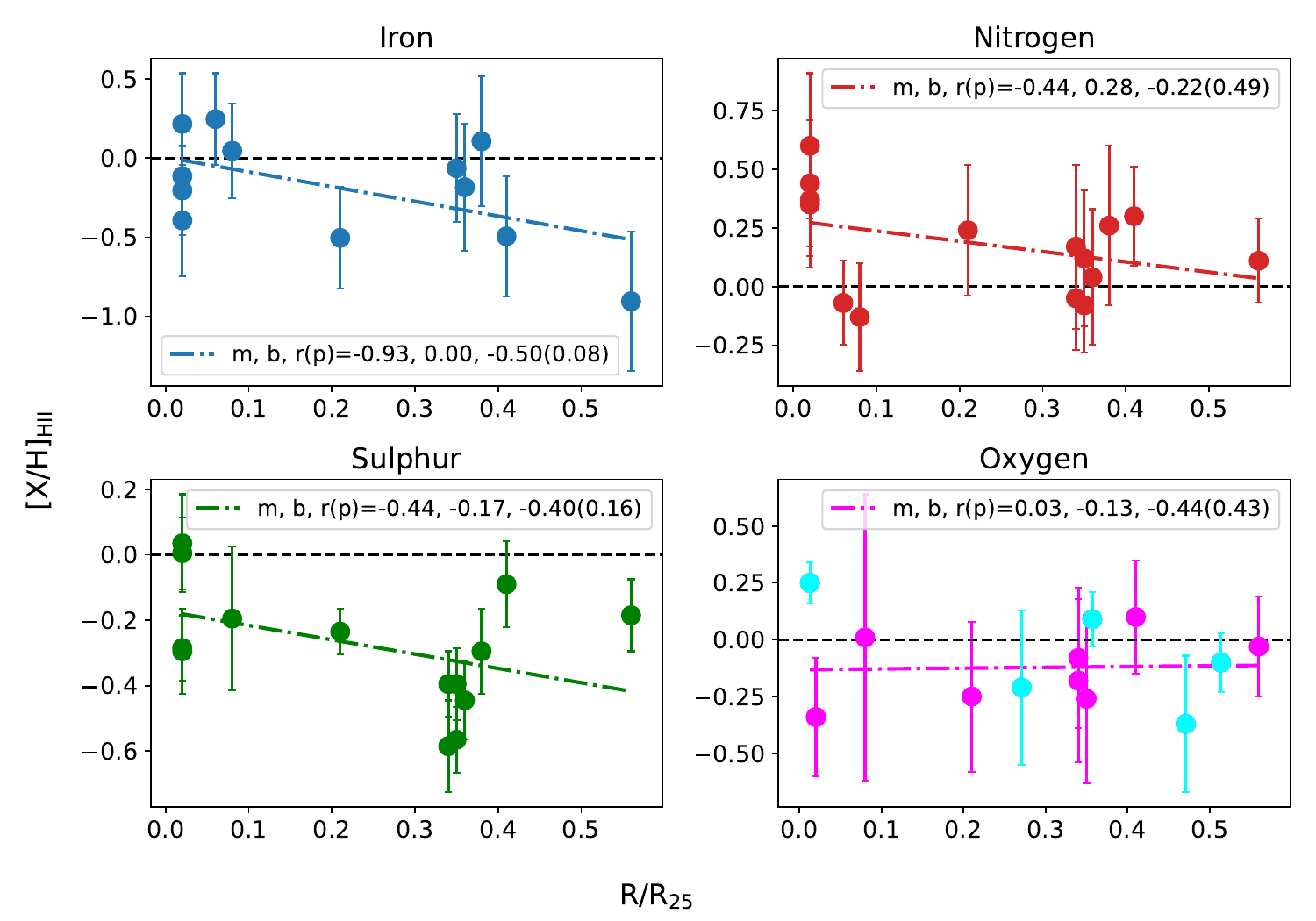}
    \end{subfigure}
    \caption{Elemental abundance (relative to solar) in neutral (top) and ionised (bottom) gas as a function of galactocentric radius, R/R$_{25}$. The horizontal black dashed line represents the solar value. The cyan points within the \HII\ region oxygen plot, taken from \citet{Bresolin2005}, are shown for comparison and included in the statistics. Only the ionised gas figure (bottom) includes points from the galactic centre (r/r$_{25}\,<0.05$, see discussion in Section.~\ref{galaxy_center_results}).}
    \label{x_h_vs_r_r25}
\end{figure}

We now examine how the absolute abundances in each phase vary with galactocentric distance, R/R$_{25}$. This comparison tests whether large-scale chemical structure shaped by mixing, feedback circulation, and galaxy assembly modulates the composition of gas surrounding young clusters. Figure.~\ref{x_h_vs_r_r25} shows the evolution in abundance for all elements (Fe, N, S, and O) as a function of galactocentric radius, R/R$_{25}$. The trends are generally weak and statistically insignificant (all $p$-values $>$ 0.05) in both gas phases. To obtain more statistically robust gradient measurements, we included direct oxygen abundance measurements of \HII\, regions from the M83 literature \citep[specifically: ][]{Bresolin2005}. What is interesting is that while ionised gas consistently shows a weak anti-correlation with galactocentric radius, neutral gas has exactly the opposite trend. This ionised gas behaviour is broadly consistent with previous studies of M83 that report the presence of a radial metallicity gradient, which may flatten at larger radii \citep[e.g.,][]{Bresolin2016}. In neutral gas, all elemental abundances are moderately correlated with galactocentric distance. A positive radial trend is indicative of inside-out growth, which is common in grand design spirals like M83, although the weak statistical significance suggests that efficient mixing and/or limited radial coverage may partially wash out large-scale gradients. While it is very interesting to see how all elemental abundances within each phase show consistent behaviour, the trend is too insignificant to investigate the cause further.

\subsection{Comparing environments of grand design spiral, M83 with a blue compact dwarf, NGC 5253}

\subsubsection{Comparing nitrogen enrichment (N/O)}

The multiphase behaviour of nitrogen relative to oxygen provides one of the clearest diagnostics of recent enrichment from different production channels: N from rapid production channels (winds from very massive and W-R stars at $\sim$2-5 Myr) and oxygen from CCSNe enrichment. The multiphase analysis of NGC 5253 by \citet{Abril-Melgarejo2024} provides a key comparison for interpreting our cluster-resolved measurements in M83, since both studies combine HST/COS absorption for the neutral gas with optical emission-line spectroscopy for the ionised phase following the co-spatial methodology of \citet{Hernandez2021}. However, the host environments differ sharply: NGC 5253 is a compact, low-mass, average-metallicity starburst, while M83 is a massive, metal-rich grand-design spiral. These structural and chemical differences shape how newly synthesised material propagates between gas phases. Hence, we look at nitrogen enrichment (N/O) within M83 and compare it with NGC 5253. 

\begin{figure}
    \centering
    \includegraphics[width=0.5\textwidth]{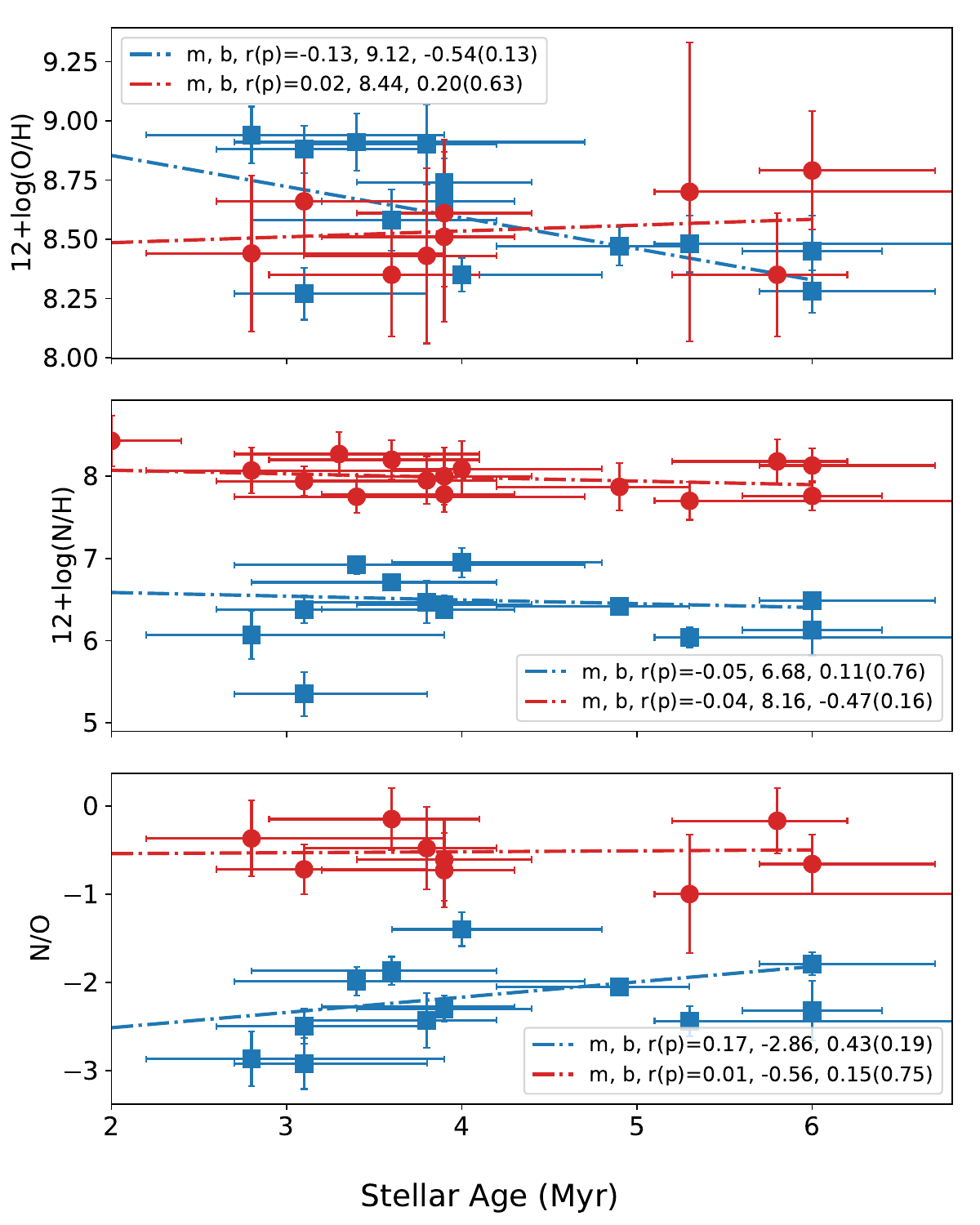}
\caption{Oxygen abundance (12+log(O/H)), nitrogen abundance (12+log(N/H)) and N/O shown within top, middle and bottom panels for neutral (as blue squares) and ionised (as red circles) phases. The trend is shown as a dashed line with the same colour code as mentioned above for both phases. Points from the galactic centre (r/r$_{25}\,<0.05$) are included for ionised gas (see discussion in Section.~\ref{galaxy_center_results}). The statistics for all trends are shown in the legend and follow the same format Figure.~\ref{hii_abundance_oxygen_vs_others}.  \label{st_age_nitrogen_enrichment}}
\end{figure}

In NGC 5253, \citet{Abril-Melgarejo2024} find that ionised gas near the youngest clusters is strongly enriched in nitrogen, with phase offsets of $\sim$0.8 dex in (N/H) and $\sim$0.6 dex in N/O, attributed to rapid WN-type WR winds \citep[see also][]{Schaerer1997, Westmoquette2013b}. Crucially, these offsets decrease with stellar age, and the neutral gas gradually approaches the ionised composition within $\sim$1-8 Myr, implying that the WR-enriched material cools and mixes into the neutral reservoir on Myr timescales \citep[][]{Lebouteiller2013, Berg2021, Abril-Melgarejo2024}. This reinforces the idea of fast enrichment followed by efficient mixing in a low-mass ISM. In M83, we instead observe a persistent and much stronger phase separation for nitrogen (See Figure.~\ref{st_age_nitrogen_enrichment}). Across all clusters in our sample (all $\lesssim$6 Myr), ionised gas shows significantly elevated nitrogen relative to the neutral gas, with typical offsets of $\Delta$N/H $\approx$ 1.3-1.6 dex, substantially larger than those in NGC 5253. In both M83 and NGC 5253, the ionised gas also shows elevated N/O very early ($\sim$2-5 Myr), consistent with rapid massive stellar wind-driven enrichment \citep[][]{Kudritzki2000, Martins2024}. In M83, however, we see that this offset does not diminish with cluster age: neither 12+log(N/H) nor N/O in the ionised phase shows evidence of approaching the neutral-phase values over $\sim$1-6 Myr, and the neutral gas remains systematically nitrogen-poor throughout. In NGC 5253, said offsets (in 12+log(N/H) and N/O) within the ionised and neutral medium are much smaller, and they do see convergence within $\sim$8 Myr. While we see oxygen to be slightly higher in the neutral gas around younger clusters, we note that the multiphase O abundances broadly agree with each other within two sigma. Since the trend is insignificant, we refrain from discussing this further.

This difference in nitrogen excess between M83 and NGC5253 could arise from environmental regulation of mixing. M83's ISM is more metal-rich and embedded within a deeper gravitational potential well, conditions known to confine feedback-driven enrichment to localised \HII\ regions to slow the transfer of metals into surrounding \HI\, regions \citep[][]{Lebouteiller2013, Emerick2019}. Nitrogen's secondary production is prominent within the older generation of stars and correlates well with the metallicity \citep[see][]{Henry2000}. The neutral gas in M83 has a mean nitrogen abundance $\sim$0.5-1 dex higher than in NGC 5253. This reflects a nitrogen enrichment in M83's high-metallicity environment from an older generation of stars. While the ionised gas traces newly ejected nitrogen from massive stars that has not yet permeated into the larger reservoir. This contrast shows that the nitrogen enrichment is strongly environment-dependent. In low-mass starbursts like NGC 5253, weaker winds enrich \HII\ regions modestly, but the shallow gravitational potential allows rapid dispersal and mixing into the larger neutral (\HI) ISM. In massive spirals like M83, strong winds drive large nitrogen enhancements locally, but the deep gravitational potential inhibits mixing, confining enrichment from the current stellar population to the ionised phase. The other possibility is that metal-rich YSCs in M83 have a higher frequency of massive stars than NGC5253, but pursuing this line of thought is beyond the scope of our study. To investigate the effects of stellar feedback further, we study the impact of outflows and stellar winds on nitrogen enrichment in both phases.

\subsubsection{Comparing outflows and stellar feedback between the two galaxies \label{comparing_outflows_and_stellar_feedback}}

In NGC 5253, enhanced ionised-neutral offsets coincide with young clusters driving strong outflows \citep{Abril-Melgarejo2024}, a pattern echoed in simulations of chemically clumpy dwarfs \citep{Corlies2018, Emerick2020}. By contrast, M83 hosts star formation within a more metal-rich, higher-density disk embedded in a deeper gravitational potential. Stellar winds are expected to be stronger, producing pronounced enrichment in the ionised gas. The efficiency of these winds to transfer these metals into the neutral phase is lower, leading to persistent phase separation between neutral and ionised gas. To ensure robustness of ISM outflow velocity, we removed the broad $p$-Cygni stellar profile for these transitions from the spectra before fitting the ISM lines. 

In this subsection, we investigate the role of stellar feedback in driving chemical enrichment within different galactic environments. We could not find any significant trend between elemental abundances in both phases and the observed outflow velocity, $v_{\rm out}$, derived from emission line fit. $v_{\rm out}$ is obtained using the fitted velocity of the wider, outflow component within strong optical emission lines, and the value is reported in Table~\ref{outflow_tab_1_new} relative to the stellar velocity of the same cluster.  

\begin{table}[]
\begin{tabular}{ll}
\hline
YSC       & $v_{\rm out}$(km s$^{-1}$) \\
\hline
\hline
M83-1     & 54.73+/-43.01 \\
M83-2     & -10.23+/-8.06 \\
M83-3     & 50.60+/-9.00  \\
M83-4     & 15.97+/-3.00  \\
M83-5     & -             \\
M83-6     & -0.30+/-7.01  \\
M83-7     & 14.94+/-17.00 \\
M83-8     & -             \\
M83-9     & -             \\
M83-10    & 8.32+/-1.31   \\
M83-11    & 17.87+/-22.00 \\
M83-12    & 39.82+/-14.55 \\
M83-13    & -             \\
M83-14    & 30.72+/-15.45 \\
M83-15    & -             \\
M83-16    & 53.15+/-4.51  \\
M83-POS-1 & -1.56+/-10.00 \\
M83-POS-2 & 142.59+/-1.5\\
\hline
\end{tabular}
\caption{Outflow velocity ($v_{\rm out}$) in km s$^{-1}$ from ionised gas emission lines (relative to stellar). \label{outflow_tab_1_new}}
\end{table}

\begin{figure}[ht]
    \centering
    \begin{subfigure}{0.48\textwidth}
        \centering
        \includegraphics[width=\linewidth]{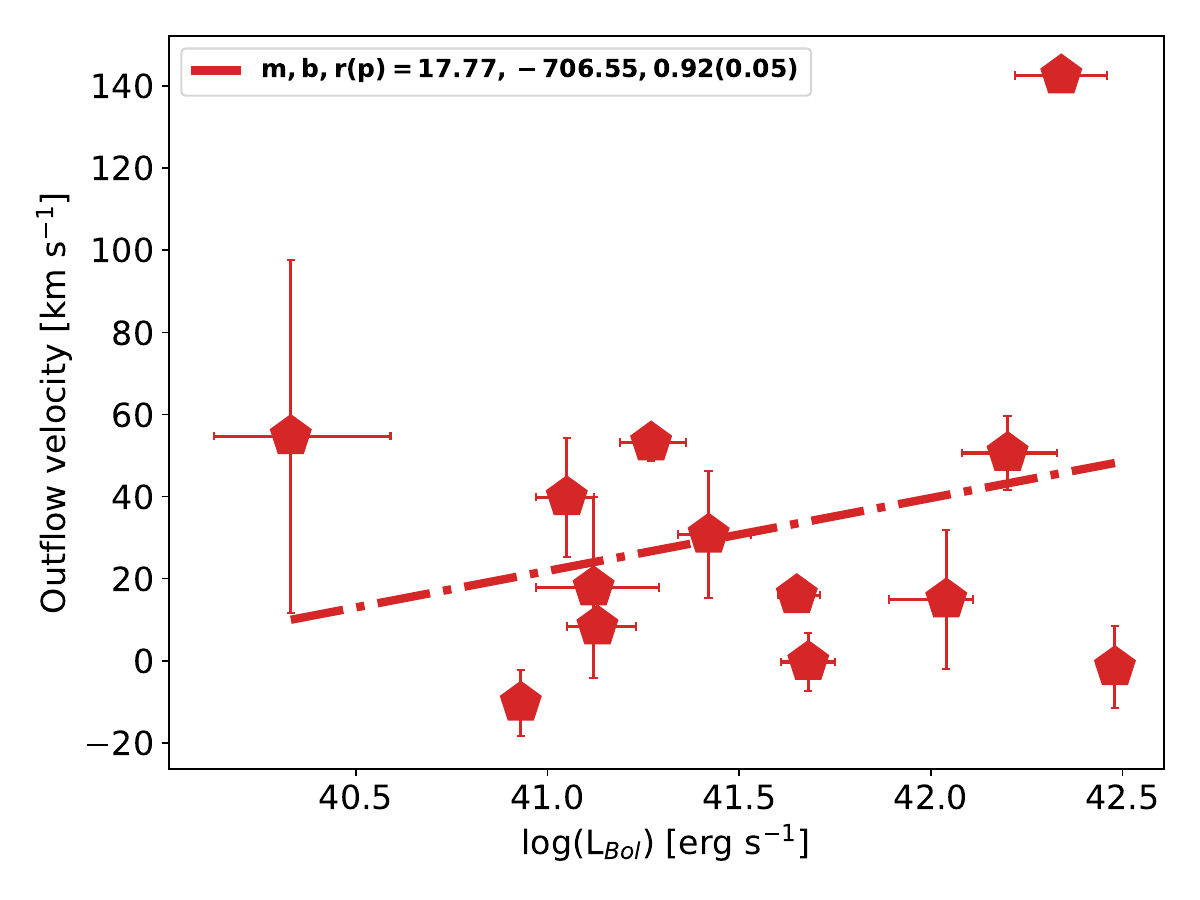}
    \end{subfigure}
    \begin{subfigure}{0.48\textwidth}
        \centering
        \includegraphics[width=\linewidth]{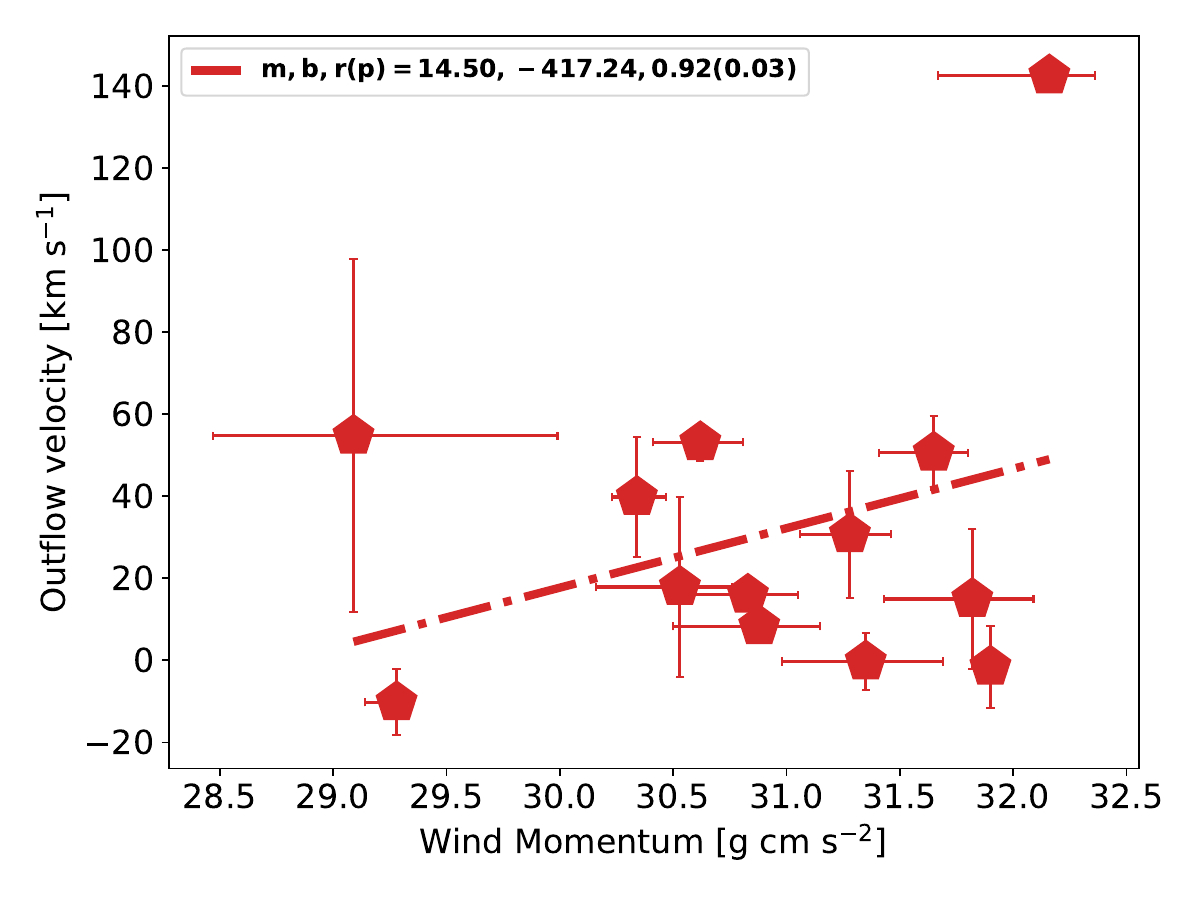}
    \end{subfigure}
    \caption{Stellar bolometric luminosity (L$_{bol}$ in [erg s$^{-1}$], top) and wind momentum (in [g cm s$^{-2}$], bottom) obtained using SB99 models plotted as a function of outflow velocity, $v_{out}$ in kms$^{-1}$ (see text within subsection.~\ref{comparing_outflows_and_stellar_feedback} for details). The statistics shown in the legend follow the same format as Figure.~\ref{hii_abundance_oxygen_vs_others}. \label{outflow_info_fig_1_new}}
\end{figure}

Hence, we further investigate whether young stellar properties are indeed having an impact on the nearby ISM gas. To this end, we compared the stellar bolometric luminosity and wind momentum obtained from stellar models (see Section.~\ref{stellar_fitting}) to see if they correlate with observed outflow velocity (See Figure.~\ref{outflow_info_fig_1_new}). We report a strong, significant correlation between the stellar luminosity and wind momentum, indicating that the impact of young stellar feedback is indeed present within outflowing gas.

\begin{figure}[ht]
    \centering
    \begin{subfigure}{0.48\textwidth}
        \centering
        \includegraphics[width=\linewidth]{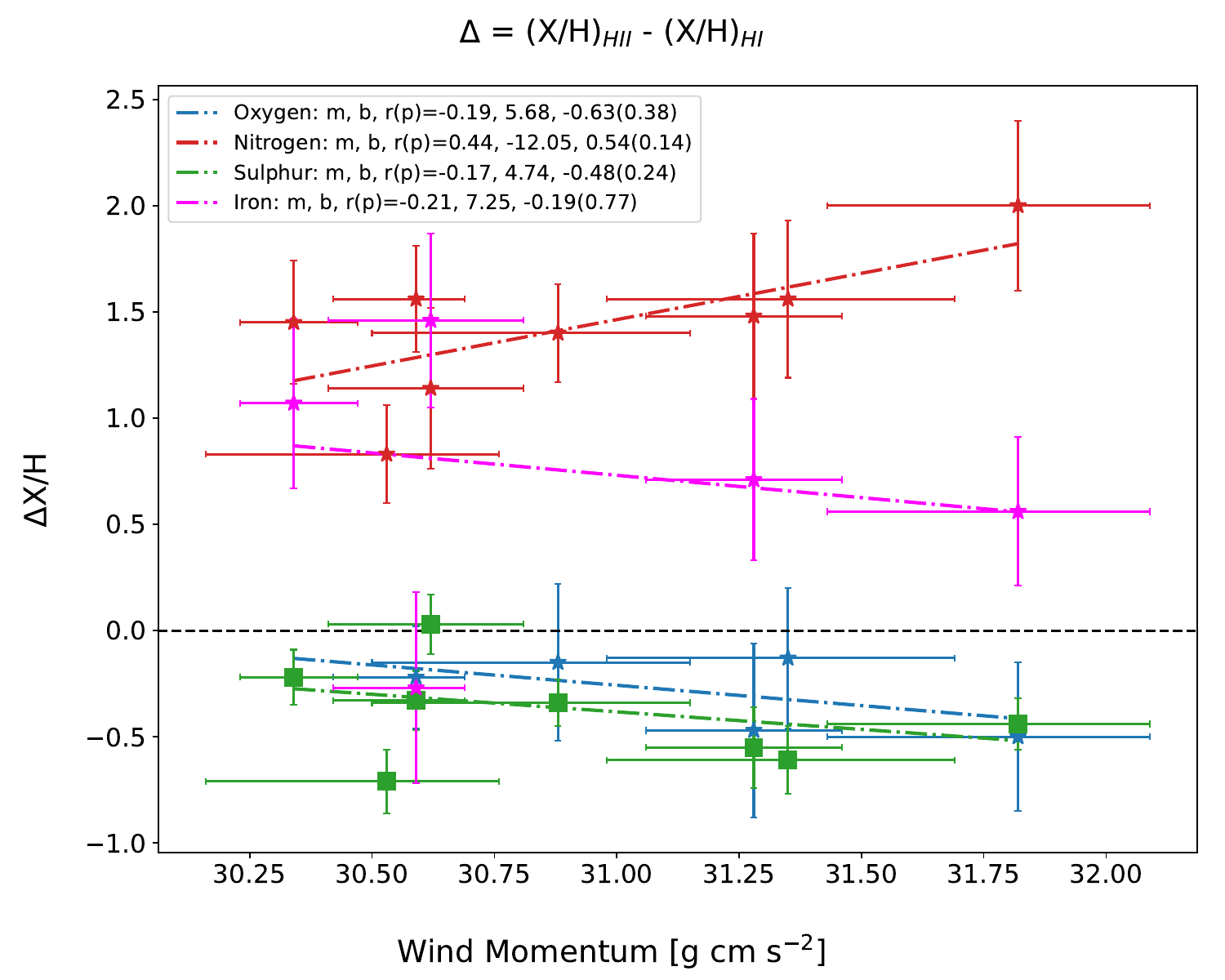}
    \end{subfigure}
    \begin{subfigure}{0.48\textwidth}
        \centering
        \includegraphics[width=\linewidth]{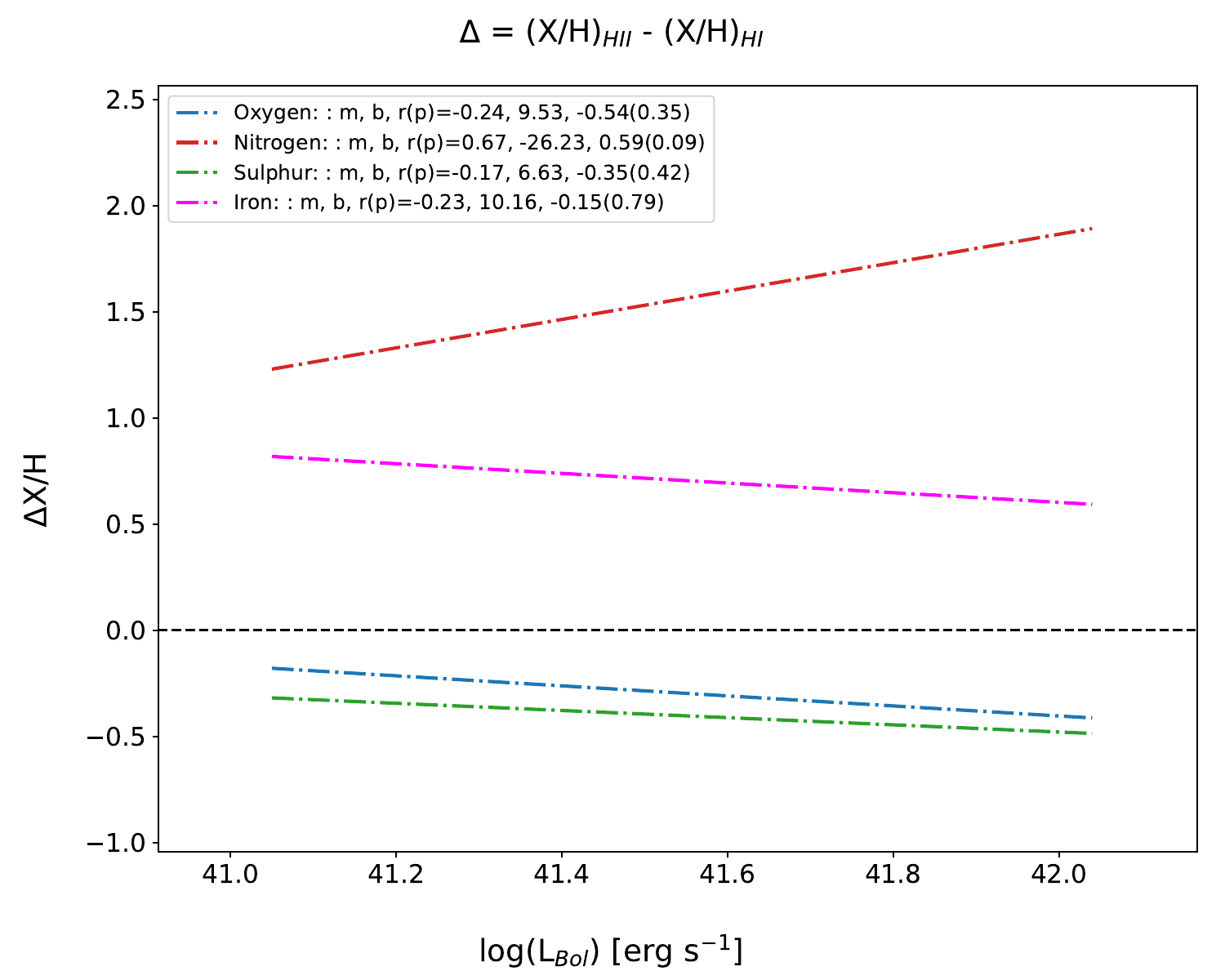}
    \end{subfigure}
    \caption{The oxygen (X=O) and nitrogen (X=N) abundance excess relative to solar, $\Delta$X/H = ((X/H)$_{ion}$ - (X/H)$_{neu}$) plotted as a function of stellar luminosity (L$_{bol}$ in [erg s$^{-1}$], bottom) and wind momentum (in [g cm s$^{-2}$], top) obtained using SB99 models. Points with high uncertainties in stellar age are also removed. For clarity, we removed the individual data points from the bottom subplot for all YSCs and just show the trend. The statistics for all trends are shown in the legend and follow the same format as Figure.~\ref{hii_abundance_oxygen_vs_others}. \label{outflow_info_fig_2}}
\end{figure}

Since we have seen that outflowing ionised gas velocity correlates strongly with stellar properties, we further investigate how elemental abundance excess varies as a function of intrinsic stellar properties. Figure.~\ref{outflow_info_fig_2} shows the elemental abundance excess ($\Delta$ (X/H) = (X/H)$_{ion}$ - (X/H)$_{neu}$) in both phases as a function of stellar luminosity (bottom) and wind momentum (top). We find that the significantly enhanced nitrogen in the ionised phase increases as a function of stellar luminosity as well as wind momentum. The correlation is moderately significant, but unique for nitrogen. The correlation with the rest of the elements is not statistically significant. This indicates that nitrogen in ionised gas abundances strongly follows stellar feedback. We can conclude that the freshly synthesised nitrogen is efficiently entrained and retained within the \HII\, regions when winds are strong. This reinforces the picture established in previous sections: in M83, strong winds enrich the ionised phase efficiently, but the deeper potential well and denser ISM inhibit the redistribution of these metals into the neutral gas reservoir on $\lesssim$6 Myr timescales.

\subsubsection{Dust content within different phases \label{dust_discussion_subsection}}

Dust plays a critical but often underappreciated role in shaping multiphase abundance measurements. It regulates how metals are partitioned between the gas and solid phases and therefore directly affects the interpretation of abundance (X/H) ratios in both ionised and neutral gas. Studying dust separately within each phase is essential to disentangle true nucleosynthetic enrichment from depletion effects and observational biases. Our unique multi-phase study has enabled us to perform a comparison of dust within young star clusters and the gas phases (\HI\, and \HII) surrounding it. 

The dust within the star clusters is estimated using stellar spectral fitting (see Section~\ref{stellar_fitting}), which probes an attenuation curve. The Balmer-line ratio (H$\alpha$/H$\beta$, assuming a \citet{Calzetti2000} reddening law) is used to estimate dust within the ionised \HII\, regions, while dust in the neutral (\HI) gas is estimated using the prescription by \citet{DeCia2018} (see Section~\ref{subsection_dust_neutral}), which is closer to an extinction-based measure along individual sightlines. The dust content, E(B-V) for star clusters and the different gas phases are given in Table~\ref{dust_content_table}. We emphasise that these diagnostics are not directly equivalent. As discussed by \citet{Salim2020}, attenuation (e.g., from SED fitting or integrated nebular emission) includes not only absorption and scattering out of the line of sight but also scattering back into the line of sight and the effects of complex star–dust geometry, whereas extinction refers to the loss of light along a single sightline due purely to absorption and scattering \citep[see also][]{Calzetti2001, Battisti2017}. Consequently, comparisons between stellar, \HII-region, and \HI-based dust estimates must be interpreted with caution, as they probe fundamentally different physical regimes and are sensitive to geometry, radiative transfer effects, and the spatial distribution of dust and stars \citep{Salim2020}.

\begin{table}[]
\begin{tabular}{llll}
\hline
YSC       & \HII\ gas & \HI\ gas & stellar \\
\hline
\hline
M83-1     & 0.34$\rm \pm$0.23  & 0.11$\rm \pm$0.05  & 0.35$^{0.38}_{0.35}$ \\
M83-2     & 0.30$\rm \pm$0.20  & 0.15$\rm \pm$0.05  & 0.3$^{0.31}_{0.29 }$ \\
M83-3     & 0.32$\rm \pm$0.03  & 0.01$\rm \pm$0.02  & 0.2$^{0.21}_{0.19 }$ \\
M83-4     & 0.37$\rm \pm$0.01  & 0.03$\rm \pm$0.04  & 0.32$^{0.33}_{0.29}$ \\
M83-5     & 0.65$\rm \pm$0.01  & 0.00$\rm \pm$0.02  & 0.4$^{0.44}_{0.37 }$ \\
M83-6     & 0.61$\rm \pm$0.19  & 0.09$\rm \pm$0.02  & 0.26$^{0.26}_{0.23}$ \\
M83-7     & 0.32$\rm \pm$0.01  & 0.07$\rm \pm$0.02  & 0.33$^{0.34}_{0.31}$ \\
M83-8     & 0.38$\rm \pm$0.03  & 0.05$\rm \pm$0.01  & 0.14$^{0.17}_{0.14}$ \\
M83-9     & 0.16$\rm \pm$0.06  & 0.05$\rm \pm$0.02  & 0.1$^{0.11}_{0.093}$ \\
M83-10    & 0.28$\rm \pm$0.24  & 0.08$\rm \pm$0.01  & 0.15$^{0.15}_{0.14}$ \\
M83-11    & 0.57$\rm \pm$0.06  & 0.02$\rm \pm$0.01  & 0.11$^{0.14}_{0.11}$ \\
M83-12    & 0.21$\rm \pm$0.18  & 0.10$\rm \pm$0.01  & 0.26$^{0.26}_{0.25}$ \\
M83-13    & --           & 0.10$\rm \pm$0.02  & 0.19$^{0.21}_{0.17}$ \\
M83-14    & 0.22$\rm \pm$0.14  & 0.06$\rm \pm$0.03  & 0.14$^{0.15}_{0.12}$ \\
M83-15    & --           & 0.04$\rm \pm$0.01  & 0.24$^{0.25}_{0.21}$ \\
M83-16    & 0.26$\rm \pm$0.24  & 0.13$\rm \pm$0.01  & 0.23$^{0.25}_{0.23}$ \\
M83-POS-1 & 0.40$\rm \pm$0.01  & 0.07$\rm \pm$0.01  & 0.32$^{0.32}_{0.32}$ \\
M83-POS-2 & 0.30$\rm \pm$0.04  & 0.20$\rm \pm$0.06  & 0.16$^{0.17}_{0.16}$ \\
\hline
\end{tabular}
\caption{The amount of dust, E(B-V) (in mag) present within ionised gas (\HII), neutral gas (\HI) and young star clusters. See text for the details of the method of dust estimation. \label{dust_content_table}}
\end{table}

\begin{figure}
\centering
\includegraphics[width=0.5\textwidth]{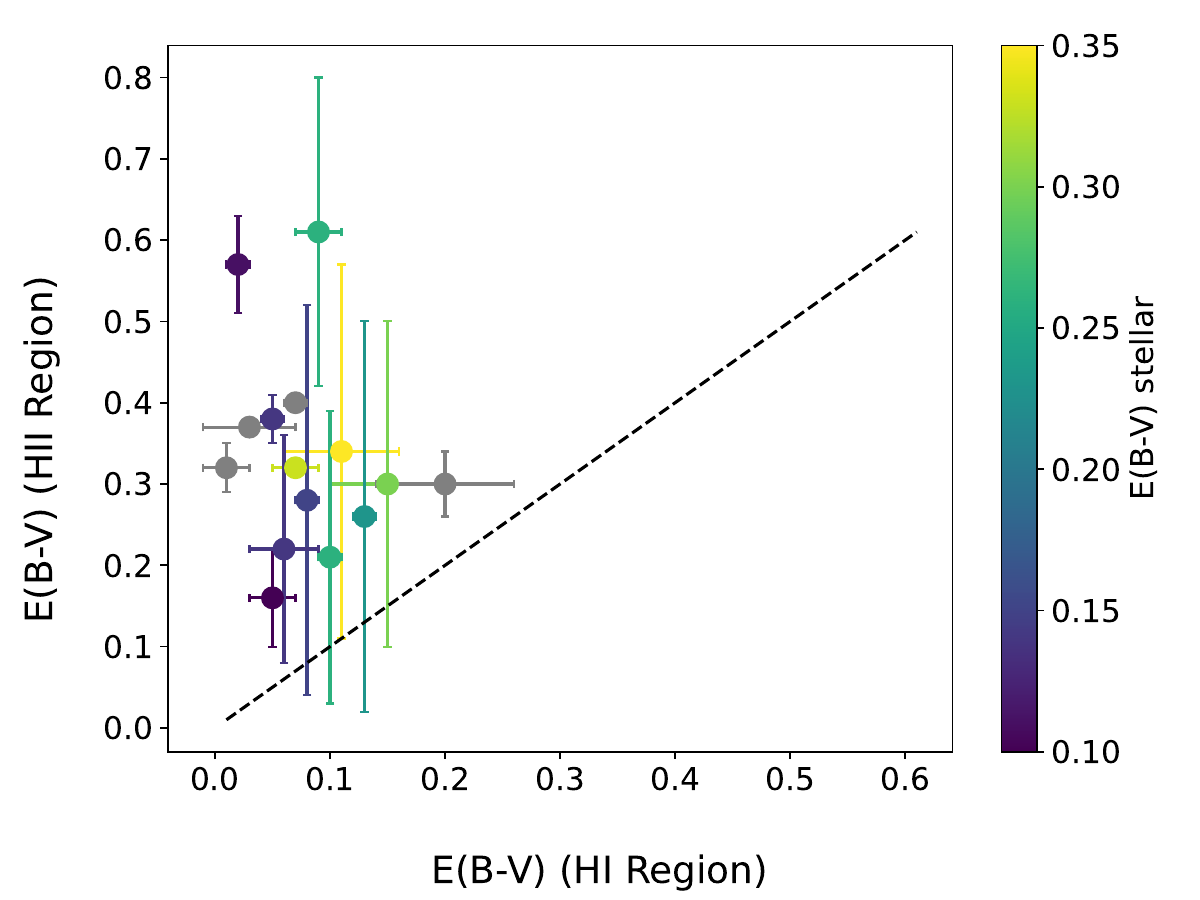}
\caption{Dust content (E(B-V)) within neutral gas (\HI\, region) plotted against the dust within ionised gas (\HII\ region). The points are colour coded as a function of dust within young stars, except for points from within the galactic centre. \label{multi_phase_dust_fig}}
\end{figure}

In M83, we find evidence that dust content is higher in the ionised gas and stellar environments than in the surrounding neutral medium (see Figure~\ref{multi_phase_dust_fig}), consistent with the presence of dense, metal-rich, and radiation-dominated regions around young star clusters. This likely reflects a combination of dust formation from the newly formed stars and efficient transport to the localised \HII\ regions by stellar winds and radiation fields. In contrast, the more diffuse neutral gas may experience lower dust content due to grain destruction, dilution over larger path lengths, or incomplete coupling to the sites of recent star formation. We note that depletion effects in iron have been seen clearly in both phases. However, comparing the amount of depletion onto dust across dust content in different phases is tricky given that the techniques used to calculate dust-excess in both phases are very different. However, we caution that part of this contrast may arise from the differing nature of attenuation versus extinction measurements, rather than purely intrinsic variations in dust content. The combination of geometry, scattering, and mixed stellar populations can lead attenuation-based estimates (e.g., SED fitting, Balmer decrement) to differ systematically from line-of-sight extinction measures \citep{Salim2020}. However, comparing the absolute level of depletion across phases remains challenging because the inferred dust content is derived using fundamentally different diagnostics and assumptions.

\section{Summary \label{Conclusion}}

We have presented a spatially resolved, multiphase study of the chemical enrichment around young star clusters (YSCs) in the nearby grand-design spiral M83, combining far-UV absorption-line spectroscopy from HST/COS with co-spatial, aperture-matched optical emission-line spectroscopy from VLT/MUSE and LBT/MODS. Due to the high metallicity of M83, the auroral lines, essential for estimating the electron temperature (T$_e$) of \HII\, regions surrounding the young star cluster, were not detected in all optical spectra. To overcome the lack of auroral-line detections in many regions, we developed and validated an empirical multiphase temperature-calibration framework based on \HII regions studied in the literature with direct-method temperature measurements, reproducing literature O/H, N/H, and S/H abundances with small residual scatter. We further use this calibration on our M83 \HII\ regions for estimating multi-phase electron temperature as well as elemental abundances. As such, we provide the most direct multi-element comparison to date of stellar feedback signatures and chemical mixing across ionised (\HII) and neutral (\HI) phases in a metal-rich spiral. We measure abundances for multiple elements: oxygen (O), sulphur (S), nitrogen (N), and iron (Fe) that trace distinct nucleosynthetic pathways and characteristic release timescales, thereby adding an explicit temporal `clock' dimension to multiphase metal-mixing studies in a metal-rich spiral.

Our key findings can be summarised as follows:

\begin{itemize}

    \item \textbf{Element-dependent enrichment:} Within the ionised phase, $\alpha$-elements (O and S) behave coherently, consistent with their shared core-collapse supernova origin, in agreement with previous studies. For the neutral phase, we assume this correlation to estimate oxygen abundances from sulphur. Iron behaves differently compared to oxygen: with weaker correlation and large scatter, consistent with its delayed dominant Type Ia SNe contribution and its strong sensitivity to dust depletion. N shows the most strikingly different behaviour: in M83 the ionised-neutral offsets reach $\rm \Delta$N/H $\approx$ 1.5 dex and persist over the several-Myr timescales probed here, pointing to localised enrichment into the ionised gas and inefficient short-timescale mixing of massive stellar wind-driven nitrogen yields into the cold neutral ISM. The magnitude and persistence of this offset in a massive spiral galaxy strongly contrast with NGC 5253 (dwarf galaxy), where nitrogen offsets diminish within $\sim$ 8Myr, highlighting the role of galactic potential wells and stellar outflows in regulating chemicals. In contrast to nitrogen, $\alpha$-elements show comparatively smaller phase offsets. Unique nitrogen enrichment in young stellar environments is seen here in metal-rich M83, as well as in the literature (in NGC 5253 and others). This suggests that nitrogen enrichment around young star clusters is a common phenomenon across a broad range of galactic metallicities. 
        
    \item \textbf{Age dependence of enrichment clocks:} The phase-offsets (ionised-neutral) of $\alpha$-elements increase with cluster age, consistent with fresh CCSNe enrichment on $\sim$3-5 Myr timescales. Iron shows no clear age dependence in the $\sim$1-6 Myr window probed here, as expected from its delayed origin in Type Ia SNe, though depletion onto dust likely induces variation within these short timescales. Nitrogen, in contrast, remains elevated in the ionised phase across $\sim$1-6 Myr, with no sign of the rapid decline as seen in NGC 5253. Together, these trends imply that while enrichment occurs rapidly within the ionised phase, the transfer of metals into the neutral gas is likely to occur on longer timescales than the $\sim$1–6 Myr window probed here.

    \item \textbf{Outflows and environment-regulated mixing:} The large, sustained $\Delta$N/O($>$1.5 dex) in M83 reflects slow cross-phase mixing in a deep potential well: M83's stronger gravitational confinement likely confines feedback-driven outflows to the local environment, so wind momentum primarily reshapes and enriches the \HII\ gas rather than dispersing metals into the larger \HI\, reservoir within $\lesssim$6 Myr. This indicates a clear decoupling between ionised and neutral gas phases on Myr timescales, with enrichment remaining localised to the ionised medium. Consistent with this, excess nitrogen abundance in ionised gas, $\Delta$N/H (=(N/H)$_{HII}$ - (N/H)$_{HI}$) shows strong positive correlations with stellar feedback (luminosity and wind momentum), while other elemental excess: $\Delta$X/H (X=O, S, Fe) show weak, insignificant trends. This reinforces the idea that stellar feedback primarily regulates enrichment within the ionised phase in M83, but also the feedback drives nitrogen enrichment much more prominently than other elements within the given stellar ages ($\sim$1-6 Myr). Together, these results indicate efficient in-situ enrichment of the ionised medium but inefficient transfer into the surrounding neutral gas on Myr timescales, contrasting with low-mass systems such as NGC5253, where winds can more readily vent and mix metals into the surrounding neutral gas.  
    
    \item \textbf{Spatial distribution and mixing:} Outside the nucleus (R/R$_{25}\gtrsim$0.05), abundance trends in both the neutral and ionised gas are generally weak and statistically insignificant ($p$-value $>$ 0.05). However, interestingly, ionised gas exhibits anti-correlations with galactocentric distance for all elements, broadly consistent with previously reported radial metallicity gradients in M83 \citep[e.g.,][]{Bresolin2016}, though these may flatten at larger radii or vary locally due to dynamical processes \citep[e.g.,][]{Sextl2025}. Neutral gas abundances show mild positive trends with radius, potentially reflecting inside-out enrichment commonly seen in star-forming galaxies, but the low statistical significance prevents us from drawing firm conclusions. Overall, the differing behaviour between phases likely reflects a combination of localised enrichment and radial mixing.

\end{itemize}

These results demonstrate that while the fundamental nucleosynthetic clocks are universal, rapid nitrogen injection(through massive stellar winds), prompt $\alpha$-element production (through CCSNe), and delayed iron enrichment (through Type Ia SNe)-their observed signatures between phases depend strongly on environment. In average-metallicity dwarfs, WR-driven yields are redistributed and mixed on relatively short timescales, producing measurable convergence between phases. In M83, by contrast, enrichment remains largely confined to the ionised phase, with large abundance offsets between \HI\, and \HII\ regions persisting over $\sim$6-8 Myr. 

Future work should refine and extend this framework along several directions. Increasing the sample to include clusters with older ages (average stellar age $\gtrsim$6 Myr) will allow us to test whether the large nitrogen offsets (between ionised and neutral phases) in M83 persist or eventually converge, but on longer timescales. Incorporating dense molecular gas tracers from archival ALMA observations, together with future JWST measurements of warm \HH, outside the galactic nucleus will provide the missing connection between chemical enrichment and the structure of the gas reservoir that fuels star formation. Extending this analysis to a wider galaxy sample-from massive spirals to extremely metal-poor galaxies will further clarify which trends reflect universal enrichment physics and which arise from environment-specific mixing efficiencies.

\begin{acknowledgments}
This work was supported in part by Director’s Research Funds from the Space Telescope Science Institute (DRF Project: D0001.82485). We thank Annalisa De Cia for invaluable guidance on dust depletion corrections in the neutral gas, Logan Jones for extensive support with the stellar fitting code SESSAMME, and Calum Hawcroft for providing the most up-to-date STARBURST99 stellar libraries as well as updated information on them. Several co-authors acknowledge support from the European Space Agency. The authors would also like to thank the anonymous referee for their careful review and invaluable comments on previous versions of the manuscript, which were instrumental in substantially improving the quality and clarity of this work.  \\

\end{acknowledgments}

\facilities{HST(COS), VLT(MUSE), LBT(MODS)}

\software{astropy \citep{astropy_2013, astropy_2018, The_Astropy_Collaboration_2022}, VPFIT \citep{Carswell2014}, SESAMME \citep{Jones2023}}

\appendix

\section{Extracting the spectral full-width at half maximum (vFWHM) of HST/COS spectra \label{varying_cos_res}}
\begin{figure}
\includegraphics[page=2,width=1.0\linewidth,trim={410 19 410 20},clip]{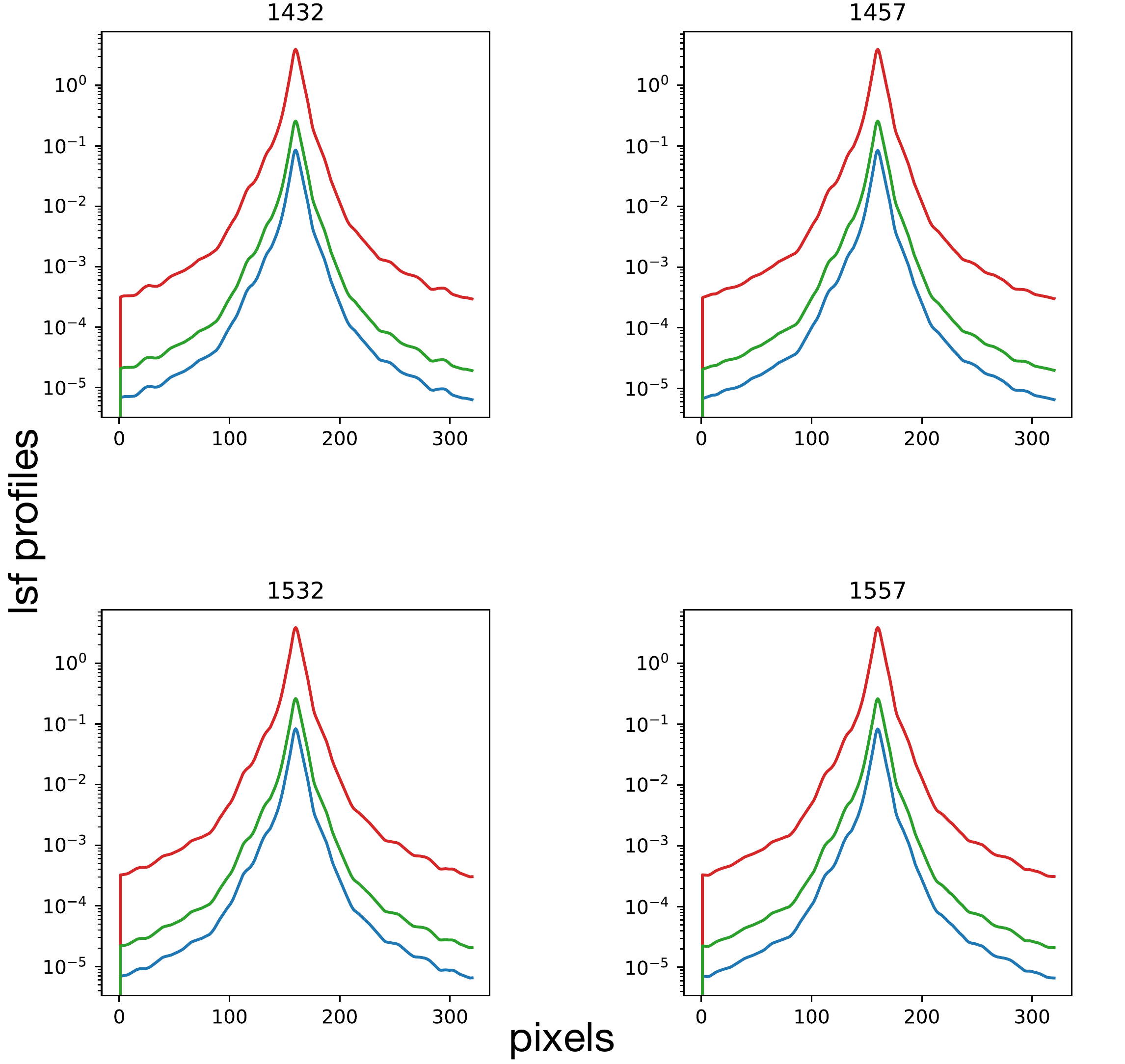}
\caption{Convolution of the COS LSF with varying spatial FWHM: original (green), narrow (blue) and wide (red) for the YSC M83-POS-1 (visit - LDN701010, PID: 15193, PI: Aloisi). The range in spatial FWHM is obtained from structure within the 1-D acquisition image. The separate sub-plots show some of the different wavelength regimes available in HST/COS spectra.}
\label{a-fig:4}
\end{figure}

Standard line spread function (LSF) profiles for point sources are available for each of the different grating/cenwave combinations of HST/COS. However, since most of our observed targets are extended, we cannot use the standard COS LSF. The programs mentioned above are separated into different visits for each target. For each visit, we fit the 1-D acquisition image (collapsed along the dispersion axis), with a multi-Gaussian model to allow multiple image components. For each visit related to every YSC, we obtain the spatial FWHM (in ") for the widest (A) and the narrowest (B) component within the 1-D image. These signify the faintest and the brightest components within the star cluster, respectively. We then convolve both these spatial FWHM (obtained from the faintest and brightest component) with the standard COS-LSF for each wavelength regime to obtain three separate LSFs, one original and two revised: wide, and narrow. Figure ~\ref{a-fig:4} shows an example of the different COS-LSFs obtained for different wavelength regimes. We use this range in LSF as a variable while convolving the observed gas absorption profiles along the line-of-sight.

Due to the variance in HST/COS LSFs and multiple position angles of the observations, we refrain from co-adding the spectra, even when they belong to the same target and have common wavelength coverage and fit them separately while considering the same absorption model for the neutral gas cloud. Then we vary the resolution within the obtained range (from above) while fitting the multi-component absorption Voigt profile.

\section{Cold diffuse molecular hydrogen, H$_2$ \label{diffuse_h2_intro}}
Due to the limitation in the HST/COS wavelength range, we cannot model the Lyman-Werner bands of cold molecular hydrogen within our gas for most clusters except M83-POS1 and M83-POS2. For these two clusters, the wavelength range (1064 - 1796 $\rm \AA$) allows us to measure column densities (or limits therein) for all rotational levels of diffuse H$_2$. For the rest of the clusters, the wavelength range (1125 - 1738 $\rm \AA$) only probes the higher rotational transitions of diffuse H$_2$ ($J>6$) and cannot inform us about the total column density of diffuse H$_2$, as most of the H$_2$ gas is in lower rotational transitional states ($J=0,1$). Hence, we refrain from obtaining the diffuse H$_2$ column density. Within the two clusters, M83-POS1 and M83-POS2, we report the total H$_2$ column density, log($N$H$_2$ [atoms cm$\rm ^{-2}$])=17.22$\rm \pm$0.11 and 17.81$\rm \pm$0.05 respectively. The total column density of hydrogen, i.e. log($N$H)=log($N$\HI\, + 2*$N$H$_2$) is given in Table.~\ref{tab1:absorption_col_den}.

\section{Information tables: Observation of neutral and ionised gas around individual clusters \label{cluster_appendix}}

This section provides detailed measurements for surrounding regions of each young star cluster (YSC) in our M83 sample. We tabulate the reddening-corrected optical emission-line intensities (I$\lambda$) measured from the surrounding \HII\ regions, derived from VLT/MUSE (and where applicable LBT/MODS) spectroscopy in Tables.~\ref{tab1:emission_line_flux} and \ref{tab2:emission_line_flux}. We also report neutral-gas column densities measured from HST/COS fUV absorption-line spectroscopy along the corresponding sightlines in Table~\ref{tab1:absorption_col_den}. Together, these tables provide the full set of ionised- and neutral- phase information used in our multiphase abundance analysis in this work.

\begin{table*}[]
\raggedright
\begin{tabular}{lllllllll}
\hline
YSC       & [\OII]$\lambda$3726   & [\OII]$\lambda$3728   & [\FeIII]$\lambda$4658     & H$\rm \beta$     & [\OIII]$\lambda$4958   & [\OIII]$\lambda$5006  & [\NII]$\lambda$5754       & [\SIII]$\lambda$6312        \\
\hline
\hline
M83-1     & -                                 & -                                 & 0.84$\rm \pm$0.20                         & 6.27$\rm \pm$0.50              & 0.36$\rm \pm$0.04                        & 1.08$\rm \pm$0.13                        & $\rm <$0.13                   & -                                  \\
M83-2     & -                                 & -                                 & 0.72$\rm \pm$0.24                         & 3.33$\rm \pm$0.31              & 0.38$\rm \pm$0.06                        & 1.14$\rm \pm$0.17                        & $\rm <$0.29                   & -                                  \\
M83-3     & -                                 & -                                 & 12.80$\rm \pm$1.03                        & 381.79$\rm \pm$5.06            & 28.07$\rm \pm$0.14                       & 82.38$\rm \pm$0.41                       & 2.28$\rm \pm$0.16                       & $\rm <$0.41                    \\
M83-4     & -                                 & -                                 & 16.20$\rm \pm$1.25                        & 449.25$\rm \pm$1.61            & 27.93$\rm \pm$0.17                       & 83.93$\rm \pm$0.47                       & $\rm \sim$2.46                        & $\rm <$0.44                    \\
M83-5     & -                                 & -                                 & 23.36$\rm \pm$2.16                        & 462.90$\rm \pm$2.08            & 24.19$\rm \pm$0.24                       & 69.21$\rm \pm$0.71                       & $\rm <$3.86                   & $\rm <$0.74                    \\
M83-6     & -                                 & -                                 & 2.78$\rm \pm$0.92                         & 20.98$\rm \pm$1.70             & 1.50$\rm \pm$0.10                        & 4.50$\rm \pm$0.28                        & $\rm <$0.79                   & -                                  \\
M83-7     & -                                 & -                                 & 1.21$\rm \pm$0.05                         & 79.01$\rm \pm$0.25             & 1.73$\rm \pm$0.01                        & 5.31$\rm \pm$0.04                        & $\rm <$0.38                   & 0.13$\rm \pm$0.01                        \\
M83-8     & -                                 & -                                 & -                                   & -                        & -                                  & -                                  & -                                 & -                                  \\
M83-9     & -                                 & -                                 & -                                   & -                        & -                                  & -                                  & -                                 & -                                  \\
M83-10    & -                                 & -                                 & 0.34$\rm \pm$0.28                         & 18.66$\rm \pm$1.77             & 0.82$\rm \pm$0.10                        & 2.39$\rm \pm$0.30                        & $\rm <$0.48                   & -                                  \\
M83-11    & -                                 & -                                 & 1.22$\rm \pm$0.18                         & 5.52$\rm \pm$0.15              & 0.59$\rm \pm$0.04                        & 1.75$\rm \pm$0.12                        & -                                 & $\rm <$0.15                    \\
M83-12    & -                                 & -                                 & 0.66$\rm \pm$0.33                         & 11.99$\rm \pm$0.81             & 0.38$\rm \pm$0.08                        & 1.14$\rm \pm$0.23                        & $\rm <$0.34                   & -                                  \\
M83-13    & -                                 & -                                 & -                                   & -                        & -                                  & -                                  & -                                 & -                                  \\
M83-14    & -                                 & -                                 & 1.23$\rm \pm$0.47                         & 20.41$\rm \pm$1.06             & 0.74$\rm \pm$0.13                        & 2.18$\rm \pm$0.39                        & $\rm <$0.36                   & $\rm <$0.29                    \\
M83-15    & -                                 & -                                 & -                                   & -                        & -                                  & -                                  & -                                 & -                                  \\
M83-16    & -                                 & -                                 & 0.82$\rm \pm$0.26                         & 10.80$\rm \pm$0.93             & 0.44$\rm \pm$0.07                        & 1.28$\rm \pm$0.22                        & $\rm <$0.3                    & $\rm <$0.19                    \\
M83-POS-1 & -                                 & -                                 & 10.34$\rm \pm$0.92                        & 748.77$\rm \pm$1.60            & 19.15$\rm \pm$0.11                       & 56.03$\rm \pm$0.32                       & $\rm \sim$3.23                        & $\rm <$0.78                    \\
M83-POS-2 & -                                 & -                                 & 13.29$\rm \pm$0.80                        & 237.32$\rm \pm$3.96            & 12.27$\rm \pm$0.12                       & 38.11$\rm \pm$0.34                       & $\rm \sim$1.97                        & -                                  \\
R(1)      & 1.54$\rm \pm$0.15                       & 1.61$\rm \pm$0.16                       & -                                   & 1.75$\rm \pm$0.03              & 0.19$\rm \pm$0.0045                      & 0.56$\rm \pm$0.014                       & $\rm <$0.08                   & $\rm \sim$0.04                         \\
R(2)      & $\rm \sim$0.43                        & 13.39$\rm \pm$1.10                      & 0.07$\rm \pm$0.12                         & 11.12$\rm \pm$0.15             & 0.1$\rm \pm$0.0069                       & 0.31$\rm \pm$0.021                       & $\rm <$0.34                   & $\rm \sim$0.10                         \\
R(6)      & 1.84$\rm \pm$0.66                       & $\rm \sim$0.62                        & 0.05$\rm \pm$0.06                         & 1.30$\rm \pm$0.13              & 0.015$\rm \pm$0.0081                     & 0.044$\rm \pm$0.024                      & $\rm <$0.20                   & -                                  \\
R(7)      & 60.65$\rm \pm$4.60                      & 44.90$\rm \pm$11.91                     & 3.11$\rm \pm$0.61                         & 121.55$\rm \pm$0.97            & 0.17$\rm \pm$0.0045                      & 0.52$\rm \pm$0.013                       & 1.66$\rm \pm$0.15                       & -                                  \\
R(8)      & $\rm \sim$1.09                        & $\rm \sim$2.81                        & 0.10$\rm \pm$0.38                         & 3.17$\rm \pm$1.41              & $\rm \sim$0.023                        & $\rm \sim$0.068                        & $\rm <$0.32                   & $\rm \sim$0.08                         \\
R(12)     & $\rm \sim$1.57                        & $\rm \sim$1.73                        & 0.12$\rm \pm$0.22                         & 5.42$\rm \pm$1.20              & $\rm \sim$0.089                        & $\rm \sim$0.27                         & $\rm <$0.30                   & $\rm \sim$0.25                         \\                        
\hline
\end{tabular}
\caption{Reddening corrected Intensity (I$\lambda$) for the optical emission lines originating from the HII regions around young star clusters (YSC) in M83. The ID of the spectra obtained from VLT/MUSE starts with `M83-', while the spectra obtained from LBT/MODS starts with an `R'. Unit: 1e$\rm ^{-15}$ erg s$\rm ^{-1}$ cm$\rm ^{-2}$. Pixel size: 0.2 pixel/arcsec for VLT/MUSE and 0.225 pixel/arcsec for LBT/MODS. \label{tab1:emission_line_flux}}
\end{table*}

\begin{table*}[]
\raggedright
\begin{tabular}{lllllllll}
\hline
YSC       & [\NII]$\lambda$6548   & [\NII]$\lambda$6583     & [\SII]$\lambda$6716   & [\SII]$\lambda$6730   & H$\rm \alpha$      & [\OII]$\lambda$7320   & [\OII]$\lambda$7329   &  [\SIII]$\lambda$9068   \\
\hline
\hline
M83-1     & 1.60$\rm \pm$0.12                       & 4.74$\rm \pm$0.36                       & 1.72$\rm \pm$0.15                       & 1.36$\rm \pm$0.11                       & 17.93$\rm \pm$1.51             & -                                 & $\rm <$0.15                   & 0.17$\rm \pm$0.02                        \\
M83-2     & 1.01$\rm \pm$0.05                       & 3.00$\rm \pm$0.14                       & 1.03$\rm \pm$0.05                       & 0.90$\rm \pm$0.05                       & 9.52$\rm \pm$0.38              & -                                 & $\rm <$0.17                   & 0.29$\rm \pm$0.02                        \\
M83-3     & 192.28$\rm \pm$2.08                     & 583.82$\rm \pm$6.10                     & 121.84$\rm \pm$1.32                     & 131.89$\rm \pm$1.40                     & 1092.05$\rm \pm$11.19          & $\rm <$2.40                   & $\rm <$3.29                   & 33.90$\rm \pm$0.34                       \\
M83-4     & 198.07$\rm \pm$0.60                     & 592.94$\rm \pm$1.81                     & 136.66$\rm \pm$0.46                     & 142.32$\rm \pm$0.48                     & 1285.03$\rm \pm$4.07           & $\rm <$1.63                   & $\rm <$2.63                   & 31.39$\rm \pm$0.15                       \\
M83-5     & 192.32$\rm \pm$0.62                     & 569.54$\rm \pm$1.83                     & 118.36$\rm \pm$0.47                     & 110.83$\rm \pm$0.44                     & 1324.24$\rm \pm$4.05           & $\rm <$0.78                   & $\rm <$2.33                   & 31.60$\rm \pm$0.19                       \\
M83-6     & 5.46$\rm \pm$0.35                       & 15.85$\rm \pm$1.01                      & 4.73$\rm \pm$0.33                       & 3.74$\rm \pm$0.23                       & 60.03$\rm \pm$3.13             & -                                 & $\rm <$0.21                   & 0.70$\rm \pm$0.06                        \\
M83-7     & 22.93$\rm \pm$0.07                      & 69.69$\rm \pm$0.21                      & 17.80$\rm \pm$0.06                      & 13.58$\rm \pm$0.05                      & 225.99$\rm \pm$0.70            & 0.15$\rm \pm$0.01                       & 0.23$\rm \pm$0.01                       & 7.01$\rm \pm$0.03                        \\
M83-8     & -                                 & -                                 & -                                 & -                                 & -                        & -                                 & -                                 & -                                  \\
M83-9     & -                                 & -                                 & -                                 & -                                 & -                        & -                                 & -                                 & -                                  \\
M83-10    & 4.65$\rm \pm$0.44                       & 14.11$\rm \pm$1.34                      & 4.53$\rm \pm$0.43                       & 3.39$\rm \pm$0.32                       & 53.37$\rm \pm$4.41             & 0.09$\rm \pm$0.04                       & 0.13$\rm \pm$0.04                       & 1.34$\rm \pm$0.14                        \\
M83-11    & 1.87$\rm \pm$0.03                       & 5.50$\rm \pm$0.08                       & 1.67$\rm \pm$0.04                       & 1.39$\rm \pm$0.04                       & 15.80$\rm \pm$0.18             & $\rm <$0.16                   & $\rm <$0.20                   & 0.25$\rm \pm$0.03                        \\
M83-12    & 3.24$\rm \pm$0.23                       & 9.88$\rm \pm$0.67                       & 3.37$\rm \pm$0.22                       & 2.43$\rm \pm$0.18                       & 34.30$\rm \pm$2.29             & $\rm <$0.11                   & $\rm <$0.27                   & 0.55$\rm \pm$0.08                        \\
M83-13    & -                                 & -                                 & -                                 & -                                 & -                        & -                                 & -                                 & -                                  \\
M83-14    & 6.24$\rm \pm$0.29                       & 18.64$\rm \pm$0.89                      & 6.69$\rm \pm$0.32                       & 4.77$\rm \pm$0.23                       & 58.38$\rm \pm$2.72             & 0.09$\rm \pm$0.05                       & 0.08$\rm \pm$0.06                       & 1.06$\rm \pm$0.09                        \\
M83-15    & -                                 & -                                 & -                                 & -                                 & -                        & -                                 & -                                 & -                                  \\
M83-16    & 3.76$\rm \pm$0.31                       & 11.00$\rm \pm$0.95                      & 4.32$\rm \pm$0.41                       & 3.23$\rm \pm$0.28                       & 30.89$\rm \pm$2.65             & $\rm <$0.11                   & $\rm <$0.11                   & 0.57$\rm \pm$0.07                        \\
M83-POS-1 & 369.33$\rm \pm$0.81                     & 1066.05$\rm \pm$2.37                    & 148.89$\rm \pm$0.32                     & 164.57$\rm \pm$0.39                     & 2141.82$\rm \pm$4.35           & $\rm <$2.22                   & $\rm <$2.42                   & 126.62$\rm \pm$0.35                      \\
M83-POS-2 & 100.81$\rm \pm$1.46                     & 300.64$\rm \pm$4.26                     & 61.27$\rm \pm$0.88                      & 63.57$\rm \pm$0.86                      & 678.80$\rm \pm$9.30            & $\rm <$0.22                   & $\rm <$1.39                   & 15.61$\rm \pm$0.15                       \\
R(1)      & 0.61$\rm \pm$0.02                       & 1.93$\rm \pm$0.05                       & 0.65$\rm \pm$0.02                       & 0.46$\rm \pm$0.02                       & 5.00$\rm \pm$0.13              & $\rm <$0.08                   & $\rm <$0.14                   & 0.37$\rm \pm$0.02                        \\
R(2)      & 3.87$\rm \pm$0.04                       & 11.84$\rm \pm$0.12                      & 3.02$\rm \pm$0.04                       & 2.17$\rm \pm$0.03                       & 31.81$\rm \pm$0.35             & 0.13$\rm \pm$0.01                       & 0.30$\rm \pm$0.03                       & 1.68$\rm \pm$0.03                        \\
R(6)      & 0.45$\rm \pm$0.04                       & 1.29$\rm \pm$0.11                       & 0.50$\rm \pm$0.05                       & 0.36$\rm \pm$0.05                       & 3.72$\rm \pm$0.30              & 0.06$\rm \pm$0.01                       & 0.16$\rm \pm$0.02                       & 0.06$\rm \pm$0.03                        \\
R(7)      & 53.17$\rm \pm$0.78                      & 153.07$\rm \pm$2.30                     & 38.40$\rm \pm$0.46                      & 36.96$\rm \pm$0.48                      & 347.73$\rm \pm$4.12            & -                                 & 1.20$\rm \pm$0.09                       & 9.91$\rm \pm$0.30                        \\
R(8)      & 1.15$\rm \pm$0.38                       & 3.43$\rm \pm$1.14                       & 0.91$\rm \pm$0.28                       & 0.67$\rm \pm$0.15                       & 9.06$\rm \pm$2.30              & $\rm <$0.12                   & $\rm <$0.51                   & 0.27$\rm \pm$0.11                        \\
R(12)     & 1.42$\rm \pm$0.54                       & 4.13$\rm \pm$1.68                       & 0.96$\rm \pm$0.21                       & 0.76$\rm \pm$0.28                       & 15.49$\rm \pm$1.42             & $\rm <$0.23                   & $\rm <$0.31                   & 0.59$\rm \pm$0.12                        \\                       
\hline
\end{tabular}
\caption{Reddening corrected Intensity (I$\rm \lambda$) for the optical emission lines (continued from table~\ref{tab1:emission_line_flux}) originating from the \HII\, regions around young star-clusters (YSC) in M83. The ID of the spectra obtained from VLT/MUSE starts with `M83-', while the spectra obtained from LBT/MODS starts with an `R'. Unit: 1e$\rm ^{-15}$ erg s$\rm ^{-1}$ cm$\rm ^{-2}$. Pixel size: 0.2 pixel/arcsec for VLT/MUSE and 0.225 pixel/arcsec for LBT/MODS. \label{tab2:emission_line_flux}}
\end{table*}

\begin{table*}[]
\raggedright
\begin{tabular}{llllllll}
\hline
YSC       & Hydrogen     & Nitrogen     & Oxygen       & Sulphur      & Iron         & Nickel       & Phosphorus   \\
\hline
\hline
M83-1     & 21.05$\rm \pm$0.08 & 15.17$\rm \pm$0.09 & 17.58$\rm \pm$0.09 & 16.01$\rm \pm$0.07 & 15.12$\rm \pm$0.08 & 15.04$\rm \pm$0.17 & 14.23$\rm \pm$0.30 \\
M83-2     & 20.71$\rm \pm$0.08 & 14.88$\rm \pm$0.30 & 17.18$\rm \pm$0.13 & 15.61$\rm \pm$0.11 & 15.03$\rm \pm$0.08 & 14.78$\rm \pm$0.28 & 14.74$\rm \pm$0.07 \\
M83-3     & 18.99$\rm \pm$0.23 & 15.03$\rm \pm$0.17 & 16.77$\rm \pm$0.11 & 15.20$\rm \pm$0.09 & 14.87$\rm \pm$0.05 & 14.52$\rm \pm$0.08 & 14.34$\rm \pm$0.12 \\
M83-4     & 19.56$\rm \pm$0.21 & 14.64$\rm \pm$0.25 & 17.18$\rm \pm$0.16 & 15.61$\rm \pm$0.15 & 15.06$\rm \pm$0.08 & 14.77$\rm \pm$0.17 & 14.43$\rm \pm$0.16 \\
M83-5     & 19.00$\rm \pm$0.30 & 13.61$\rm \pm$0.72 & 16.94$\rm \pm$0.13 & 15.37$\rm \pm$0.12 & 15.01$\rm \pm$0.13 & 14.75$\rm \pm$0.19 & 13.58$\rm \pm$0.56 \\
M83-6     & 20.60$\rm \pm$0.06 & 15.09$\rm \pm$0.09 & 17.37$\rm \pm$0.08 & 15.80$\rm \pm$0.05 & 15.05$\rm \pm$0.06 & 14.55$\rm \pm$0.10 & 14.51$\rm \pm$0.06 \\
M83-7     & 20.65$\rm \pm$0.09 & 14.72$\rm \pm$0.28 & 17.59$\rm \pm$0.08 & 16.02$\rm \pm$0.05 & 15.09$\rm \pm$0.11 & 14.75$\rm \pm$0.23 & 14.45$\rm \pm$0.10 \\
M83-8     & 20.54$\rm \pm$0.03 & 15.06$\rm \pm$0.08 & 16.84$\rm \pm$0.09 & 15.27$\rm \pm$0.07 & 15.08$\rm \pm$0.02 & 14.43$\rm \pm$0.15 & 14.30$\rm \pm$0.08 \\
M83-9     & 20.45$\rm \pm$0.07 & 14.86$\rm \pm$0.16 & 17.34$\rm \pm$0.07 & 15.77$\rm \pm$0.04 & 15.34$\rm \pm$0.06 & 14.54$\rm \pm$0.20 & 14.44$\rm \pm$0.08 \\
M83-10    & 20.62$\rm \pm$0.02 & 15.05$\rm \pm$0.07 & 17.31$\rm \pm$0.07 & 15.74$\rm \pm$0.03 & 15.02$\rm \pm$0.03 & 14.65$\rm \pm$0.12 & 14.24$\rm \pm$0.11 \\
M83-11    & 20.05$\rm \pm$0.07 & 14.99$\rm \pm$0.09 & 16.98$\rm \pm$0.10 & 15.41$\rm \pm$0.08 & 15.05$\rm \pm$0.03 & 14.56$\rm \pm$0.18 & 14.49$\rm \pm$0.07 \\
M83-12    & 20.85$\rm \pm$0.02 & 15.34$\rm \pm$0.05 & 17.36$\rm \pm$0.08 & 15.79$\rm \pm$0.05 & 15.14$\rm \pm$0.03 & 14.80$\rm \pm$0.16 & 14.38$\rm \pm$0.08 \\
M83-13    & 20.63$\rm \pm$0.02 & 15.39$\rm \pm$0.09 & 17.24$\rm \pm$0.13 & 15.67$\rm \pm$0.11 & 15.06$\rm \pm$0.12 & 14.64$\rm \pm$0.15 & 14.54$\rm \pm$0.07 \\
M83-14    & 20.43$\rm \pm$0.13 & 14.89$\rm \pm$0.23 & 17.33$\rm \pm$0.11 & 15.76$\rm \pm$0.09 & 15.17$\rm \pm$0.10 & 14.89$\rm \pm$0.65 & 14.42$\rm \pm$0.16 \\
M83-15    & 20.91$\rm \pm$0.05 & 14.33$\rm \pm$0.27 & 17.22$\rm \pm$0.10 & 15.65$\rm \pm$0.08 & 14.91$\rm \pm$0.08 & 14.69$\rm \pm$0.13 & 14.01$\rm \pm$0.18 \\
M83-16    & 21.05$\rm \pm$0.02 & 16.08$\rm \pm$0.18 & 17.45$\rm \pm$0.07 & 15.88$\rm \pm$0.04 & 15.25$\rm \pm$0.04 & 14.98$\rm \pm$0.09 & 14.55$\rm \pm$0.05 \\
M83-POS-1 & 19.93$\rm \pm$0.03 & 15.15$\rm \pm$0.02 & 17.19$\rm \pm$0.06 & 15.62$\rm \pm$0.02 & 14.80$\rm \pm$0.02 & 14.28$\rm \pm$0.12 & 14.41$\rm \pm$0.02 \\
M83-POS-2 & 19.02$\rm \pm$0.03 & 14.85$\rm \pm$0.04 & 16.79$\rm \pm$0.30 & 15.22$\rm \pm$0.29 & 14.75$\rm \pm$0.01 & 13.51$\rm \pm$0.21 & 14.14$\rm \pm$0.11 \\
\hline
\end{tabular}
\begin{minipage}{\textwidth}
\footnotesize
$^{a}$ For M83-POS-1, \HH does not add any meaningful contribution to the total log($N(H)$) of that region. \\
$^{b}$ For M83-POS-2, the total column density of hydrogen, i.e.\ log($N(H)$)=log($N$(\HI)\,+\,2$N$(\HH)) is $19.08\pm0.03$.
\end{minipage}
\caption{Observed column density of elements log(N(X)) where X=Hydrogen, iron, nitrogen, oxygen, sulphur, Nickel and Phosphorus observed using HST/COS spectra in the neutral gas towards the young star clusters in M83. Unit: [atoms cm$\rm ^{-2}$]. \label{tab1:absorption_col_den}}
\end{table*}

\section{Ionisation Correction Factors for neutral (\HI) gas abundances \label{icf_neutral_appendix_section}}

The ionic column densities measured from UV absorption lines do not necessarily correspond to the total elemental abundances of the neutral interstellar medium. In particular, two ionisation-related effects must be considered: 1. contamination from ionised gas along the line of sight and 2. the presence of unobserved ionisation stages within the predominantly neutral gas. To account for these effects, we apply ionisation correction factors (ICFs) following the methodology of \citet{Hernandez2021}. These corrections are derived using tailored \textsc{CLOUDY} photoionisation models run initially by \cite{James_2014} for M83 POS-1 and POS-2 for all elements and then later extended by \citet{Hernandez2021} for all M83 clusters. The CLOUDY runs by \citet{Hernandez2021} reproduce the physical conditions of each sightline and quantify both the ionised-gas contamination and the contribution from unobserved ions. The total ionisation correction is defined as - 

\begin{equation}
ICF_{TOTAL} = ICF_{ionised} + ICF_{neutral}
\end{equation}

where ICF$_{ionised}$ accounts for contaminating absorption arising in ionised gas and ICF$_{neutral}$ accounts for unobserved ionisation stages within the neutral medium. The corrected elemental column density is then given by - 

\begin{equation}
log N(X)_{corr} = log N(X)_{obs} + ICF_{TOTAL}]
\end{equation}

where (N(X)$_{obs}$) is the observed ionic column density and (N(X)$_{corr}$) is the ionisation-corrected column density used for abundance determinations. Depending on the element and local ionisation conditions, these corrections can reach several tenths of a dex and are therefore essential for obtaining accurate neutral-gas abundances. The total ICF values adopted for each element and sightline are listed in Table~\ref{tab1:absorption_icf_table} and are applied before calculating the ICF-corrected elemental abundances given in Table.~\ref{tab:data_icf}.

\begin{table*}[]
\raggedright
\begin{tabular}{llllllll}
\hline
YSC       & Hydrogen & Nitrogen & Oxygen & Sulphur & Iron   & Nickel & Phosphorus \\
\hline
\hline
M83-1     & -0.120   & -0.044   & -0.074 & -0.101  & -0.050 & -0.078 & -0.073     \\
M83-2     & -0.116   & -0.068   & -0.089 & -0.084  & -0.062 & -0.090 & -0.047     \\
M83-3     & 0.044    & 0.013    & 0.064  & 0.396   & 0.439  & 0.422  & 0.472      \\
M83-4     & 0.013    & 0.004    & 0.020  & 0.170   & 0.187  & 0.175  & 0.225      \\
M83-5     & -0.123   & -0.057   & -0.085 & -0.101  & -0.062 & -0.092 & -0.070     \\
M83-6     & -0.116   & -0.074   & -0.092 & -0.079  & -0.061 & -0.089 & -0.041     \\
M83-7     & -0.002   & -0.003   & -0.001 & 0.031   & 0.026  & 0.019  & 0.067      \\
M83-8     & -0.116   & -0.078   & -0.094 & -0.077  & -0.061 & -0.088 & -0.038     \\
M83-9     & -0.113   & -0.083   & -0.096 & -0.069  & -0.059 & -0.081 & -0.031     \\
M83-10    & -0.118   & -0.073   & -0.093 & -0.083  & -0.062 & -0.091 & -0.044     \\
M83-11    & -0.149   & -0.130   & -0.134 & -0.072  & -0.062 & -0.087 & -0.027     \\
M83-12    & -0.123   & -0.061   & -0.088 & -0.097  & -0.062 & -0.093 & -0.061     \\
M83-13    & -0.117   & -0.072   & -0.091 & -0.082  & -0.062 & -0.090 & -0.043     \\
M83-14    & -0.004   & -0.005   & -0.002 & 0.042   & 0.038  & 0.028  & 0.079      \\
M83-15    & -0.123   & -0.059   & -0.088 & -0.099  & -0.062 & -0.093 & -0.064     \\
M83-16    & -0.131   & -0.047   & -0.081 & -0.111  & -0.055 & -0.084 & -0.082     \\
M83-POS-1 & 0.011    & 0.000    & 0.016  & 0.154   & 0.162  & 0.167  & 0.203      \\
M83-POS-2 & 0.106    & 0.022    & 0.159  & 0.651   & 0.657  & 0.702  & 0.699       \\    
\hline
\end{tabular}
\caption{Total ionisation correction factor (ICF) for observed neutral (\HI) gas elements. \label{tab1:absorption_icf_table}}
\end{table*}

\section{Dust depletion correction for neutral gas abundances \label{subsection_dust_neutral}}
\begin{figure}
\includegraphics[width=\linewidth]{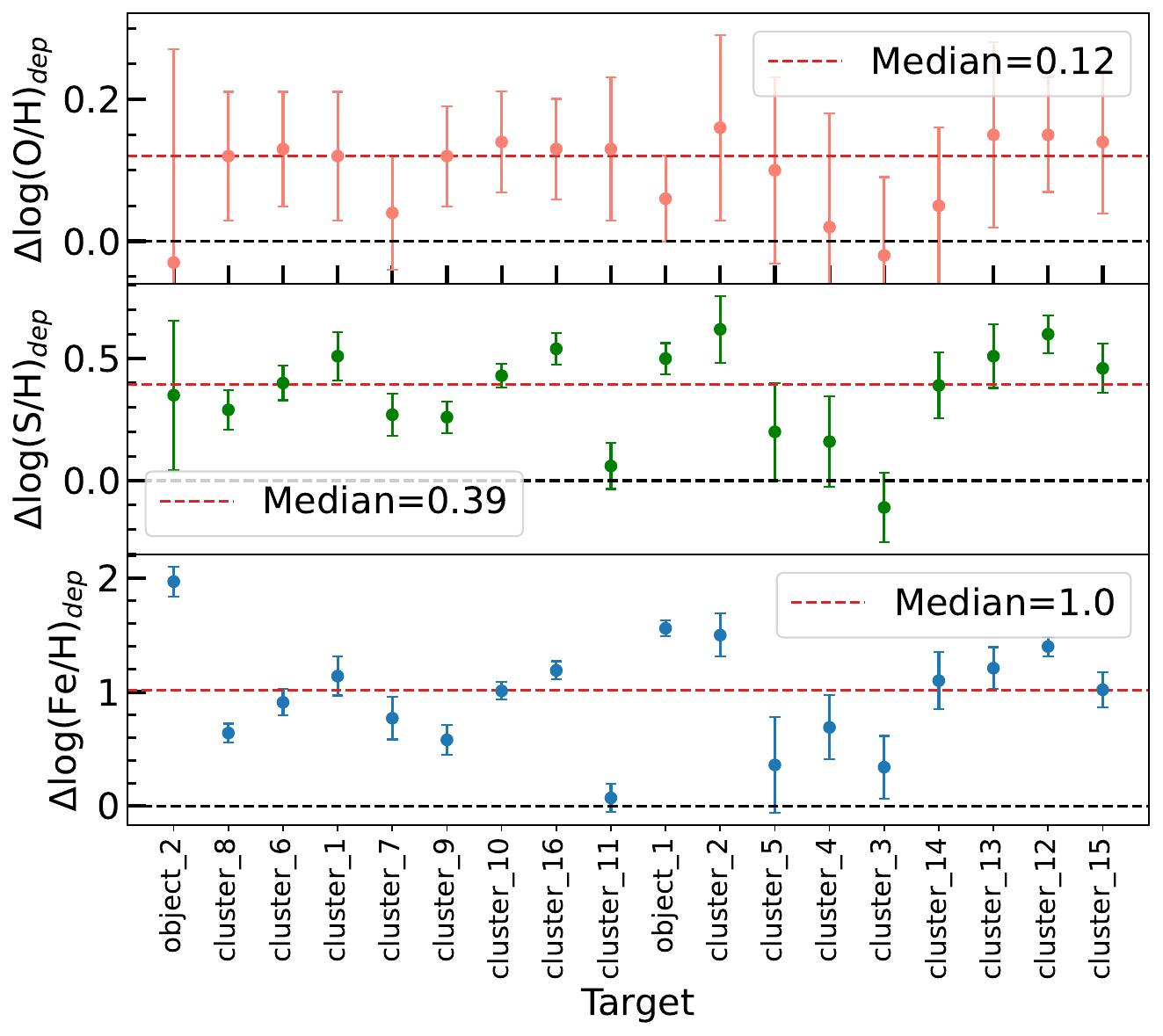}
\caption{Plotting change in abundance due to depletion($\Delta$log(X/H)$_{dep}$ = (X/H)$_{depleted}$ - (X/H)$_{ICF}$) for each observed neutral gas element, X, for each YSC in M83 observed with VLT/COS. The three subplots from bottom to top show values for X = Fe, S, and O. The median $\Delta$log(X/H)$_{dep}$ is plotted as a horizontal orange line in each subplot.}
\label{fig:data5}
\end{figure}

While elements are abundantly present within the gas phase in a cloud, a fraction of them $\lesssim$10\% could be locked within dust, a phenomenon widely known as dust depletion \citep[see][and references therein]{De_Cia_2024}. The amount of dust depletion varies from one element to another depending on the physical properties of dust as well as the refractory index of the element. For our sample of metals within neutral gas, we estimate the amount of dust using a combination of different refractory metals. We use our ionisation-corrected elemental column densities with the prescriptions described in detail in \citet[][]{De_Cia_2024, Konstantopoulou2024a, Konstantopoulou2024b} to obtain the dust content and the fraction of those elements depleted onto dust. Instead of the standard [Zn/Fe] depletion historically used in the literature \citep[see e.g.][]{Ledoux2002}, we use a combination of different elements with varying refractory index to estimate dust depletion. For our study, we primarily find corrections for iron, sulphur, and oxygen. Figure.~\ref{fig:data5} shows the change in abundance due to depletion($\Delta$log(X/H)$_{dep}$ = (X/H)$_{depleted}$ - (X/H)$_{ICF}$) measured for element, X. We note that iron is strongly depleted with the largest correction (median$\sim$0.57) while sulphur is mildly depleted (median$\sim$0.21). There is almost no dust depletion seen for oxygen (median$\sim$0.03). This is consistent with what is found in the literature \citep[see e.g.][]{Konstantopoulou2024b}. We note that there is no prescription for dust depletion correction for nitrogen in the literature.

\section{Constraining the properties of young star clusters \label{stellar_fitting}}

\subsection{Literature studies}
\citet{Hernandez2019} employed the integrated-light (IL) spectral-fitting framework introduced by \citet{Larsen2012} (hereafter the L12 method) to derive overall metallicities for the young star clusters (YSCs) analysed in this work (except M83-POS-1 and M83-POS-2). While M83-5 was a part of their selection, it was excluded from their analysis due to the low S/N of the HST/COS spectra. In their approach, metallicity is constrained by fitting ensembles of metal absorption features in the IL spectrum. A key strength of applying the method in the far-UV is that diagnostic stellar absorption and wind features help mitigate the well-known degeneracies among age, metallicity, and reddening. In particular, the P-Cygni profiles of \NV, \SiIV, and \CIV\, are jointly sensitive to population age, chemical content, and attenuation, enabling a more robust characterisation of the underlying stellar populations \citep{Chisholm2019}. Operationally, the L12 method requires estimates of the ages and metallicities for the selection of the initial models. Stellar parameters of the contributing stars (e.g., $M$, $T_{\rm eff}$, and $\log g$) are obtained from theoretical isochrones or, where available, from colour-magnitude diagrams (CMDs). Isochrones provide a theoretical mapping between stellar mass, luminosity, and effective temperature at fixed age and metallicity, while CMDs place resolved stars onto the Hertzsprung-Russell (HR) diagram observationally. In the implementation adopted by \citet{Hernandez2019}, the initial isochrone selection was performed using cluster ages derived from photometric techniques similar to those of \citet{Chandar2010} and \citet{Whitmore2011}, while assuming a metallicity representative of the host galaxy. From these inputs, model atmospheres are generated for each stellar type, synthetic spectra are computed and coadded into a single IL model spectrum, and the resulting model is compared to the observed spectrum. The comparison is repeated iteratively while varying the global metallicity (and associated abundances within the assumed pattern) to obtain the best-fitting solution. An important practical consideration is that the adopted photometric stellar ages carry substantial uncertainties (of order $\sim$50\%; see Table~2 of \citet{Hernandez2019}). While the L12 framework can efficiently refine metallicity given the adopted population parameters, these large age uncertainties propagate into the IL modelling and introduce additional systematic uncertainty, making direct comparisons between their metallicity estimates and our independently derived results less straightforward.

\subsection{UV-spectral fitting of young stellar population \label{SESSAME_intro}}
Hence, we decided to fit the stellar continuum and line features in our HST/COS spectra to obtain simultaneous information about the stellar age, metallicity, mass, and dust in the observed YSCs. The fitting is performed on the flux-calibrated spectra (i.e., not continuum-normalised), allowing simultaneous constraints from both continuum shape and line features. In particular, stellar age is primarily constrained by the strength and morphology of the P-Cygni profiles (which evolve rapidly over $\sim$1–10 Myr), while metallicity is constrained through the overall strength of metal line absorption and wind features, which scale with metal abundance. We use \href{https://github.com/astrolojo/SESAMME}{SESAMME} (Simultaneous Estimates of Star-cluster Age, Metallicity, Mass, and Extinction) - a Python-based software that performs full spectral fitting of the integrated-light spectrum of star clusters using the Markov Chain Monte Carlo (MCMC) method \citep[][]{Jones2023}. We used both Starburst99 \citep[SB99; ][]{Leitherer2014, Hawcroft2025} and BPASS \citep[][]{Eldridge2017PASA} stellar models, and we ended up adopting SB99 given its higher spectral resolution suitable for our HST/COS spectra. The SB99 models used in SESAMME are based on the GENEC stellar evolutionary tracks and associated stellar atmosphere libraries. The metallicities are defined relative to the solar abundance scale of \citet{Asplund+09}, adopting (Z$_{\odot}$=0.014). The initial heavy-element mixtures are approximately solar-scaled, following the GENEC prescriptions implemented in STARBURST99. Surface abundances are not held fixed during stellar evolution but evolve self-consistently through the effects of nuclear processing, mass loss, and internal mixing included in the adopted evolutionary tracks. We do not introduce additional age-dependent modifications to individual elemental abundance ratios (e.g., enhanced N/C from CN-processed material) beyond those already incorporated in the underlying stellar evolution and atmosphere models. The age range in the SB99 grids we used for analysis goes from 1 to 50 Myr with a stepsize of $\sim$1 Myr and metallicity, Z$_{SB99}$=[0.001,0.004,0.008,0.02,0.04]. In SESAMME, metallicity is represented by [Z](=log10(Z/Z$\rm _{\odot}$), where Z$\rm _{\odot}$=0.0134 from \citet{Asplund2009}). SB99 models adopt a Kroupa initial mass function (IMF) with slopes $\rm \alpha$=1.3 for 0.1-0.5 M$_{\odot}$ and $\rm \alpha$=2.3 for 0.5-100 M$_{\odot}$. The stellar population models assume a lower and upper mass limit of 0.1 and 100 M$_{\odot}$, respectively, consistent with standard implementations in population synthesis models used to derive YSC masses and luminosities. We note that uncertainties in the modelling of stellar winds (e.g., ionisation structure, clumping, and X-ray emission from shocks) can affect the detailed shape of UV wind lines. These effects are incorporated in the STARBURST99 framework, but may introduce systematic uncertainties in the derived parameters \citep[see][]{Hawcroft2025}. For dust extinction fitting, we assume the \citet{Calzetti2000} law, which is useful for dereddening the spectra of galaxies where massive stars dominate the radiation output, with R$_V$=4.05. We emphasise that we do not combine additional spectroscopic data sets (e.g., from VLT/MUSE or LBT/MODS) with the HST/COS spectra for the stellar population analysis. In the case of LBT/MODS, the aperture size differs from that of HST/COS, leading to mismatched spatial sampling of the young star clusters and their immediate environments. Although the VLT/MUSE spectra were extracted using apertures matched to HST/COS, the instruments remain subject to independent flux calibrations and systematics. As a result, directly combining these data for joint spectral fitting would introduce inconsistencies in the absolute and relative flux scales, potentially biasing the derived stellar parameters. For these reasons, and following similar approaches adopted in previous UV studies of young star clusters \citep[e.g.,][]{Sirressi2022a, Sirressi2022b}, we restrict our stellar population modelling to the internally consistent HST/COS data set.

In SESAMME, dust attenuation and stellar mass estimates are primarily constrained by the shape of the stellar continuum and a flux normalisation parameter, whereas stellar age and metallicity are driven by diagnostic stellar absorption and wind features. Although a broad UV-optical spectro-photometric SED fit based on archival HST imaging would, in principle, provide tighter constraints on dust attenuation, the far-UV wavelength range covered by HST/COS is intrinsically highly sensitive to even modest variations in dust content. Consequently, our UV-based fitting provides sufficiently robust constraints on the dust attenuation affecting the young stellar populations, ensuring that both the stellar mass and dust content are reliably determined within the framework of our analysis. While a comprehensive UV–optical SED modelling approach would further refine the dust estimates, such detailed treatment lies beyond the scope of this work, which focuses on the multiphase (\HI-\HII) gas properties in the immediate surroundings of these young star clusters. We note that dust in the surrounding ionised and neutral gas phases is independently estimated using established techniques widely adopted in the literature, including Balmer decrement measurements for \HII\ regions and dust-depletion corrections based on UV absorption-line abundances in the \HI\, phase. \\

\begin{figure}
    \centering
    \includegraphics[width=0.5\textwidth]{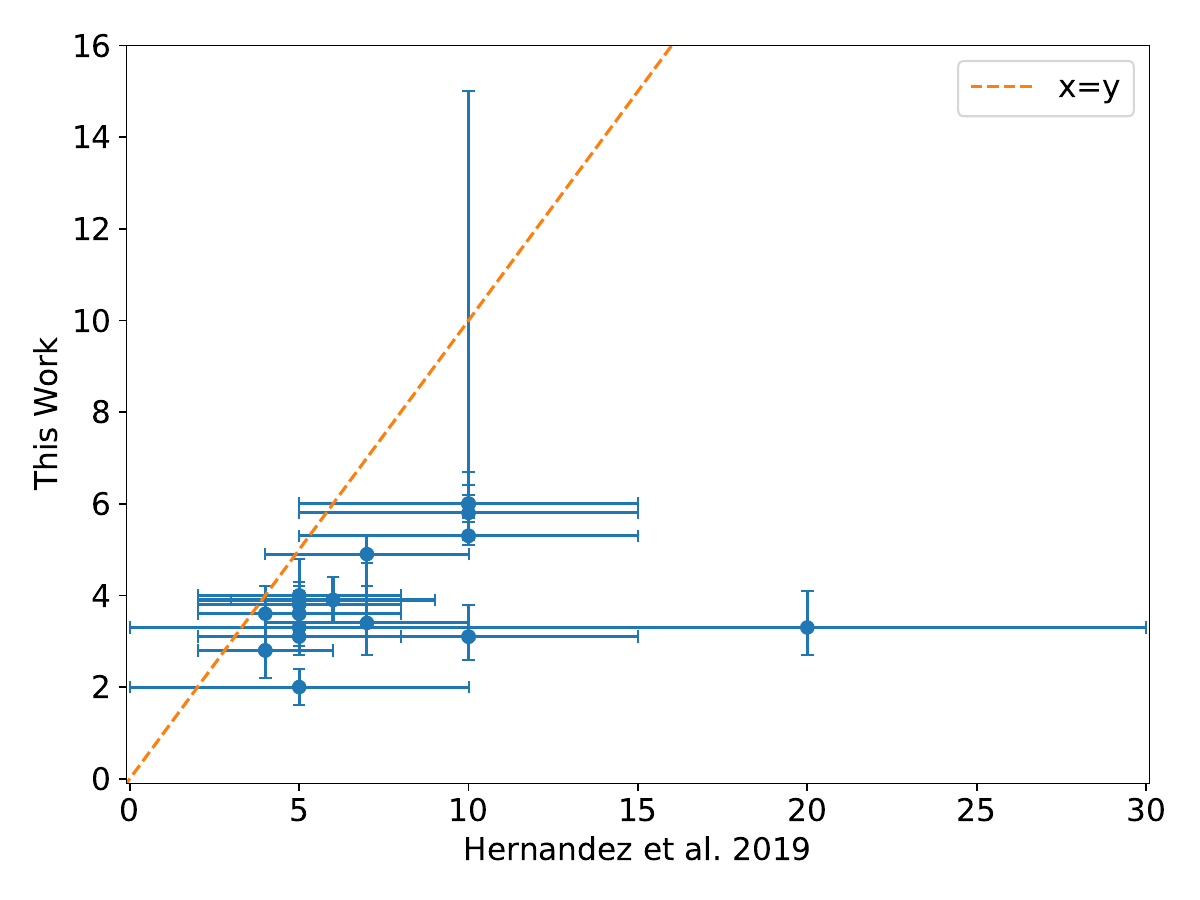}
    \includegraphics[width=0.5\textwidth]{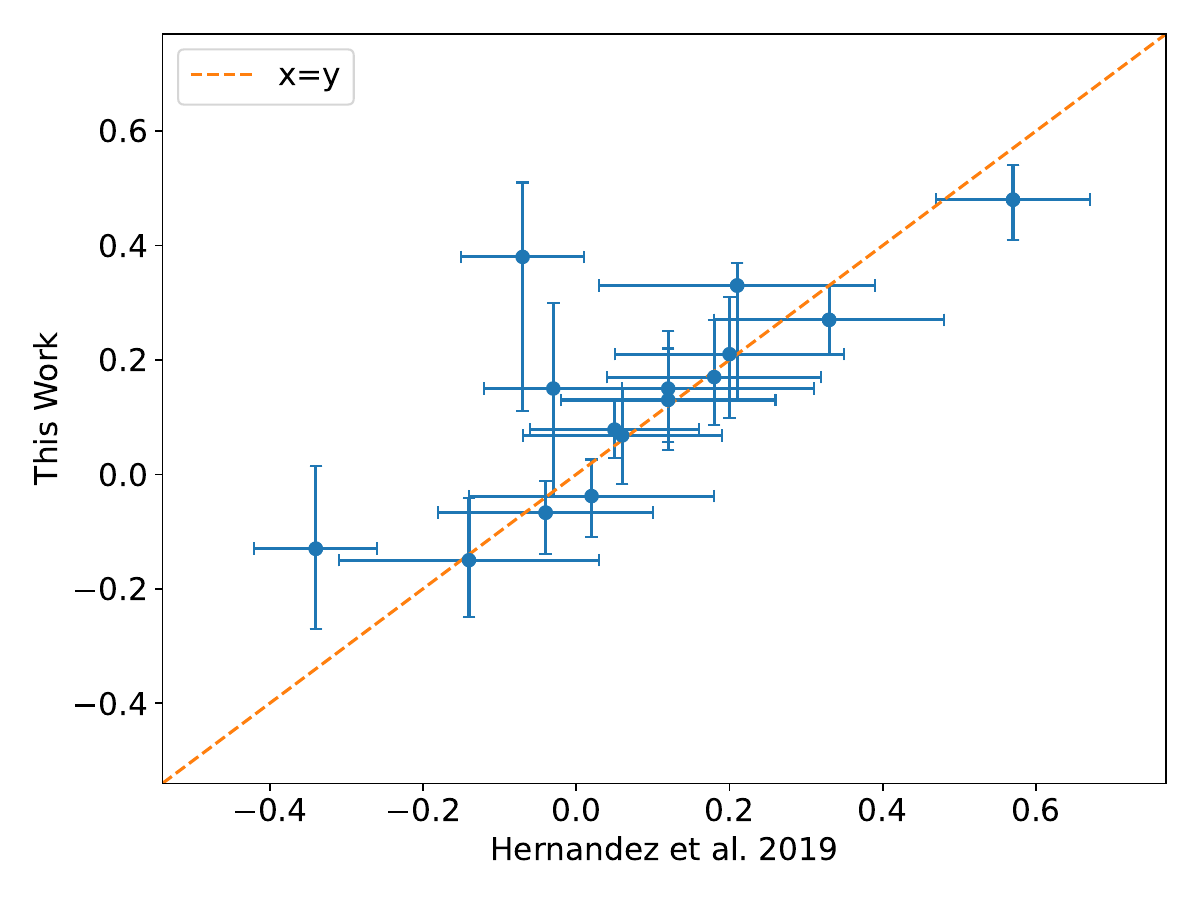}
    \caption{Comparison of fitted stellar ages (top) and metallicity (bottom) for YSCs in M83 using SESSAMME (see subsection.~\ref{SESSAME_intro}) shown on the x-axis against the values from \citet{Hernandez2019}. The red dashed line shows the 1:1 relation. \label{fig_st_fit_comparison}}
\end{figure}

We used the estimates of stellar age and metallicities from \citet{Hernandez2019} to set boundaries on the priors, allowing for a broad range of values: metallicity: -1.1$\rm <\,[Z]\,<$0.6, stellar age: 5.5 $\rm <\,log(Age)\,<$7.5, stellar mass: (4.0 $\rm <\,log(M_{\odot})\,<\,$7.0) and dust: 0.0 $\rm <\,E(B-V)(mag)\,<$1.0). We found good fits for all YSCs except M83-9, M83-11, and M83-15. We fitted these separately (in addition to M83-5, M83-POS-1 and M83-POS-2) using a median stellar age and metallicity estimate from other YSCs as the initial guess. Fig.~\ref{fig_st_fit_comparison} shows the comparison of both stellar ages and metallicity in the YSCs with \citet{Hernandez2019}. We note that the average metallicity derived by \citet{Hernandez2019} is consistent with our results for all YSCs except M83-POS-15, where the deviation is close to $\sim$0.1 dex. We find that, while the average age reported by \citet{Hernandez2019} is older than our estimates, the ages from both studies are consistent within uncertainties, except for M83-9. It is worth noting that the adopted ages in the work of \citet{Hernandez2019} were inferred through photometric techniques which are known to suffer from age-metallicity degeneracies \citep[see also][]{Kaviraj2007}.

\begin{figure}
\includegraphics[width=1.0\linewidth,trim={0 0 0 50},clip]{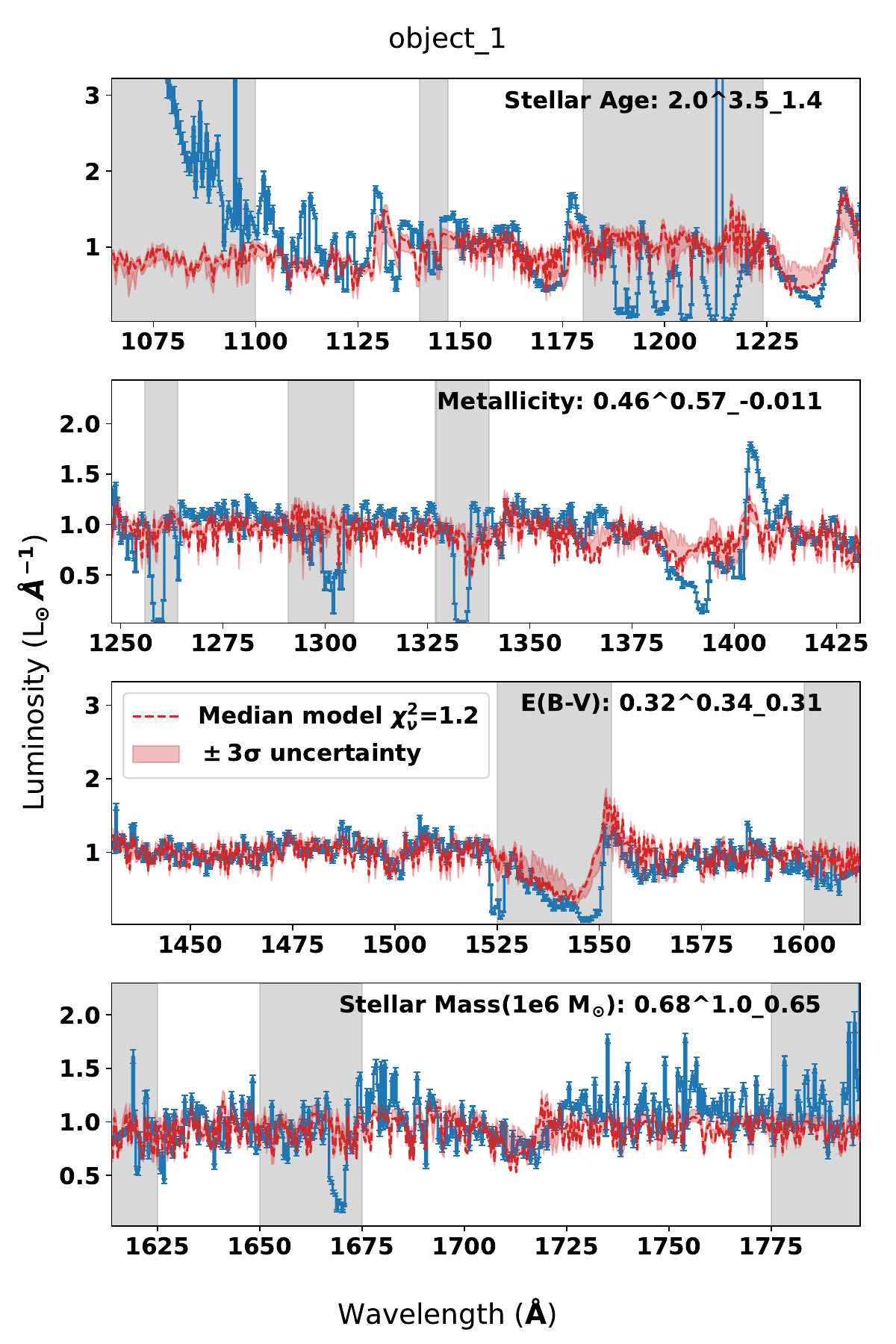}
\caption{HST/COS spectra (binned to fit SB99 models) of M83-POS-1 showed in blue along with best-fit (with $\rm \pm\,3-\sigma$ uncertainty) stellar model showed in red. The best-fit stellar parameters are also shown as legends. Regions shown in grey are masked out due to contamination from prominent nebular emission, ISM absorption features, having instrument-based issues or generally low SNR. \label{fig:4}}
\end{figure}

Figure.~\ref{fig:4} shows an example of the stellar spectral fitting done for the YSC: M83-POS-1. We note that fitting M83-POS-1 and M83-POS-2 is especially difficult given the numerous ISM absorption features within the spectra. The ISM absorption, especially within the stellar p-cygni profiles (e.g. \NV$\lambda\lambda$1238,1242 doublet, \SiIV$\lambda\lambda$1393,1402 doublet, and \CIV$\lambda\lambda$1548,1550 doublet), is very strong. We note that it could be possible that these specific YSC spectra also represent more than one young stellar population. However, since we have robust measurements of metallicity and age, which are both consistent with the spectra (\NV$\lambda$1240) as well as values from the literature, we refrain from doing further multiple-stellar population fitting. We also note that since this paper is focused on the properties of the metal-mixing around the YSCs, such an in-depth analysis for these two spectra specifically is beyond the scope of the study.       

\subsection{Calculating stellar bolometric luminosity and wind momentum from stellar fitting}

Bolometric luminosity (L$_{Bol}$, the total radiant energy output per unit time from a star cluster integrated over all wavelengths) quantifies the total radiative power the stellar population emits. Wind power in massive stars arises from their strong, radiation-driven stellar winds, which carry kinetic energy and momentum into the surrounding medium. Studying both quantities in M83-YSCs, is important. The bolometric luminosity of massive stars sets the radiation field that shapes gas clearing, cluster evolution, and potential feedback-driven outflows on galactic scales. Additionally, higher metallicity enhances radiative driving of stellar winds and can increase mass and energy injection into the interstellar medium. \\

\begin{figure}
    \centering
    \includegraphics[width=0.5\textwidth]{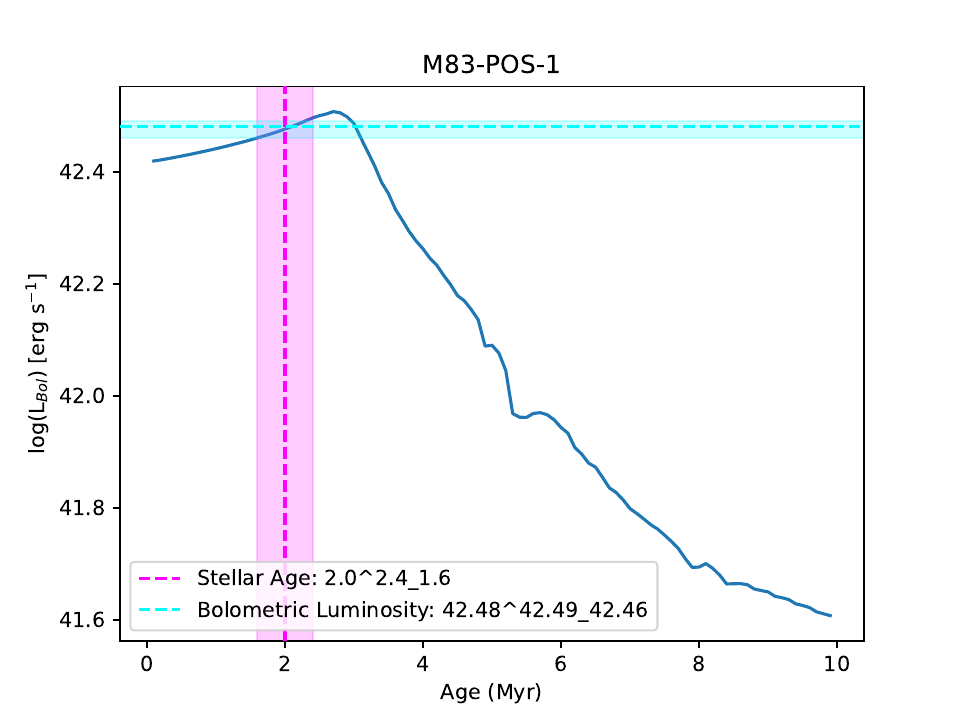}
    \includegraphics[width=0.5\textwidth]{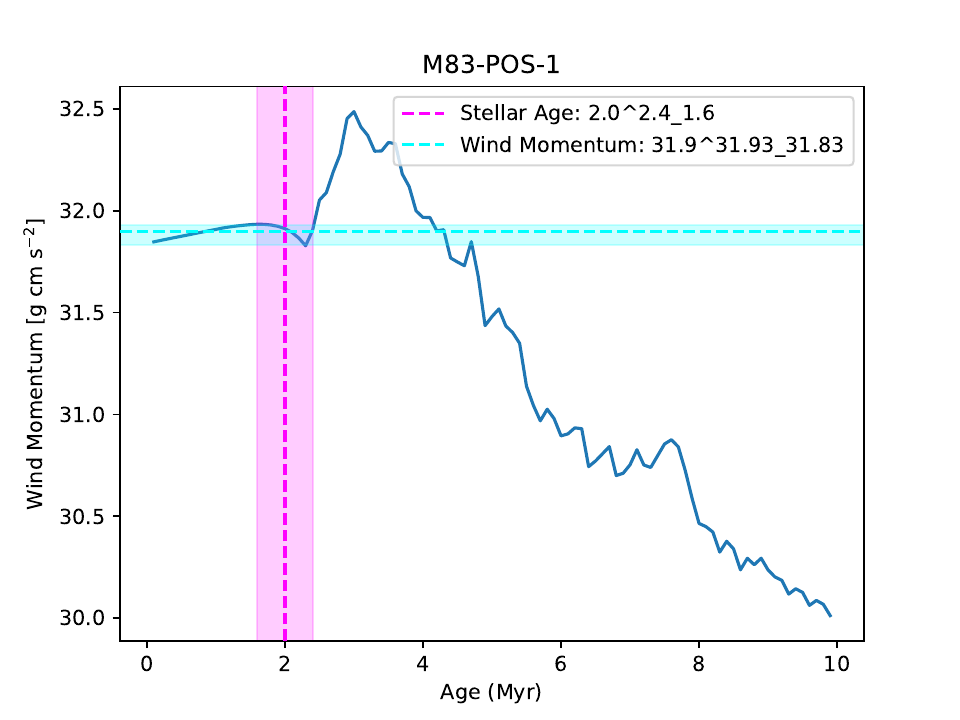}
    \caption{Stellar luminosity and wind momentum derived from the SB99 model with inference of stellar ages from SESAMME UV-spectral fitting of YSC: M83-POS-1. The purple vertical line (and shaded regions) indicate the stellar age of the YSCs (and $\pm\,1-\sigma$ uncertainty). A mask of this region along the y-axis is used to derive both stellar luminosity (top) and wind momentum (bottom). \label{wind_lum_power_fig}}
\end{figure}

We use our fitted stellar ages to obtain the stellar luminosity (L$_{Bol}$ in [erg s$^{-1}$]) and wind momentum injection rate (in [g cm s$^{-2}$]) using the SB99 models. From the inferred ages and their corresponding uncertainties, we determine the predicted range in luminosity and wind momentum from the theoretical SB99 models. Fig.~\ref{wind_lum_power_fig} shows an example of stellar luminosity and wind momentum derived from stellar ages for M83-POS-1. We caution that uncertainties in wind-derived quantities are likely dominated by systematic effects (e.g., wind clumping, ionisation structure), and may be larger than formal fitting uncertainties.  

\section{Constraining temperature for ionised gas surrounding M83 star-clusters \label{temperature_fitting}}

Determining electron temperatures in metal-rich \HII\ regions is inherently difficult because metals efficiently cool the ionised gas. As metallicity increases, cooling through forbidden metal lines lowers the electron temperature, causing the temperature-sensitive auroral lines used for direct $T_e$ measurements to become progressively weaker \citep[see][]{Stasinska2002}. Since abundances derived from collisionally excited lines depend strongly on the adopted electron temperature, reliable $T_e$ estimates remain essential for accurate abundance determinations, even though the auroral lines required to measure them become increasingly inaccessible in high-metallicity environments. \\

To overcome the difficulty of measuring auroral lines in metal-rich environments, several studies have developed strong-line methods that estimate chemical abundances without requiring direct electron-temperature measurements. For example, the latest version of HII-CHI-mistry employs grids of photoionisation models to derive oxygen and sulphur abundances from strong optical emission lines, providing results consistent with the direct method even when auroral lines are unavailable \citep[see][]{Perez-Montero2025}. While such approaches are valuable for estimating global abundances, they are not suitable for this work, since our analysis requires independent determinations of the electron temperatures associated with multiple ionisation zones, specifically $T_e$(\OII), $T_e$(\OIII), and $T_e$(\SIII), to derive self-consistent multiphase elemental abundances. We therefore avoid relying on empirical temperature-conversion relations. Although relations between different ionic temperatures have been extensively explored in both theoretical and observational studies \citep[see e.g.,][]{Garnett1992, Izotov2006b, Rickards_Vaught2024, Scholte2026}, the adopted prescriptions differ among authors and can exhibit significant intrinsic scatter. To minimise systematic uncertainties arising from temperature transformations, we proceed to instead derive individual $T_e$s for different phases. \\

To construct temperature estimates appropriate for our analysis, we first compile literature measurements of nearby star-forming \HII\ regions with available emission-line intensities, electron temperatures, densities, and direct elemental abundances, focusing on the metal-rich samples of \citet{Bresolin2002, Bresolin2005, Bresolin2009}. For each \HII\ region in the literature, we compute a suite of strong-line diagnostics and compare them separately against the measured $T_e$(\OII), $T_e$(\OIII), and $T_e$(\SIII), restricting the calibration to the high-metallicity regime relevant for M83: 12+log(O/H)$>$8.0. We first fit a linear relation between each diagnostic and each ionic temperature over this limited metallicity range. This initial fit is then used to identify and remove outliers through iterative sigma clipping, where residuals are normalised by the combined uncertainty from both the temperature and diagnostic axes. To prevent densely populated regions of diagnostic space from dominating the fit, we additionally apply inverse-density weighting in the diagnostic variable. The final calibration is obtained by refitting only the retained points after clipping, yielding a more robust empirical estimate of $T_e$ in each phase. Finally, we derive direct elemental abundances and assess, for each predicted temperature zone, which strong-line calibrator produces abundances closest to the one-to-one direct-abundance scale. \\

As an initial test, we applied our methodology to the suite of strong-line metallicity diagnostics summarised by \citet{Kewley2019}. These diagnostics span a broad range of empirical, theoretical, and hybrid calibrations and therefore provide a useful benchmark for evaluating the robustness of the temperature-abundance relations. For the literature \HII\ regions, the resulting abundances generally reproduce the direct-method values with good agreement around the one-to-one relation. However, the inferred electron temperatures exhibit substantial scatter between different \HII\ regions within a specific strong-line prescription. When propagated through the direct abundance calculations, these temperature variations produce large uncertainties in the derived elemental abundances, limiting the predictive power of the calibrations. We therefore explored an alternative approach based on the recently developed DESIRED strong-line calibrations \citep[][]{Rosales-Ortega2026}. The DESIRED project was specifically designed to provide empirical metallicity relations anchored to the largest homogeneous compilation of direct-method measurements currently available, comprising more than $\sim$2300 \HII\ regions and star-forming galaxies with electron-temperature determinations. In contrast to many previous calibrations, the DESIRED relations were explicitly optimised to reproduce direct-method oxygen abundances (within $\sim$0.15-0.35 dex uncertainty), including temperature inhomogeneities over a wide metallicity range, using a large set of commonly observed strong-line diagnostics (see their table.~2). This feature is particularly advantageous for the present study, as our primary goal is not simply to estimate metallicity, but rather to establish empirical relations between strong-line diagnostics, electron temperatures, and direct elemental abundances. By construction, the DESIRED calibrations provide strong-line metallicity estimates that remain closely tied to the direct-method abundance scale, making them an ideal framework for identifying the temperature diagnostics that best reproduce the multiphase abundance structure of metal-rich \HII\ regions.\\

\begin{figure}
    \centering
    \includegraphics[width=0.4\textwidth]{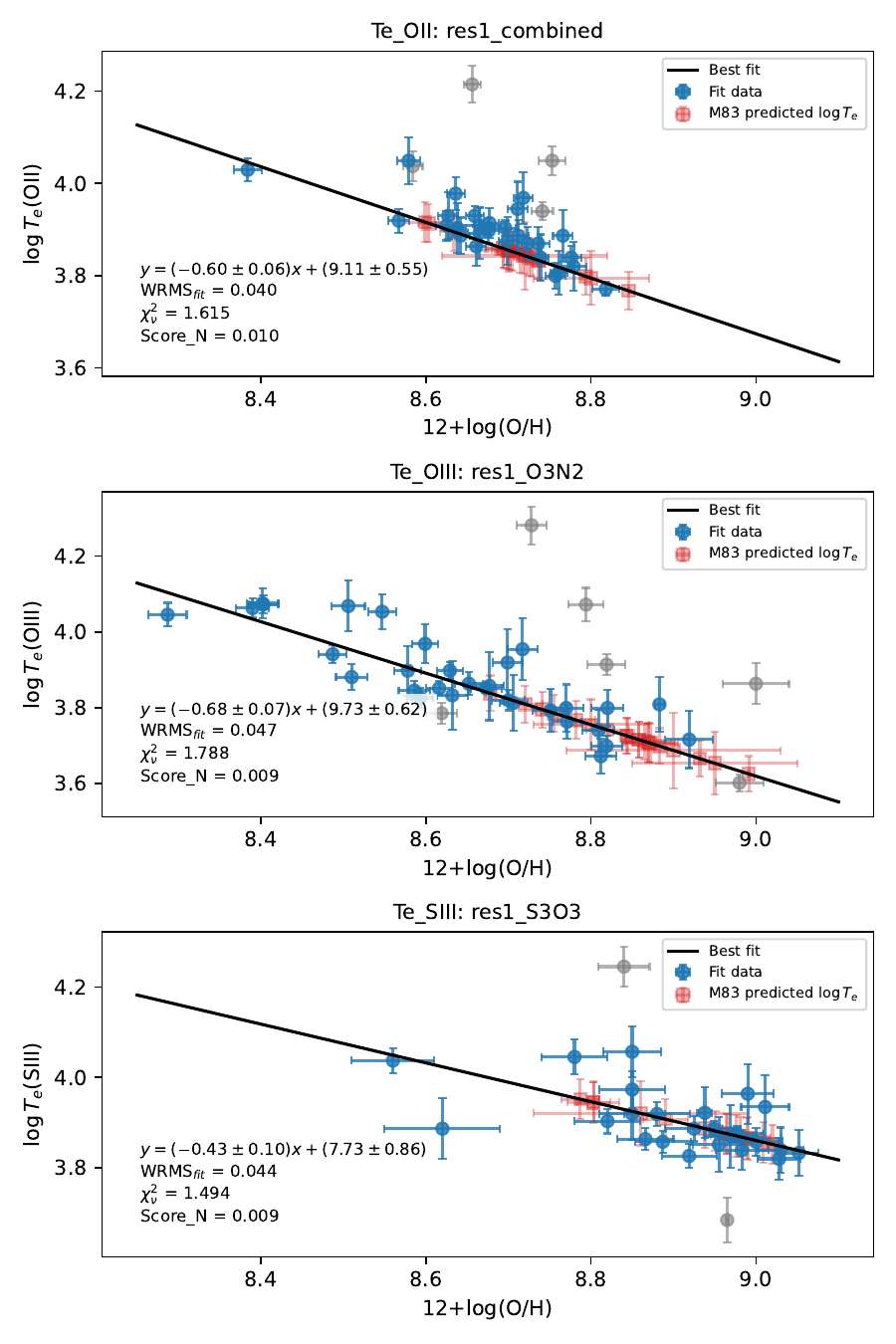}
    \caption{Strong-line method abundances (Top to bottom: combined(see text), O3N2, S3O3) plotted as a function of electron temperatures (Top to bottom, $T_e$: \OII, \OIII, and \SIII). The fitted data from \HII\ regions in SF galaxies (see text for details), are shown in blue, while the outlier points excluded from the line fit are greyed out. The `res\_1' label indicates that we are using the `upper' version suitable for the metal-rich regime for each calibrator (see table.~4 or \citet{Rosales-Ortega2026} for reference). The red points show the predicted $T_e$ obtained from the fit for each M83 cluster (MUSE+LBT) studied in this work. \label{DESIRED_on_lit_Te}}
\end{figure}

Among the various strong-line metallicity diagnostics tested, we find that the highest-ionisation temperature tracers exhibit the tightest and most well-behaved relations. In particular, $T_e$(\OIII) shows a strong linear correlation with the O3N2 diagnostic, while $T_e$(\SIII) is most tightly correlated with the S3O3 diagnostic (See Fig.~\ref{DESIRED_on_lit_Te}). The best-fitting relations are - 

\begin{equation}
   [T_e({\rm O,III}) = (0.68 \pm 0.07),{\rm O3N2} + (9.73 \pm 0.62),] 
\end{equation}

and

\begin{equation}
    [T_e({\rm S,III}) = (0.43 \pm 0.10),{\rm S3O3} + (7.73 \pm 0.86).]
\end{equation}

To quantitatively compare the performance of different strong-line diagnostics, we evaluate each fit using three complementary statistical metrics. The weighted root-mean-square residual, WRMS$_{\rm fit}$, measures the scatter of the data around the best-fitting relation while accounting for the observational uncertainties; smaller values indicate tighter correlations and therefore greater predictive power. The reduced $\chi^2$ statistic evaluates whether the observed scatter is consistent with the quoted measurement uncertainties, with values close to unity indicating an acceptable fit, values significantly larger than unity suggesting either intrinsic scatter or underestimated uncertainties, and values substantially below unity indicating potentially overestimated errors. Finally, we define a normalised ranking statistic, Score$\rm \_N$, 

\begin{equation}
{\rm Score}_{N} =
\frac{\sqrt{\overline{\Delta}^{\,2} + {\rm WRMS}_{\rm fit}^{\,2}}}{\sqrt{N}},
\end{equation}

where $\overline{\Delta}$ is the mean residual of the fit, WRMS$_{\rm fit}$ is the weighted root-mean-square residual, and $N$ is the number of calibration points. This metric combines the goodness-of-fit and residual scatter into a single metric for comparing different diagnostics. Lower values of Score$\rm \_N$ correspond to calibrators that simultaneously minimise scatter and provide statistically consistent fits, allowing the most reliable temperature predictors to be identified objectively. Both relations exhibit low weighted residual scatter (WRMS$_{\rm fit}$ = 0.047 and 0.044, respectively), reduced $\chi^2$ values close to unity (1.79 and 1.49), and among the lowest normalised scores of all tested calibrators. These results indicate that O3N2 and S3O3 provide robust empirical predictors of the temperatures associated with the high-ionisation zones of metal-rich \HII\ regions. \\

The low-ionisation temperature $T_e$(\OII) proved considerably more challenging to calibrate. No individual strong-line diagnostic produced a relation with scatter comparable to those obtained for $T_e$(\OIII) and $T_e$(\SIII). We therefore adopted a combined metallicity estimator based on the DESIRED prescriptions, where the oxygen abundance is calculated independently from all available diagnostics and then combined through an inverse-variance weighted average using the intrinsic calibration dispersions reported by \citet{Rosales-Ortega2026}. This approach effectively reduces the influence of any individual diagnostic and provides a more stable estimate of the underlying metallicity. Using this combined calibrator, we obtain a significantly improved relation - 

\begin{equation}
    [T_e({\rm O,II}) = (0.60 \pm 0.06),{\rm Combined} + (9.11 \pm 0.55),]
\end{equation}

with WRMS$_{\rm fit}$ = 0.040, reduced $\chi^2$ = 1.62, and Score$\rm \_N$ = 0.010. The improvement relative to the individual diagnostics suggests that the low-ionisation temperature is more sensitive to the systematic differences among strong-line calibrations and benefits from averaging over multiple abundance indicators.

\begin{figure}
    \centering
    \includegraphics[width=0.4\textwidth]{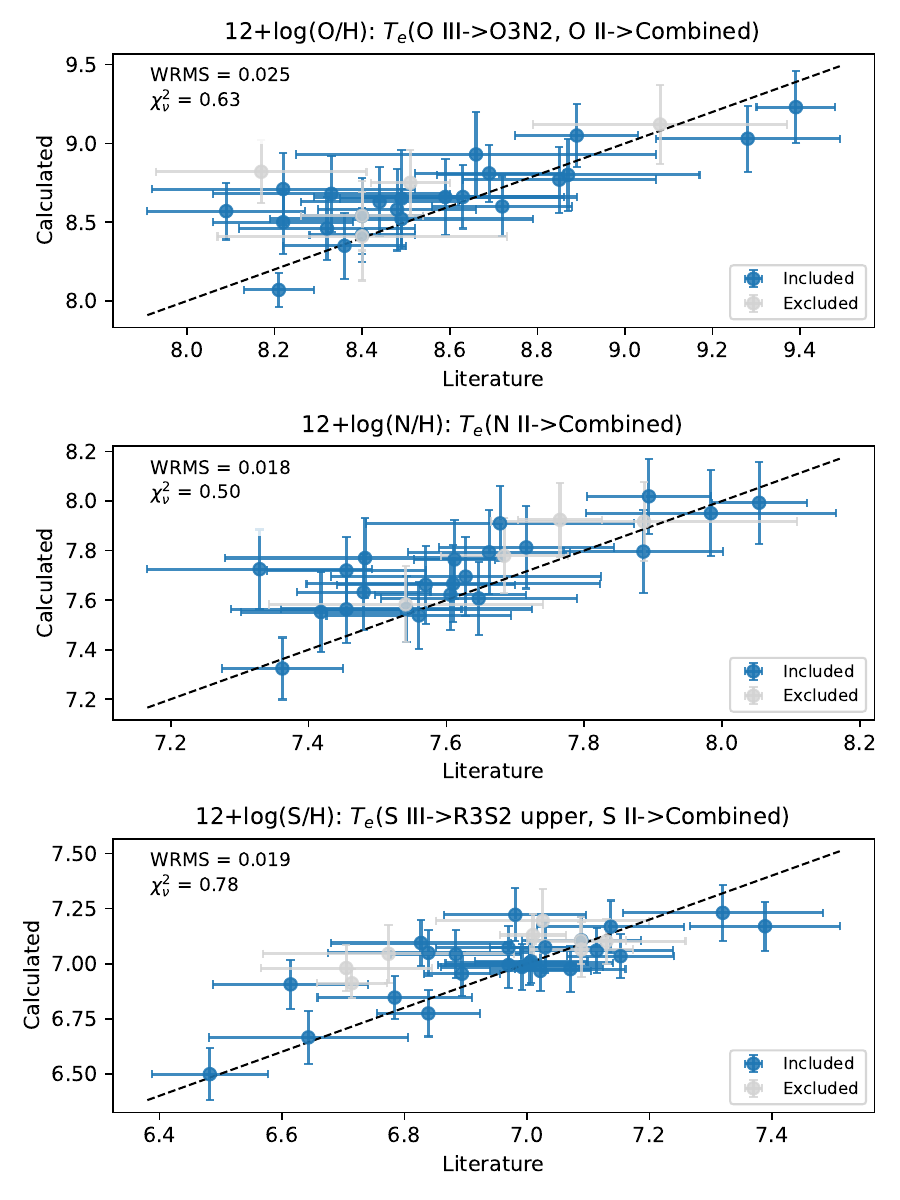}
    \caption{Comparing direct-method elemental abundances (12+log(X/H), where X=O,N,S from top to bottom) of \HII\ regions in star-forming galaxies. The x-axis shows the abundances from the literature, and the y-axis is the abundance calculated using predicted multi-phase $T_e$. The greyed-out points were excluded from the original $T_e$-fit and shown here separately for reference. \label{DESIRED_on_lit_abund}}
\end{figure}

To assess the performance of the derived temperature relations, we applied the predicted values of $T_e$(\OII), $T_e$(\OIII), and $T_e$(\SIII) to all literature \HII\ regions and recalculated their elemental abundances using the direct method. These abundances were then compared to the corresponding direct-method abundances reported in the literature (See Figure. ~\ref{DESIRED_on_lit_abund}). For oxygen, we find excellent agreement, with a weighted root-mean-square residual of WRMS = 0.026 dex and a reduced $\chi^2 = 0.68$ relative to the one-to-one relation. Similarly, nitrogen abundances are recovered with WRMS = 0.020 dex and reduced $\chi^2 = 0.58$, while sulphur abundances show WRMS = 0.019 dex and reduced $\chi^2 = 0.78$. We emphasise that these statistics are not derived from a fitted relation; rather, they quantify the deviations of the abundances predicted using our temperature prescriptions from the direct-method abundances along the ideal one-to-one line. The small residual scatter and reduced $\chi^2$ values close to unity demonstrate that the empirically predicted temperatures successfully reproduce the abundance scale of the original direct-method measurements across the literature sample.

These three relations are subsequently adopted to predict $T_e$(\OII), $T_e$(\OIII), and $T_e$(\SIII) for the M83 \HII\ regions (See Figure.~\ref{DESIRED_on_lit_Te}). For regions where direct auroral-line measurements are available, such as [\SIII] $\lambda6312$ or [\NII] $\lambda5755$, the corresponding directly measured temperatures are used in the abundance calculations. Otherwise, the temperatures are inferred from the empirical strong-line relations presented above. This procedure allows us to derive a self-consistent set of ionic temperatures for the multiphase abundance analysis while remaining anchored to the direct-method temperature scale.

\section{ICF corrections for ionised gas abundances \label{hii_icf_estimation}}

\begin{figure}
    \centering
    \includegraphics[width=0.4\textwidth]{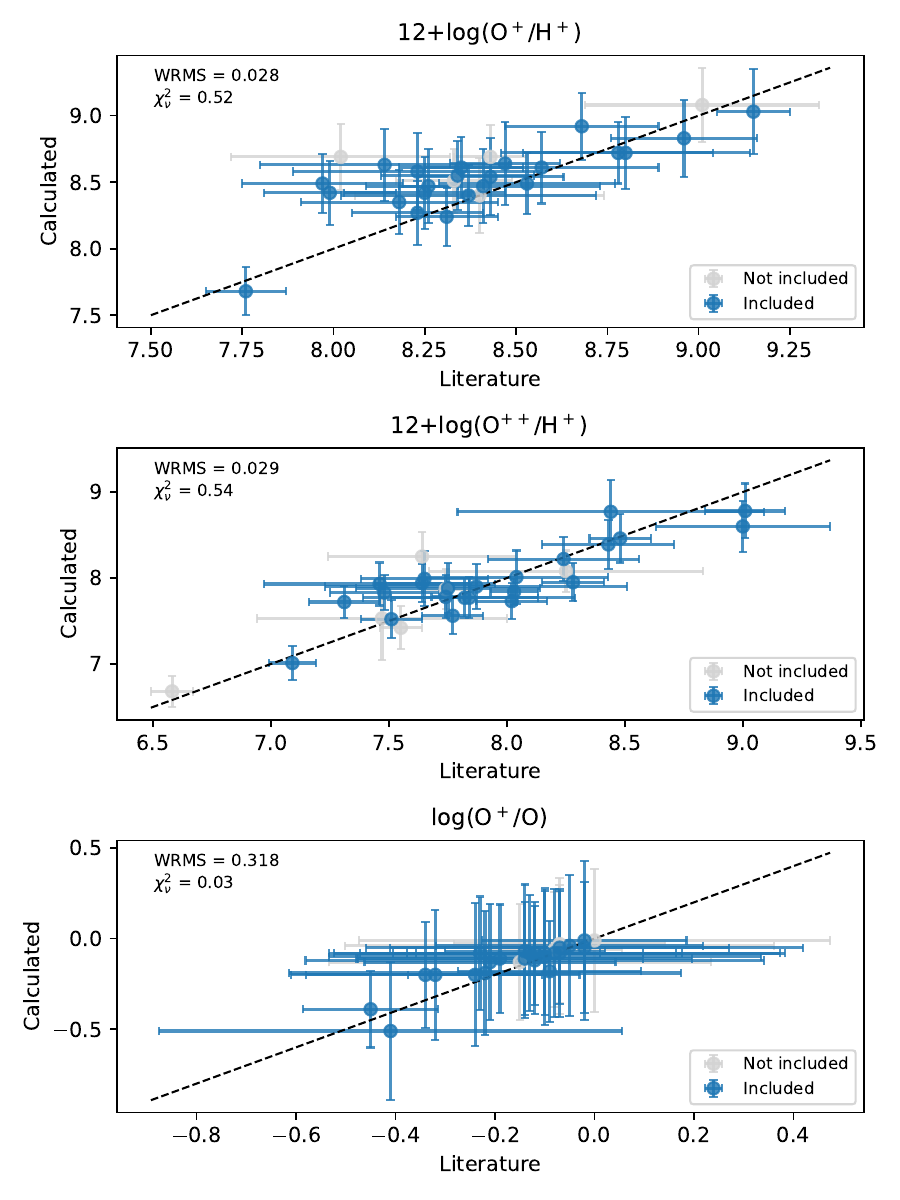}
    \caption{Comparing ionic oxygen abundances: 12+log(O$^+$/H$^+$), 12+log(O$^{++}$/H$^+$) and ratio: log(O$^+$/O) derived using temperature in literature (x-axis) against the predicted temperatures from the method explained in Section~\ref {temperature_fitting} for \HII\ regions in literature. The greyed-out points indicate that those were excluded from the line fit that led to the temperature predictions. \label{DESIRED_ionic_oxygen_lit_abund}}
\end{figure}

The ionic abundances derived from the observed emission lines do not necessarily represent the total elemental abundances of an \HII\ region, since a fraction of each element may reside in unobserved ionisation stages. Ionisation correction factors (ICFs) are therefore required to convert ionic abundances into total elemental abundances. In this work, ICF corrections were applied to nitrogen, sulphur, and iron. Numerous prescriptions have been proposed in the literature, including the photoionisation-model-based formulations of \citet{Izotov2006b}, the iron-specific corrections of \citet{Rodriguez_Rubin_2005}, and the more recent prescriptions of \citet{Amayo2021}. These ICFs are generally parameterised as functions of the ionic oxygen abundances, O$^+$/H$^+$ and O$^{++}$/H$^+$, or equivalently the ionisation degree O$^+$/O, and in some cases also depend on the metallicity regime being considered \citep[see e.g.,][]{Izotov2006b}. Since our ICF calculations rely directly on these quantities, it is important to verify that they can be accurately recovered using the predicted temperatures. We therefore compared the oxygen ionic abundances derived using the predicted temperatures against those obtained from temperatures in the literature (See Figure~\ref {DESIRED_ionic_oxygen_lit_abund}). We find excellent one-to-one agreement, with WRMS = 0.028 dex and reduced $\chi^2 = 0.52$ for 12+log(O$^+$/H$^+$), WRMS = 0.029 dex and reduced $\chi^2 = 0.54$ for 12+log(O$^{++}$/H$^+$), and WRMS = 0.318 dex with reduced $\chi^2 = 0.03$ for log(O$^+$/O). These results demonstrate that the ionic quantities required by the various ICF prescriptions are robustly reproduced by our temperature calibration framework, providing confidence in the subsequent elemental abundance determinations.

\begin{figure}
    \centering
    \includegraphics[width=0.4\textwidth]{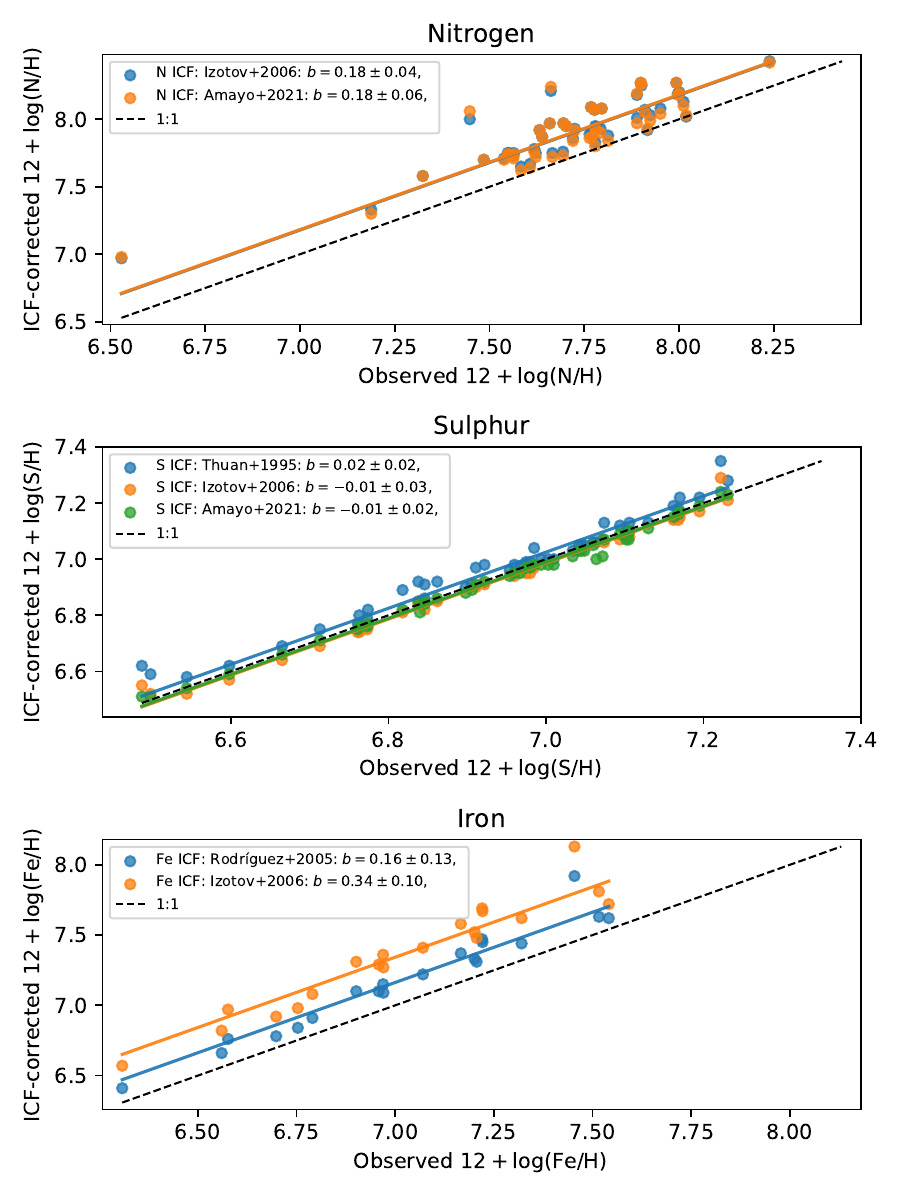}
    \caption{Plotting the original observed elemental abundance vs the ICF-corrected elemental abundances for Nitrogen, Sulphur, and Iron for all M83 \HII\ regions studied in this work (limits+detections). The uncertainties are removed for clarity. \label{DESIRED_icf_correction_m83}}
\end{figure}

To further assess the systematic impact of different ionisation-correction prescriptions, we computed elemental abundances using several commonly adopted ICF formulations from the literature for all \HII\ region points in M83 (See Figure.~\ref{DESIRED_icf_correction_m83}). For Iron, we compared the prescriptions of \citet{Rodriguez_Rubin_2005} and \citet{Izotov2006b}, which produce mean corrections of $\sim$ 0.2 dex and 0.3 dex, respectively, for all the M83 \HII\ regions studied in this work. For sulphur, we considered the prescriptions of \citet{Izotov2006b}, \citet{Thuan1995}, and \citet{Amayo2021}, yielding typical corrections of $\sim$0.01, 0.02, and 0.01 dex. For nitrogen, we compared the prescriptions of \citet{Izotov2006b} and \citet{Amayo2021}, both of which result in typical corrections of $\sim$0.2 dex.

A detailed investigation of the physical differences between these ICF schemes is beyond the scope of this work. We therefore provide the abundances obtained using all individual prescriptions in Table~\ref{tab:emission_abundance_icf}, allowing future studies to assess the systematic uncertainties associated with the choice of ICF. For consistency throughout the remainder of this paper, however, we adopt the \citet{Izotov2006b} corrections, which provide a homogeneous set of prescriptions for multiple elements and are widely used in abundance studies of ionised nebulae.

\begin{sidewaystable*}[]
\flushleft
{\setlength{\tabcolsep}{1pt}} % apply only inside this group
\begin{tabular}{lllllllllll}
\hline
YSC       & (O/H)           & (Fe/H)          & (Fe/H)          & (Fe/H)          & (S/H)           & (S/H)           & (S/H)           & (S/H)           & (N/H)          & (N/H)        \\
\hline
          & obs        & obs        & ICF\_i06b       & ICF\_r05        & obs        & ICF\_i06b       & ICF\_t95        & ICF\_a21        & obs       & ICF\_i06b    \\
\hline
\hline
M83-1     & 8.70$\rm \pm$0.63*    & 7.201$\rm \pm$0.238   & 7.52$\rm \pm$0.30     & 7.33$\rm \pm$0.45     & 6.960$\rm \pm$0.216*  & 6.94$\rm \pm$0.22*    & 6.98$\rm \pm$0.26*    & 6.95$\rm \pm$0.22*    & 7.485$\rm \pm$0.141  & 7.70$\rm \pm$0.23  \\
M83-2     & $\rm <$9.30 & 7.541$\rm \pm$0.282   & 7.72$\rm \pm$0.29     & 7.62$\rm \pm$0.48     & $\rm <$6.94 & $\rm <$6.94 & $\rm <$6.94 & $\rm <$6.92 & 7.694$\rm \pm$0.156  & 7.76$\rm \pm$0.18  \\
M83-3     & 8.35$\rm \pm$0.26*    & 6.970$\rm \pm$0.223   & 7.27$\rm \pm$0.28     & 7.09$\rm \pm$0.45     & 7.195$\rm \pm$0.145   & 7.17$\rm \pm$0.15     & 7.22$\rm \pm$0.16     & 7.19$\rm \pm$0.15     & 8.001$\rm \pm$0.163  & 8.20$\rm \pm$0.24  \\
M83-4     & 8.35$\rm \pm$0.26*    & 6.969$\rm \pm$0.218   & 7.36$\rm \pm$0.31     & 7.15$\rm \pm$0.49     & 6.862$\rm \pm$0.100   & 6.85$\rm \pm$0.10     & 6.92$\rm \pm$0.12     & 6.86$\rm \pm$0.10     & 7.889$\rm \pm$0.144  & 8.18$\rm \pm$0.27  \\
M83-5     & 8.35$\rm \pm$0.26*    & 7.221$\rm \pm$0.215   & 7.67$\rm \pm$0.35     & 7.45$\rm \pm$0.51     & 6.818$\rm \pm$0.101   & 6.81$\rm \pm$0.12     & 6.89$\rm \pm$0.12     & 6.82$\rm \pm$0.11     & 7.901$\rm \pm$0.151  & 8.25$\rm \pm$0.30  \\
M83-6     & 8.61$\rm \pm$0.31*    & $\rm <$7.64 & $\rm <$8.53 & $\rm <$8.38 & 6.487$\rm \pm$0.089   & 6.55$\rm \pm$0.14     & 6.62$\rm \pm$0.15     & 6.51$\rm \pm$0.09     & 7.448$\rm \pm$0.140  & 8.00$\rm \pm$0.35  \\
M83-7     & 8.44$\rm \pm$0.33     & 6.576$\rm \pm$0.207   & 6.97$\rm \pm$0.32     & 6.76$\rm \pm$0.49     & 6.911$\rm \pm$0.066   & 6.90$\rm \pm$0.07     & 6.97$\rm \pm$0.10     & 6.91$\rm \pm$0.07     & 7.777$\rm \pm$0.146  & 8.07$\rm \pm$0.28  \\
M83-8*    & 8.79$\rm \pm$0.25*    & 6.753$\rm \pm$0.370*  & 6.98$\rm \pm$0.38*    & 6.84$\rm \pm$0.46*    & 7.045$\rm \pm$0.131*  & 7.045$\rm \pm$0.131*  & 7.05$\rm \pm$0.13*    & 7.03$\rm \pm$0.13*    & 8.012$\rm \pm$0.181* & 8.13$\rm \pm$0.21* \\
M83-9*    & 8.66$\rm \pm$0.22*    & 6.308$\rm \pm$0.429*  & 6.57$\rm \pm$0.44*    & 6.41$\rm \pm$0.49*    & 6.980$\rm \pm$0.105*  & 6.95$\rm \pm$0.11*    & 6.99$\rm \pm$0.11*    & 6.97$\rm \pm$0.11*    & 7.782$\rm \pm$0.142* & 7.94$\rm \pm$0.18* \\
M83-10    & 8.51$\rm \pm$0.36     & $\rm <$6.98 & $\rm <$7.27 & $\rm <$7.25 & 6.761$\rm \pm$0.098   & 6.74$\rm \pm$0.10     & 6.77$\rm \pm$0.11     & 6.75$\rm \pm$0.10     & 7.618$\rm \pm$0.145  & 7.78$\rm \pm$0.22  \\
M83-11    & $\rm <$8.79 & $\rm <$7.54 & $\rm <$7.88 & $\rm <$7.85 & 6.598$\rm \pm$0.097   & 6.57$\rm \pm$0.10     & 6.62$\rm \pm$0.11     & 6.59$\rm \pm$0.10     & 7.548$\rm \pm$0.133  & 7.75$\rm \pm$0.20  \\
M83-12    & $\rm <$8.84 & 6.958$\rm \pm$0.324   & 7.29$\rm \pm$0.40     & 7.10$\rm \pm$0.60     & 6.713$\rm \pm$0.115   & 6.69$\rm \pm$0.12     & 6.75$\rm \pm$0.14     & 6.71$\rm \pm$0.12     & 7.639$\rm \pm$0.163  & 7.87$\rm \pm$0.29  \\
M83-13    & --              & --              & --              & --              & --              & --              & --              & --              & --             & --           \\
M83-14    & 8.43$\rm \pm$0.37     & 7.070$\rm \pm$0.257   & 7.41$\rm \pm$0.34     & 7.22$\rm \pm$0.54     & 6.763$\rm \pm$0.108   & 6.74$\rm \pm$0.11     & 6.80$\rm \pm$0.13     & 6.76$\rm \pm$0.11     & 7.702$\rm \pm$0.166  & 7.95$\rm \pm$0.29  \\
M83-15    & --              & --              & --              & --              & --              & --              & --              & --              & --             & --           \\
M83-16    & $\rm <$8.71 & 7.166$\rm \pm$0.268   & 7.58$\rm \pm$0.41     & 7.37$\rm \pm$0.63     & 6.846$\rm \pm$0.125   & 6.84$\rm \pm$0.13     & 6.91$\rm \pm$0.15     & 6.85$\rm \pm$0.13     & 7.768$\rm \pm$0.156  & 8.09$\rm \pm$0.34  \\
M83-POS-1 & $\rm <$9.40 & 6.790$\rm \pm$0.258   & 7.08$\rm \pm$0.35     & 6.91$\rm \pm$0.63     & 7.162$\rm \pm$0.107   & 7.14$\rm \pm$0.11     & 7.19$\rm \pm$0.13     & 7.15$\rm \pm$0.11     & 8.239$\rm \pm$0.180  & 8.43$\rm \pm$0.31  \\
M83-POS-2 & $\rm <$8.64 & 7.220$\rm \pm$0.201   & 7.69$\rm \pm$0.32     & 7.47$\rm \pm$0.44     & 6.838$\rm \pm$0.114   & 6.84$\rm \pm$0.13     & 6.92$\rm \pm$0.13     & 6.85$\rm \pm$0.12     & 7.899$\rm \pm$0.141  & 8.27$\rm \pm$0.27  \\
\hline
\end{tabular}
\caption{Elemental abundance, (X/H)$\rm _{HII}$ = 12+log(X/H) of the element, X, in \HII\ regions (ionised gas) around YSCs in M83. Most measurements use spectra from VLT/MUSE. We use VLT/MUSE spectra throughout, other than those indicated by an asterisk(*), for which we use LBT/MODS measurements. The elemental abundances are given for observed: `obs' and ionisation corrected: `ICF\_', with correction IDs representing ICF correction from different works in the literature: 1. t95-\citet{Thuan1995}, 2. r05-\citet{Rodriguez_Rubin_2005}, 3. i06b-\citet{Izotov2006b}, 4. a21-\citet{Amayo2021}. \label{tab:emission_abundance_icf}}
\end{sidewaystable*}

\bibliography{bibliography}{}
\bibliographystyle{aasjournalv7}



\end{document}